\RequirePackage{luatex85}
\documentclass[12pt]{article}
\usepackage{a4wide}
\usepackage[centertags]{amsmath}
\usepackage{amssymb}
\usepackage[sort&compress,numbers]{natbib}
\usepackage{ifpdf}
\usepackage{setspace}
\usepackage{amsfonts}
\usepackage{color}
\usepackage{threeparttable}
\usepackage{cancel}
\usepackage{tikz}
\usetikzlibrary{decorations.pathmorphing}
\usepackage{feynmf}
\usepackage[compat=1.1.0]{tikz-feynman}
\tikzfeynmanset{compat=1.1.0}

\usepackage{lscape}

\usepackage{blindtext, rotating}

\usepackage{yfonts}

\usetikzlibrary{decorations.markings}
\ifpdf
\usepackage[colorlinks=true,
linkcolor=black,
citecolor=black,
urlcolor=blue,
filecolor=blue,
pdfstartview=FitV,
pdftitle={},
pdfauthor={},
pdfsubject={Hydrodynamics},
pdfkeywords={hydrodynamics, AdS/CFT},
pdfpagemode=None,
bookmarksopen=true
]{hyperref}
\else
\usepackage[hypertex]{hyperref}
\fi
\usepackage{rotating}
\usepackage{rotating}
\makeatletter
\@addtoreset{equation}{section}

\newcommand{\qn}{{\mathfrak{q}}}
\newcommand{\wn}{{\mathfrak{w}}}

\makeatletter
\renewcommand\section{\@startsection {section}{1}{\z@}%
                                   {-3.5ex \@plus -1ex \@minus -.2ex}
                                   {2.3ex \@plus.2ex}%
                                   {\normalfont\large\bfseries}}
\renewcommand\subsection{\@startsection{subsection}{2}{\z@}%
                                     {-3.25ex\@plus -1ex \@minus -.2ex}%
                                     {1.5ex \@plus .2ex}%
                                     {\normalfont\bfseries}}

\renewcommand{\Re}{\mathfrak{Re}}

\def\sec#1{\S \;\ref{#1}}

\title{{Long-time tails  in the SYK chain  from the effective field theory with a large number of derivatives}}
\author{Navid Abbasi\footnote{abbasi@lzu.edu.cn} \\
	\small{\emph{}}\\
	\small{\emph{School of Nuclear Science and Technology, Lanzhou University,}}\\
	\small{\emph	{ 
			222 South Tianshui Road, Lanzhou 730000, China }} \\
	\small{\emph{}}\\
}

\begin{document}

\setlength{\baselineskip}{16pt}
\begin{titlepage}
\maketitle

\vspace{-36pt}

\begin{abstract}
We study the nonlinear energy diffusion through the  SYK chain in the framework of Schwinger-Keldysh effective field theory. We analytically construct the interacting effective Lagrangian up to $40^{th}$ order in the derivative expansion. According to this effective Lagrangian, we calculate the first order loop correction of the energy density response function, the pole of which is the dispersion relation of energy diffusion. As expected, we see that the standard derivative expansion of that dispersion relation,   $\omega=-i D_{(1)} k^2- i D_{(2)} k^4+\mathcal{O}(k^6)$, breaks down  due to the long-time tails. However, we find that the nonlinear contribution of order  $n$ to the self-energy  is proportional to $\left(k^{2}\right)^{n+1/2}$. This suggests to modify the dispersion relation by splitting it into two dispersion relations and double the number of transport coefficients at any order as $\omega=-i k^2\big( D_{(1,1)}\pm i D_{(1,2)} \left(k^2\right)^{1/2}\big)-i k^4\big( D_{(2,1)}\pm i D_{(2,2)} \left(k^2\right)^{1/2}\big)+\mathcal{O}(k^6)$. We find that the modified series, which include the effect of long-time tails, are convergent. The radius of convergence is proportional to the ratio of thermal conductivity to diffusion constant.

  \end{abstract}
\thispagestyle{empty}
\setcounter{page}{0}
\end{titlepage}

\renewcommand{\baselinestretch}{1}  
\tableofcontents
\renewcommand{\baselinestretch}{1.2}  
\section{Introduction}
The classical picture of hydrodynamics, as a set of deterministic conservation equations, can only describe the low-energy long-wavelength dissipation processes, without considering fluctuations \cite{Landau:fluid}. 
In order to include the effect of fluctuations, one method is to place noise terms on the right side of the conservation equations. Then fluctuation-dissipation theorem determines the strength of the noise \cite{Landau:stat}.
 However, in this traditional ``stochastic"  picture, the interaction between the noise fields is ignored. The recently developed effective field theory (EFT) of hydrodynamics solves this problem by considering the effects of non-Gaussian noise \cite{Crossley:2015evo,Haehl:2015pja,Jensen:2017kzi,Haehl:2018lcu,Jensen:2018hse} \footnote{See \cite{Dubovsky:2011sj,Endlich:2012vt,Grozdanov:2013dba,Kovtun:2014hpa,Harder:2015nxa} for earlier works and also \cite{Glorioso:2018wxw} for a review.}.  In this framework, one can systematically derive the well known stochastic effects, such as long-time tails \cite{Chen-Lin:2018kfl}. 
 
The EFT of hydrodynamics  has also  some  new predictions. An example is the discovery of a positive contribution to the magneto-resistance in a $U(1)$ anomalous system \cite{Sogabe:2021wqk}. Another interesting example  is the prediction of stochastic transport \cite{Jain:2020fsm}. In two pioneering works, this EFT has also been applied to quantum chaotic systems \cite{Blake:2017ris,Blake:2021wqj}. The pole-skipping phenomenon as a prediction of such EFT in Ref.\cite{Blake:2017ris} reveals the hydrodynamic origin of the quantum chaos, at least in maximally chaotic systems \cite{Maldacena:2015waa}\footnote{See \cite{Grozdanov:2018kkt} for the first observation of pole-skipping and the relation with hydrodynamics in holography. See also \cite{Blake:2019otz,Grozdanov:2018kkt,Natsuume:2019sfp,Natsuume:2019xcy,Natsuume:2019vcv,Wu:2019esr,Ahn:2019rnq,Li:2019bgc,Ceplak:2019ymw,Das:2019tga,Abbasi:2019rhy,Liu:2020yaf,Ahn:2020bks,Ahn:2020baf,Kim:2020url,Sil:2020jhr,Yuan:2020fvv,Abbasi:2020xli,Ceplak:2021efc,Jeong:2021zhz,Yuan:2021ets,Blake:2021hjj,Kim:2021xdz} and \cite{Haehl:2018izb,Ramirez:2020qer} for the holographic  and CFT extensions, respectively. }. 
In another direction, the EFT of hydrodynamics has been recently applied to study the critical fluctuations near the QCD critical point \cite{Chao:2020kcf,Sogabe:2021svv} \footnote{See also \cite{Baggioli:2020haa} for the application to the quasicrystals systems and \cite{Delacretaz:2021qqu} to a system in a phase with a spontaneously broken U(1) symmetry.}.

On the other hand, an important problem in classical hydrodynamics is to study the large-order behavior of derivative expansion. In holographic systems,  this problem has been studied in real space and in momentum space \cite{Heller:2013fn}\cite{Withers:2018srf,Grozdanov:2019kge,Grozdanov:2019uhi}. In both cases it turns out that the large-order derivative expansion  encodes some information about the lowest non-hydrodynamic modes in the system. What we want to do in this paper is to go one step further and include the effect of fluctuations on the large-order derivative expansion within the framework of EFT of hydrodynamics.

 As is well known from stochastic \cite{Kovtun:2003vj,Kovtun:2011np,Kovtun:2012rj,Shukla:2021ksb} and EFT \cite{Chen-Lin:2018kfl} studies, the derivative expansion breaks down due to the hydrodynamic fluctuations, and the long-time tails. However,  all these studies are based on the consideration of the first order derivative expansion \footnote{In the language of EFT, the above works focus on the statistical fluctuations.}. Paraphrasing the discussion in the previous paragraph, we would like to determine how much information can be extracted from the large-order derivative expansion in the presence of long-time tails.

The EFT of hydrodynamics allows us to study the  above issue, systematically. We can construct EFT in the derivative expansion, and then compute the loop corrections.
In a general system, these two expansions should be truncated at the same order. But in a large $N$ theory, due to suppression of coupling constants in the $\frac{1}{N}$ expansion, the situation is different. One can go through the derivative expansion while dropping out higher-than-one loops \cite{Delacretaz:2020nit}\footnote{We would like to thanks Luca Delacr\'etaz for pointing this out.}. Therefore, , in a theory with a large number of degrees of freedom, we can study the effect of one-loop interactions on the classical Green's functions at large-order in the derivative expansion.

The system that we choose to study is the low energy SYK chain with a limit of $N\gg1$.  In the continuum limit of the chain, the effective action of the system  is found to be a local Schwarzian \cite{Gu:2016oyy}. The system is not CFT and in contrast to CFT in 1+1 dimension, it has a well-defined  hydrodynamic regime \cite{Herzog:2007ij,Delacretaz:2020nit}. In fact, energy is conserved in the system and diffuses through the chain \cite{Gu:2016oyy,Choi:2020tdj}. Thus we can describe the low energy dynamics of the system by the EFT of energy diffusion. In order to capture the effect of long-time tails in this system, we assume that the diffusion constant is  a function of energy fluctuations.  However, since the thermodynamic effective action of the system includes quantum effects, we should use the quantum hydrodynamic EFT, which is valid at finite $\hbar$ and to all order in derivatives \cite{Blake:2017ris}.  Then by putting two copies of the system on the two legs of closed time path (CTP) contour, we construct the fluctuating EFT of hydrodynamics for the energy diffusion in the chain.  Although in the previous studies the EFT of energy diffusion was constructed up to first \cite{Chen-Lin:2018kfl} or second \cite{Jain:2020fsm} order in derivatives, we analytically construct it up to $40^{th}$ order in derivatives.

In the next step, we use the above EFT to compute the energy density response function, whose pole is the diffusion dispersion relation. In the absence of loop corrections, the ``classical" dispersion relation is given in a derivative expansion in momentum space, $\wn=\wn(\qn^2)$. This series is convergent in the entire range of momenta allowed by the EFT.  When including the interactions and loop effects, however, the derivative expansion breaks down. Nonetheless, we show that one can still derive the diffusion pole in a power series as $\wn=\wn(|\qn|)$.  Compared to the classical dispersion relation, two new things would appear. First, the number of transport coefficients in this series is twice of that in the classical dispersion relation. In other words, the dispersion relation at order $n$ in derivatives is characterized by $2n$ transport coefficients. Second, we find that the dispersion relation series converges in momentum space. The radius of convergence is given by the ratio of the thermal conductivity to the diffusion constant.
In the rest of the paper, we first briefly review the EFT of hydrodynamics in \sec{EFT}. Then in \sec{model} we discuss the relation between Schwarzian theory and EFT of   hydrodynamics. We then explain how to add fluctuations to this picture. In \sec{fluc_hydro} we construct the EFT of fluctuating energy diffusion in the SYK chain to $40^{th}$ order in the derivative expansion. \sec{loop} is devoted to compute the loop corrections of the energy density response function. We then introduce the idea of modified dispersion relation and calculate the corresponding series coefficients in \sec{modified}. Finally, in \sec{conclusion} we end with review of the results, mentioning possible applications and discussing some followup directions.

\section{Effective field theory around thermal equilibrium}
\label{EFT}
The main quantity that we want to compute is the energy density response function in a quantum chaotic system at finite temperature. We aim to do this via computing the  energy density correlation function
\begin{equation}\label{EE}
\langle \mathcal{E}(t,x)\mathcal{E}(t',x')\rangle_{\beta}=\text{Tr}\big(\rho_0 \, \mathcal{E}(t,x)\mathcal{E}(t',x')\big)
\end{equation}
where $\rho_0$ is the thermal density matrix. Note that the energy density is a macroscopic dynamical variable. For this reason, we would like to compute the above correlator in the framework of effective field theory (EFT). 
We will construct the EFT we want according to ref. \cite{Glorioso:2018wxw}.
Below, we start from a microscopic theory and briefly review the steps of ref. \cite{Glorioso:2018wxw} to construct EFT around thermal equilibrium.

The microscopic analogue of \eqref{EE} can be written as  $\langle \psi(t,x)\psi(t',x')\rangle_{\beta}$ where $\psi$ is a microscopic dynamical variable. 
Such an out of equilibrium observable corresponds to insertion of microscopic fields $\psi_1$ and $\psi_2$ on the two legs of a closed time path (CTP), namely on the   Schwinger-Keldysh contour \footnote{Later on, we will elaborate on the connection between $\psi$ and $\mathcal{E}$.}.
Then by coupling the fields to external sources $\phi_1$ and $\phi_2$, one can compute the generating functional
\begin{equation}\label{W_0}
W[\phi_1,\phi_2]=\,\int_{\rho_0}D\psi_1 D\psi_2\, e^{i I_0[\psi_1, \phi_1]-i I_0[\psi_2, \phi_2]}\,,
\end{equation}
where $I_0[\psi_1, \phi_1]$ is the microscopic action in the presence external source $\phi$. 

In the language of Wilsonian RG, one may integrate out the UV dynamical variables, the so-called fast modes, in \eqref{W_0}, to find an effective action describing the low energy dynamics. Let us ideally assume that $\psi$ fields can be separated into UV and IR variables, $\varphi$ and $\chi$, respectively. Integrating out $\varphi$'s, \eqref{W_0} takes the following form:
\begin{equation}\label{W_eff}
W[\phi_1,\phi_2]=\,\int_{\ell_{\text{mic}}} D\chi_1 D\chi_2\, e^{i I_{\text{EFT}}[\chi_1,  \chi_2; \,\phi_1, \phi_2; \, \rho_0]}\,.
\end{equation}
Here, $I_{\text{EFT}}$ is the effective action of slow variables $\chi$ and is valid at length scales larger than the microscopic scale  $\ell_{\text{mic}}$, or equivalently at energy scales much smaller than the UV cutoff $1/\ell_{\text{mic}}$.

What are the slow modes $\chi$ in \eqref{W_eff}? In order to answer the question, let us limit the discussion to the situation in which the length (and time) scale of perturbations in the thermal system is much larger than the microscopic length (and time) of relaxation, i.e. $L\gg \ell$ (and $t_{L}\gg \tau$). In this case, \textit{non-conserved quantities} relax back to equilibrium very fast with the rate of $\sim1/\tau$. However, for \textit{conserved quantities}, which  cannot be locally destroyed, the only way to relax to equilibrium is transport on scales of order $L$ (and $t_L$).
One concludes that at such macroscopic scales, which denote the so-called IR limit  in \eqref{W_eff}, the only relevant variables $\chi$ are those \textit{associated} with conserved quantities.
The EFT of $\chi$ modes is indeed hydrodynamics.

\subsection{Effective field theory of energy diffusion}
\label{}
Let us consider the special case in which the only conserved quantity in the system is energy. In this EFT of energy diffusion, the $\chi$ field in \eqref{W_eff} is associated with the energy conservation $\partial_{\mu}J^{\mu}=0$, with $J^{\mu}$ being the energy current coupled to background gauge field $A_{\mu}$. So, we should replace $\phi_{1,2}$ in  \eqref{W_eff} with $A_{s\mu}$, $s=1,2$
\begin{equation}\label{W_eff_A}
W[A_{1\mu}, A_{2\mu}]=\,\int D\chi_1 D\chi_2\, e^{i I_{\text{EFT}}[\chi_1,  \chi_2; \,A_{1\mu}, A_{2\mu}]}\,.
\end{equation}
Then the conservation of $J^{\mu}_{1,2}$ translates to gauge invariance of $W$:
\begin{equation}\label{gauge_sym}
W[A_{1\mu}, A_{2\mu}]=\,W[A_{1\mu}+\partial_{\mu}\lambda_1, A_{2\mu}+\partial_{\mu}\lambda_2]\,.
\end{equation}
In addition, one can also show that the generating functional on CPT satisfies the following properties: 
\begin{eqnarray}\label{conditions_1}
&&W[A_{\mu}, A_{\mu}]=1\\
\text{reflectivity:}&&W^*[A_{1\mu}, A_{2\mu}]=W[A_{2\mu}, A_{1\mu}]\\
\text{Cauchy-Schwarz ineq.:}&&\Re W[A_{1\mu}, A_{2\mu}]\le0\\\label{conditions_last}
\text{KMS condition:}&&W[A_{1\mu}(x), A_{2\mu}(x)] =W[A_{1\mu}(-t, -\vec{x}), A_{2\mu}(-t - i \beta , -\vec{x})] 
\end{eqnarray}
Now one may ask in what sense $\epsilon$ field is associated with  $\partial_{\mu}J^{\mu}=0$? Let us denote that the above-mentioned $I_{EFT}$ should be such that the equations of motion of $\chi_{1,2}$ are equivalent to conversations of $J^{\mu}_{1,2}$. This condition is fixed if $\chi_{1,2}$ always appear with external fields through the combinations
\begin{equation}\label{}
B_{1\mu}=\,A_{1\mu}+\partial_{\mu}\chi_{1},\,\,\,\,\, B_{2\mu}=\,A_{2\mu}+\partial_{\mu}\chi_{2}\,.
\end{equation}
Therefore the $\chi_{1,2}$ are in the fact Stueckelberg fields associated with the symmetry \eqref{gauge_sym}. The generating functional \eqref{W_eff_A} then can be rewritten as
\begin{equation}\label{W_eff_B}
W[A_{1\mu}, A_{2\mu}]=\,\int D\chi_1 D\chi_2\, e^{i I_{\text{EFT}}[\,B_{1\mu}, B_{2\mu}]}\,.
\end{equation}
Now the main question is how to construct $I_{\text{EFT}}
$. The separation of scales $L\gg \ell$ motivates to construct  $I_{\text{EFT}}$ in a derivative expansion with the expansion parameter $\ell \partial_{\mu}\sim \frac{\ell}{L}\ll 1$. However, even writing down the most general local derivative expansion of $I_{\text{EFT}}[B_{1\mu},B_{2\mu}]$ subject to conditions \eqref{gauge_sym}-\eqref{conditions_last} is not enough \footnote{Let us suppose $J^{\mu}$ current was related to an internal  $U(1)$ symmetry. Then writing  down the most general local derivative expansion of $I_{\text{EFT}}[B_{1\mu},B_{2\mu}]$ would  describe a super fluid phase with the spontaneously broken $U(1)$ symmetry \cite{Glorioso:2018wxw}.}. One further symmetry has to be imposed. To understand it, it is convenient to define
\begin{equation}\label{sigma_X_a}
\sigma=\,\frac{1}{2}(\chi_1+\chi_2)\,,\,\,\,\,\,\,\,\,X^a=\chi_1-\chi_2\,.
\end{equation}
Calculating the equations of motion from local $I_{EFT}$, it turns out that $\partial_t\sigma(t,\vec{x})$ is related to the energy density $\mathfrak{h}+\mathcal{E}(t,\vec{x})$, and $X^{a}(t,\vec{x})$ corresponds to the noise field\footnote{$\mathfrak{h}$ denotes the energy density in thermal equilibrium.}. This has two direct consequences:
\begin{enumerate}
	\item The derivative counting scheme should be such that $\partial_t\sigma\sim X^a$ \cite{Chen-Lin:2018kfl}.
	\item Since only $\partial_t\sigma$ is physical, $\sigma$ must always appear with at least one time derivative in the effective action. It is equivalent to say that $I_{\text{EFT}}$ is required to be invariant under the so-called \textit{diagonal shift symmetry} \cite{Crossley:2015evo}:
	\begin{equation}\label{}
\sigma(t, \vec{x})\rightarrow \sigma(t, \vec{x})+ a(\vec{x})\,,\,\,\,\,\,\,\,X(t,\vec{x})\rightarrow X^a(t,\vec{x})\,.
	\end{equation}
	\end{enumerate}
\subsection{Effective Lagrangian and perturbation strategy }
\label{Effective_L}
So far, we have discussed all the elements that should be included when constructing EFT of energy diffusion around thermal equilibrium.
Considering $I_{\text{EFT}}=\int dx^{d+1}\mathcal{L}_{\text{eff}}$,
  the most general \textit{nonlinear} Lagrangian satisfying the above-mentioned conditions, to second order in $a$-fields, and in the absence of external sources, can be written as \cite{Crossley:2015evo,Glorioso:2018wxw}
\begin{equation}\label{The_most_general_L}
\mathcal{L}_{\text{eff}}[\sigma, X_a]=\,-H\partial_t X_a-G_i \partial_i X_a+\, i\,\partial_t X_aM_1 \partial_t X_a+\,i\,\partial_iX_aM_2\partial_iX_a+\mathcal{O}(X_a^3)\,.
\end{equation}
In this equation, the coefficient functions $H$ and  $G_i$ are (in general nonlinear) functions of $\dot{\sigma(t,x)}$ and its partial derivatives. Similarly, $M_{1,2}$ are  differential operators constructed out of $\dot{\sigma(t,x)}$ and differential operators $\partial_i$ and $\partial_t$.
At the quadratic level ($\mathcal{L}_2$), however, $M_{1,2}$ are purely differential operators  constructed out of $\partial_t$ and $\partial_i$ acting on the $X_a$ field sitting in their right side.

Another point with \eqref{The_most_general_L} is that the classical equation of motion of $X^a$ is simply the equation of conservation of energy:
\begin{equation}\label{Hydro_eq}
\partial_t H+\partial_i G_i=\,0\,.
\end{equation}
Thus $H$ and $G_i$ are classical parts of the energy density and energy flux, respectively \footnote{By classical here, we mean non-noisy part of the quantities.}.
In any particular system of interest, $H$ and $G_i$ should be used as input data; then by applying KMS conditions to \eqref{The_most_general_L}, one specifies $M_1$ and $M_2$, as well.

Before introducing the system we are interested in in this work, let us recall that our final goal is to compute one-loop corrections to \eqref{EE} from \eqref{The_most_general_L}, to high orders in the derivative expansion. By doing this and obtaining the corresponding energy diffusion pole, we will be able to study the large-order behavior of derivative expansion in the presence of fluctuations. 

Let us first mention our perturbation strategy for constructing Lagrangian functions.
We will include \textit{three expansions}.
\begin{enumerate}
	\item  \underline{\textbf{expansion around equilibrium:}}
	The equilibrium state corresponds to $\sigma(t,x)=t$ and $X_a(t,x)=0$. By considering the small deviations as
	\begin{equation}\label{out_of_equ}
	\sigma(t,x)= t+\,\epsilon(t,x),\,\,\,\,\,\,\,\,X_{a}(t,x)=-\epsilon_a(t,x)\,,
	\end{equation}
	we construct our EFT to third order in the above out-of-equilibrium fields.  We will see that the cubic Lagrangian is sufficient to find the finite one-loop  correction to \eqref{EE}.
	\item  \underline{\textbf{noised field expansion:}} We limit our study to include the effect of thermal fluctuations up to second order in $a-$fields.
	\item  \underline{\textbf{derivative expansion:}} In order to study the large-order behavior of the hydrodynamic derivative expansion in the presence of fluctuations, we need to construct $\mathcal{L}$ to the high-order derivatives. On the other hand, we would like to find all terms in $\mathcal{L}$ analytically.  By implementing a systematic method, we will construct the Lagrangian to $40^{th}$ order in the derivative expansion. Although due to the large size of analytic expressions, we only explicitly display the Lagrangian terms to $20^{th}$ order, in the appendices \footnote{It should be noted that, in principle, our method can be implemented to any higher order in the derivative. However, we found that for our current purposes, keeping terms to $40^{th}$ order is sufficient.}. 
\end{enumerate}
Before concluding this section, let us comment on the size of the spatial derivative in our system. Since our expected EFT should describe a diffusion process controlled by \eqref{Hydro_eq}, we are interested in the disturbance of $\omega\sim k^2$. This together with $\partial_t\sigma\sim X^a$ (see below \eqref{sigma_X_a}) provide our derivative counting  scheme in this work.

\section{The model}
\label{model}
So far, our discussion about EFT of hydrodynamics has been general, without mentioning any special physical systems. When restricting the study to the lower orders in the derivative expansion, it would be feasible to construct $\mathcal{L}_{EFT}$ for the general system. For example, Ref.\cite{Chen-Lin:2018kfl} constructs the energy diffusion EFT in a general thermal system up to first order in the derivative expansion. But at higher orders of the derivative, the calculation will become more complicated because more possible terms may contribute to $H$ and $G_i$. Therefore, since our goal is to explore the large-order behavior of the derivative expansion in this paper, we choose to use a specific system with the well-known $H$ function.

Our system of interest is the SYK chain. The $(0+1)$ dimensional SYK model \cite{Kitaev,Sachdev} and also the $(1+1)$ dimensional SYK chain \cite{Gu:2016oyy} have been widely studied in the literature \cite{Maldacena:2016hyu}. However, in order to be clear about the problem that we want to address, we need to recall some aspects of the SYK model.  In particular, it is well known that the infrared theory in the strong coupling limit is described by hydrodynamics \cite{Jensen:2016pah}. In \sec{SYK}, we first briefly review the model.
Then following \cite{Blake:2017ris} and by using the language of Schwinger-Keldysh EFT, we revisit the mentioned hydrodynamic picture and specify the $H$ function associated with the SYK model.
  Then in \sec{SYK_chain}, after a quick look at the  SYK chain model,  we will review its connection with the EFT of hydrodynamics and specify the corresponding $H$ and $G_i$ functions in \eqref{The_most_general_L} \cite{Blake:2017ris}.

Everything we will discuss in the following two subsections will be devoted to classical aspects of EFT of hydrodynamics. But our ultimate goal is to couple classical EFT with fluctuations, which is the subject of the next section.

\subsection{SYK model and hydrodynamics}
\label{SYK}
The SYK model consists of $N$ Majorana fermions $\chi_{j}(\tau),\,j=1,2,\cdots,N$\footnote{Here $\tau$ is the Euclidean time coordinate. At finite temperature, we have $\tau\sim \tau+\beta$. } with a random four-fermion interaction. The random coupling has zero mean $\overline{J_{jklm}}=0$ and non-zero variance $\frac{1}{3!} N^3\overline{J^2_{jklm}}=J^2$. The model is non-local in space in the sense that the interaction is all-to-all. This model should be actually considered as a (0 + 1)-dimensional quantum mechanical system.

The  model is solvable in the large N limit. Specifically in the strong coupling limit, $N\gg \beta J \gg 1$, the model shows up a conformal symmetry in the infrared limit, that is the invariance under reparameterization $\tau\rightarrow f(\tau)$. However, it turns out that the corresponding conformal      
correlators, $G_{c}$, are only invariant under  $SL(2,R)$ subgroup of this conformal symmetry.   In other words, the conformal symmetry is spontaneously broken by the conformal solution $G_c$. As a result, the infrared effective action of the model is zero when evaluated on fluctuations that are reparameterizations of $G_c$.

Beyond the conformal limit, the picture above will no longer be the case. In particular, due to explicitly breaking of the conformal symmetry, the effective action becomes non-zero when evaluated  on fluctuations  of $\delta G_c$, the reparameterizations of $G_c$.   For an infinitesimal reparameterization $\tau\rightarrow  \tau + \epsilon(\tau)$, one finds \cite{Maldacena:2016hyu}
\begin{equation}\label{SYK_EFT}
\frac{S}{N}=\frac{\alpha}{J}\int_{0}^{\beta}d\tau \frac{1}{2}\bigg[(\epsilon'')^2-\lambda^2(\epsilon')^2\bigg],\,\,\,\,\,\,\lambda=\frac{2\pi}{\beta}\,,
\end{equation}
with $\alpha$ being a numerical constant. Now let us take $\mathcal{L}[\epsilon(t)]= C \big((\epsilon'')^2+\lambda^2(\epsilon')^2\big)$ with the Lorentzian time coordinate $t$, and put two copies of it on the two legs of CTP contour. The Schwinger-Keldysh action, with real time coordinate $t$,  then reads
\begin{equation}\label{SYK_L_EFT}
\mathcal{L}[\epsilon,\epsilon_a]=\,\mathcal{L}[\epsilon_1]-\mathcal{L}[\epsilon_2]=\,C\,\left(\lambda^2-\partial_t^2\right)\partial_t \epsilon\,\partial_t\epsilon_a\,.
\end{equation}
Considering \eqref{out_of_equ}, equation \eqref{SYK_L_EFT} is exactly the first term in the effective Lagrangian of hydrodynamics \eqref{The_most_general_L}, specific to a system in $(0+1)$ dimension with  
\begin{equation}
H(\sigma)= C \left(\lambda^2-\partial^2_t \sigma\right)\partial_t \sigma,\,\,\,\,\,\,\,\,\,\,H_{0}=C\,\lambda^2 \,.
\end{equation}
Since $H_0$ represents the energy density in thermal equilibrium, $\sigma=t$,  it is convenient to take $C=\frac{\mathfrak{h}}{\lambda^2}$, where $\mathfrak{h}$ is the thermodynamic energy density.
The classical equation of motion is then simply the equation of conservation of energy, i.e. $\partial_t H=0$.
We conclude that the EFT describing  dynamics of the reparameterization mode $\epsilon$ in the SYK model is actually hydrodynamics \footnote{See \cite{Jensen:2016pah} for another way of describing the hydrodynamic origin of $\epsilon$.}.

We can extend the above-mentioned hydrodynamic action beyond the second order, that is, the regime of nonlinear hydrodynamics. It can be done by generalizing the action  \eqref{SYK_EFT} to a finite reparameterization $\tau \rightarrow \tau+f(\tau)$. One finds \cite{Maldacena:2016hyu}
\begin{equation}\label{Shc}
\frac{S}{N}=-\frac{\alpha}{J}\int_{0}^{\beta}\,d\tau\, \text{Sch}(f(\tau),\tau),\,\,\,\,\,\,\,\text{Sch}(f(\tau),\tau)=\left(\frac{f''}{f'}\right)'-\frac{1}{2}\frac{f''^2}{f'^2}
\end{equation}
with $f(\tau)=\tan\left(\frac{\pi \tau}{\beta}\right)$. We can have further reparameterization on the Euclidean time circle, $\tau\rightarrow \sigma(\tau)$. Then writing $\sigma(\tau)=\tau+\epsilon(\tau)$ and expanding $\text{Sch}(f(\sigma(\tau)),\tau)$ in \eqref{Shc}, we get exactly  the two quadratic terms of \eqref{SYK_EFT}.

In Lorentzian coordinate, $f(\tau)$ transforms to $f(t)=\tanh\left(\frac{\pi \sigma(t)}{\beta}\right)$. Inverting this equation, we define $u(t)$ as
\begin{equation}\label{}
\frac{1}{2}e^{-\lambda\, \sigma(t)}=\,\frac{1}{2}\,\frac{1-f(t)}{1+f(t)}\equiv \frac{1}{2}u(t)\,.
\end{equation}
Since $\frac{1}{2}u$ is $SL(2,R)$ transformed of $f$, we immediately write: $\text{Sch}(f(t), t)=\text{Sch}(\frac{1}{2}u(t), t)$. Thus in Lorentzian coordinate, \eqref{Shc} takes the following form 
\begin{equation}\label{Shc_Lorentz}
S=\,-i \int dt\, \frac{\mathfrak{h}}{\lambda^2}\text{Sch}(u(t),t),\,\,\,\,\,\,\,\,\,u(t)=e^{-\lambda\,\sigma(t)}\,,
\end{equation}
where we have used $\frac{1}{2}\frac{N\,\alpha}{J}=C=\frac{\mathfrak{h}}{\lambda^2}$. 

We can repeat the discussion around \eqref{SYK_L_EFT} to find the Schwinger-Keldysh analogue of \eqref{Shc_Lorentz}.  Doing so, we find  that  the reparameterization mode in the SYK model, in general, is described by nonlinear hydrodynamics  with the corresponding $H$ function given by
\begin{equation}\label{H}
H=- \frac{\mathfrak{h}}{\lambda^2}\text{Sch}(u(t),t),\,\,\,\,\,\,\,\,\,u(t)=e^{-\lambda\,\sigma(t)}\,.
\end{equation}
Let us recall that equilibrium state corresponds to $\sigma(t)=t$.

 It is worth  nothing that this Schwarzian action goes actually beyond ordinary long distance hydrodynamics. The reason is it includes modes whose time variation rate is comparable to the temperature \cite{Maldacena:2016upp}. Correspondingly, the framework of \cite{Crossley:2015evo} can be used to any large-order in derivatives. Therefore, it can be well applied to study the dynamics of the reparametrization mode in the SYK model.
\subsection{SYK chain and classical EFT of energy diffusion}
\label{SYK_chain}
The SYK chain describes a coupled array of SYK model sites. Each site contains $N\gg1$ Majorana fermions with four-fermion random coupling $J_{jklm,x}$. Here $j=1,2,\cdots,N$ and $x=1,2,\cdots,M$ where $M$ is the number of sites. Then there is further coupling between each pair of neighboring sites through another four-fermion random coupling $J'_{jklm,x}$. The random couplings  $J_{jklm,x}$ and $J'_{jklm,x}$ are drawn independently and 
\begin{equation}\label{}
\overline{J_{jklm,x}}=\overline{J'_{jklm,x}}=0,\,\,\,\,\,\,\,\,\,\,\frac{1}{3!} N^3\overline{J^2_{jklm,x}}=J_{0}^2,\,\,\,\,\,\,N^3\overline{J'^2_{jklm,x}}=J_1^2\,.
\end{equation}
One can show that the Schwinger-Dyson equations of the model reduce to exactly the
same form as those of a $(0+1)$-d SYK model with the coupling constant $J=\sqrt{J_0^2+J_1^2}$.

As in the SYK model, the effective action of the SYK chain admits an
approximate reparametrization symmetry of time in the IR limit. For small deformations of time $\tau\rightarrow \tau +\epsilon_x(\tau)$, the quadratic effective action is found to be \cite{Gu:2016oyy}
\begin{equation}\label{EFT_chain}
\frac{S}{N}=\sum_{n, k}\, \frac{\alpha}{2 J}\,\epsilon_{n,k}\bigg(\omega_n^2(\omega_n^2-\lambda^2)+\,D\,k^2\,|\omega_n|(\omega_n^2-\lambda^2)\bigg)\,\epsilon_{-n,-k}\,,
\end{equation}
where $\epsilon_{n,k}$ is the Fourier transform of $\epsilon_x(\tau)$ and $D\sim J_1^2/J$ is the diffusion constant. 
The first term in \eqref{EFT_chain} is exactly the EFT of SYK model \eqref{SYK_EFT}, if we remove the dependence on $k$. Recalling \eqref{Shc_Lorentz}, we find that in the limit of an infinite lattice, $M\rightarrow \infty$, this term can be interpreted as a quadratic approximation of the local Schwarzian action $\text{Sch}(u(t,x),t)$. What about the second term in \eqref{EFT_chain}? Let us consider two copies of the system placing on the two legs of CTP contour. Then we find that
\begin{equation}\label{SYK_chain_L_EFT}
\mathcal{L}[\epsilon,\epsilon_a]=\,\mathcal{L}[\epsilon_1]-\mathcal{L}[\epsilon_2]=\,C\,\bigg(\left(\lambda^2-\partial_t^2\right)\partial_t \epsilon\, \partial \epsilon_a+D \left(\lambda^2-\partial_t^2\right)\partial_x\partial_t\epsilon\,\partial_x\epsilon_a\bigg)\,.
\end{equation}
The above Lagrangian  is actually the \textit{quadratic} approximation to \eqref{The_most_general_L} in $(1+1)$ dimension  with\footnote{In our setup, there is only one spatial direction along the Chain. However, we use subscript $i$ to display the spatial derivative so that we can discuss the generalization of things to higher dimensions later.}
\begin{equation}\label{H_G_i}
H=- \frac{\mathfrak{h}}{\lambda^2}\text{Sch}(u(t,x),t)\,\,\,\,\,\,\,\,\,\bigg(u(t,x)=e^{-\lambda\,\sigma(t,x)}\bigg),\,\,\,\,\,\,\,G_i=-D\,\partial_i H\,.
\end{equation}
Thus the second term in \eqref{SYK_chain_L_EFT} denotes the diffusion of energy through the SYK chain with $D$ being the diffusion constant.

In summary, given $H$ and $G_i$  by \eqref{H_G_i}, the linearized version of equation  \eqref{Hydro_eq} represents the classical \textit{linear} hydrodynamics of the diffusive reparametrization mode in the SYK chain. In next section,
we discuss the \textit{nonlinear} hydrodynamic description of the energy diffusion  through the SYK chain, taking into account the effects of fluctuations generated by going beyond the quadratic order in \eqref{The_most_general_L}.

\section{SYK Chain and EFT of nonlinear energy diffusion}
\label{fluc_hydro}
 Let us recall that for the system to be specified in the framework of EFT of hydro, the $H$ and $G_i$ associated with it need to be determined. As was discussed earlier, our system of interest is the SYK chain with the $H$ and $G_i$ functions given by \eqref{H_G_i}. On the other hand, In order to include the interaction between the hydrodynamic field $\epsilon$ and the fluctuation field $\epsilon_a$, we have to take into account nonlinear terms in the effective action. As mentioned before, we aim to construct a cubic interacting Lagrangian. Considering $u=e^{-\lambda (t+\epsilon(t,x))}$, there are two potential sources of nonlinearity in \eqref{H_G_i}:
 \begin{enumerate}
\item $H$ itself is a nonlinear function of $\epsilon$; in order to construct the cubic Lagrangian, we expand $H$ about its equilibrium value $\mathfrak{h}$ and keep terms to second order:
\begin{equation}\label{H_original}
\begin{split}
H&=\,-\frac{\mathfrak{h}}{\lambda^2}\,\text{Sch}(e^{-\lambda(t+\epsilon(t,x))},t)\\
&=\,\,\,\frac{\mathfrak{h}}{2}+\,\mathfrak{h}\left(\partial_t \epsilon-\frac{1}{\lambda^2}\partial_t^3\epsilon\right)+\,\mathfrak{h}\left(\frac{1}{2}(\partial_t \epsilon)^2+\frac{3}{2\lambda^2}(\partial^2_t\epsilon)^2+\frac{1}{\lambda^2}\partial_t\epsilon\,\partial^3_t\epsilon\right)+\mathcal{O}(\epsilon^3)\,.
\end{split}
\end{equation}
Note that although we aim to perform our computations to high orders in the derivative expansion, the above expression is by construction truncated at $\partial_t^2$. Let us remind that as mentioned below \eqref{sigma_X_a}, $\partial_t\delta\sigma\sim\partial_t \epsilon\sim \mathcal{O}(\partial_t^0)$.
 
\item  The diffusion coefficient $D$ can also be a nonlinear function of energy fluctuations $\delta H=H-\frac{\mathfrak{h}}{2}$. We may write
\begin{equation}\label{G_i_M_2}
\begin{split}
G_i=&\,-D(\delta H)\,\partial_i H\\
=&\,-D_0\mathfrak{h}\left(\partial_i\partial_t \epsilon-\frac{1}{\lambda^2}\partial_i\partial_t^3\epsilon\right)-D_1\mathfrak{h}\left(\partial_i\partial_t \epsilon-\frac{1}{\lambda^2}\partial_i\partial_t^3\epsilon\right)\left(\partial_t \epsilon-\frac{1}{\lambda^2}\partial_t^3\epsilon\right)+\mathcal{O}(\epsilon^3)
\end{split}
\end{equation}
where we have used
\begin{equation}\label{}
D(\delta H)=\,D_0+\,\frac{D_1}{\mathfrak{h}}\,\delta H+\frac{D_2}{2\mathfrak{h}^2}(\delta H)^2+\cdots
\end{equation}
with $D_n=\,\mathfrak{h}^n\,\frac{\partial^n D}{\partial H^n}$.
 Note that coefficients $D_n: n>1$ do not appear in the cubic Lagrangian. It is also worth noting that in general, any of the $D_n$ coefficients can be a differential operator, constructed out of $\partial_t$ and $\partial_i$. However, in this work, we assume that these coefficients are constant. One important consequence of this assumption is that the \textit{classical} dispersion relation of  energy diffusion, namely the dispersion obtained from linear hydrodynamic equations, will be exact in the derivative expansion and is given by $\omega=D_0 k^2$. In \sec{modified} we will discuss the general case in which $D_0$ itself has a derivative expansion. Finally, one of the problems we will solve in this work is how this simple dispersion relation changes when nonlinear effects are included.
  \end{enumerate}
Having specified the ingredients of the model, we should now compute $M_1$ and $M_2$ terms in \eqref{The_most_general_L}. Since there are two $X_a$ fields, each of the terms $M_1$ and $M_2$ is at least quadratic. So in order to construct a cubic Lagrangian, we need to determine $M_1$ and $M_2$ up to first order in $\epsilon$.
Then we find that the appropriate ansatz in the most general form is as follows
  \begin{eqnarray}\label{M_1_general}
  M_1&=&\,\,\rho\,+\alpha_1\delta H\,\alpha_2+\cdots=\,\rho+\,\alpha_1\mathfrak{h}\left(\partial_t \epsilon-\frac{1}{\lambda^2}\partial_t^3\epsilon\right)\alpha_2+\mathcal{O}(\epsilon^2)\,,\\\label{M_2_general}
  M_2&=&\,\,\chi\,+\,\xi_1\delta H\,\xi_2+\cdots\,=\,\chi+\,\xi_1\mathfrak{h}\left(\partial_t \epsilon-\frac{1}{\lambda^2}\partial_t^3\epsilon\right)\xi_2+\mathcal{O}(\epsilon^2)\,.
  \end{eqnarray}
The coefficients in these two equations are all differential operators, each of which acting on the first function on the right. Now the task is to apply the KMS conditions to the effective action and specify these coefficients, perturbatively in the derivative expansion, in terms of $D_0$, $D_1$, $\beta=T^{-1}$ and  $\mathfrak{h}$.

By defining $\mathcal{F}\in\{\rho(\partial_t),\,\alpha_j(\partial_t)\,\cdots\}$ and $\mathcal{G}\in\{\chi(\partial_t),\,\xi_j(\partial_t),\cdots\}$, we want to find coefficients $\mathcal{F}^{(n)}$ in the following derivative expansions:
\begin{equation}\label{Taylor_chi}
\mathcal{F}(\partial_t)=\,\sum_{n=3} \mathcal{F}_{n}\times\left(\frac{\partial_t}{\lambda}\right)^{n-2},\,\,\,\,\,\,\,\,\,\,\,\mathcal{G}(\partial_t)=\,\sum_{n=1} \mathcal{G}_{n}\times\left(\frac{\partial_t}{\lambda}\right)^{n-1}.
\end{equation}
In the following subsections we will find equations constraining $\mathcal{F}_{n}$'s and $\mathcal{G}_{n}$'s.

\subsection{KMS condition}
The KMS condition \eqref{conditions_last} is actually a constraint equation on the generating functional in the presence of external sources $A_{\mu}$. It is shown that a sufficient condition for invariance of $W[A_{1\mu}, A_{2\mu}]$ under \eqref{conditions_last} is that $I_{EFT}[B_{1\mu},B_{2\mu}]$ be invariant under
\begin{equation}\label{}
B_{1\mu}(x)\rightarrow B_{1\mu}(-t,-\vec{x}),\,\,\,\,\,B_{2\mu}(x)\rightarrow B_{2\mu}(-t-i \beta,-\vec{x})
\end{equation}
which is the same as transformation of background $A_{\mu}$ field (see \cite{Crossley:2015evo,Gao:2018bxz} for details). These constraints should be imposed to quadratic and  cubic action separately. Here, we do not derive the constraints and just follow Ref.\cite{Crossley:2015evo} and quote their results (see also \cite{Wang:1998wg} for an earlier derivation).
\subsubsection{quadratic action}
For a quadratic effective Lagrangian of the form
\begin{equation}
\mathcal{L}^{(2)}_{\text{eff}}[B^1,B^2]=\,K_{ra}^{\mu_1\mu_2}B_{a\mu_1}B_{r\mu_2}+\,\frac{i}{2}\,G_{rr}^{\mu_1\mu_2}B_{a\mu_1}B_{a\mu_2}\,,
\end{equation}
where $\mu_1, \mu_2\in\{t,i\}$,
it is found that the KMS constraint in momentum space is given by
\begin{equation}\label{KMS_2}
i\,G_{rr}(\omega)=\,\frac{1}{2}\coth\left(\frac{\beta \omega}{2}\right)\bigg(K_{ra}(\omega)-K_{ar}(\omega)\bigg)\,.
\end{equation}
Here, $K_{ar}(\omega)=K_{ra}(-\omega)$ and for simplicity we have dropped the superscripts $\mu_1$ and $\mu_2$. In the classical limit $\hbar\rightarrow0$, the $\coth$ factor in the right side is simply replaced with $(\beta \omega)^{-1}$.  However, since quantum effects as well as the  higher order derivatives are important in our study, we  continue to use the original form \eqref{KMS_2}.

The corresponding $G_{rr}$ and $K_{ra}$ functions associated with  our system, in the absence of external sources, can be written as
\begin{eqnarray}\label{K_ra}
K_{ra}^{\mu_1\mu_2}&=&\,\mathfrak{h}\,\left(1+\frac{\omega^2}{\lambda^2}\right)\,\delta^{\mu_1}_{t}\delta^{\mu_2}_{t}+\,(-D_0\,\mathfrak{h})\,(-i \omega)\,\left(1+\frac{\omega^2}{\lambda^2}\right)\,\delta^{\mu_1}_{i}\delta^{\mu_2}_{i}\,,\\
G_{rr}^{\mu_1\mu_2}&=&\,2\, \rho_{-\omega}\,\delta^{\mu_1}_{t}\delta^{\mu_2}_{t}+\,2\,\chi_{-\omega}\,\delta^{\mu_1}_{i}\delta^{\mu_2}_{i}\,.
\end{eqnarray}
Note that we take the Fourier components of $B_{\mu_1}$ and $B_{\mu_2}$ as $B_{\mu_1}^{\omega}$ and $B_{\mu_2}^{-\omega}$, respectively.
Plugging above $G_{rr}$ and $K_{ra}$ into \eqref{KMS_2} and expanding it in the derivative expansion, we can specify $\rho_n$ and $\chi_n$ coefficients, order by order in the derivative expansion. 
We have computed these coefficients to $40^{th}$ order in derivatives. The $\chi_n$ coefficients with $n\le 20$ have been given in Appendix \ref{chi_n}. It is worth noting that $\rho$ vanishes to all order in the derivative expansion. The reason is $K^{tt}_{ra}$ in \eqref{K_ra}, is an even function of $\omega$. So the right side of \eqref{KMS_2}  vanishes, causing $G^{tt}_{rr}$ to be identically zero.

Putting the above-mentioned parts together, the quadratic Lagrangian takes the following form:
\begin{equation}
\mathcal{L}^{(2)}_{\text{eff}}=\mathfrak{h}\,\partial_t \epsilon_a\,\bigg( 1-\frac{\partial_t^2}{\lambda^2}\bigg)\partial_t\epsilon-\,D_0\,\mathfrak{h}\,\partial_{i}\epsilon_{a}\,\bigg(1-\frac{\partial_t^2}{\lambda^2}\bigg)\partial_i\partial_t\epsilon+\,i\frac{D_0\,\mathfrak{h}}{\beta}\partial_i\epsilon_a\sum_{n=1}\chi_n\left(\frac{\partial_t}{\lambda}\right)^{n-1}\partial_i\epsilon_a\,.
\end{equation}
See Appendix \ref{chi_n} for the $\chi_n$ coefficients.

\subsubsection{cubic action}
For a cubic effective Lagrangian of the form:
\begin{equation}\label{L_EFT_3}
\mathcal{L}^{(3)}_{\text{eff}}[B^1,B^2]=\,\,\frac{1}{2}\,G_{raa}^{\mu_1\mu_2\mu_3}\,B_{a\mu_1}B_{r\mu_2}B_{r\mu_3}+\,\frac{1}{2}\,G_{rra}^{\mu_1\mu_2\mu_3}\,B_{a\mu_1}B_{a\mu_2}B_{r\mu_3}\,,
\end{equation}
the KMS constraint in momentum space is given by \cite{Crossley:2015evo,Chen-Lin:2018kfl,Wang:1998wg}
\begin{equation}\label{KMS_3}
2\,G_{rra}=-\big(N_1+N_2\big)G^{*}_{aar}+N_1\,G_{ara}+N_2\,G_{raa}\,,\,\,\,\,\,\,\,\,\,N_i=\coth\left(\frac{\beta\,\omega_i}{2}\right)\,.
\end{equation}
Note that any of the above $G$'s is actually a function of $\omega_1$, $\omega_2$ and $\omega_3$, where $\omega_1+\omega_2+\omega_3=0$. And again, for simplicity, we have dropped the superscripts $\mu_1$, $\mu_2$ and $\mu_3$.

The corresponding $G_{raa}$ and $G_{rra}$ functions associated with  our system, in the absence of external sources, can be written as
\begin{equation}\label{KMS_3_right}
\begin{split}
G_{raa}^{\mu_1\mu_2\mu_3}=&\,-\mathfrak{h}\,(- i \omega_2)\bigg[D_0+D_1+3D_0\prod_{23}+(D_0-D_1)\big(\prod_{33}+\prod_{22}\big)+D_1\prod_{22}\prod_{33}\bigg]\,\delta_{i}^{\mu_1}\delta_{i}^{\mu_2}\delta_{0}^{\mu_3}\\
&\,\,\,\,+\frac{\mathfrak{h}}{2}\,\left[1+3\prod_{23}+\,2\prod_{22}\right]\delta_{0}^{\mu_1}\delta_{0}^{\mu_2}\delta_{0}^{\mu_3}\,+\,\big(2\leftrightarrow 3\big)\,,
\end{split}
\end{equation}
\begin{equation}\label{KMS_3_left}
G_{rra}^{\mu_1\mu_2\mu_3}=\,\left(1+\prod_{3}\right)\,\bigg[\alpha_2^{\omega_2,k_2}\alpha_1^{\omega_3,k_3} \delta_{0}^{\mu_1}\delta_{0}^{\mu_2}\delta_{0}^{\mu_3}+\,\xi_2^{\omega_2,k_2}\xi_1^{\omega_3,k_3} \delta_{i}^{\mu_1}\delta_{i}^{\mu_2}\delta_{0}^{\mu_3}\bigg]+\,(1\leftrightarrow 2)\,.
\end{equation}
In the above expressions: $\omega_1+\omega_2+\omega_3=0$ and we have also defined
\begin{equation}
\prod_{j_1 j_2 \cdots}=\,\left(\frac{-i\, \omega_{j_1}}{\lambda}\right)\left(\frac{-i\, \omega_{j_2}}{\lambda}\right)\cdots\,.
\end{equation}
Substituting \eqref{KMS_3_right} and \eqref{KMS_3_left} into \eqref{KMS_3} and expanding the two sides in the derivative expansion, we find equations between the  expansion coefficients  of $\alpha_1$ and  $\alpha_2$, defined in \eqref{Taylor_chi}, and also equations between of expansion coefficients of $\xi_1$ and $\xi_2$. Two comments about these equations are as follows.
\begin{enumerate}
	 \item Using KMS condition, it is clear that $\alpha_{1}$ and $\alpha_{2}$ are related to the first line of \eqref{KMS_3_right} while $\xi_{1}$ and $\xi_{2}$ can be found from the first expression in the second line of  \eqref{KMS_3_right}. 
	 \item For concreteness, let us consider $\xi_1$ and $\xi_2$ at $n^{th}$ order. The corresponding  expansion coefficients are $\xi^{(n)}_1$ and $\xi^{(n)}_2$ (see \eqref{Taylor_chi}). It is clear that these coefficients always appear in pairs as 
	 \begin{equation}\label{}
	\xi^{(j)}_1\xi^{(k)}_2\equiv\,T_{j,k}\,.
	 \end{equation}
	 We find that at $n^{th}$ order in the derivative expansion, there are $\left[\frac{n-1}{2}\right]+1$  equations between $n$ independent $T_{j,k}$ coefficients \footnote{It can be understood as follows. Equation \eqref{KMS_3} is symmetric with respect to $\omega_1$ and $\omega_2$. While the symmetry in the right side is obvious, in the left side it is due to \eqref{KMS_3_left}. Thus at any order in the derivative expansion, after imposing $\omega_1+\omega_2+\omega_3=0$, equation \eqref{KMS_3} becomes a symmetric polynomial of $\omega_1$ and $\omega_2$, giving a set of equations. The number of independent equations then is equal to the number of independent coefficients in the expansion of $(\omega_1+\omega_2)^n$, i.e. $\left[\frac{n-1}{2}\right]+1$. These equations describe all KMS constraints of order $n$.  On the other hand there are $n$ independent variables: $T_{n,1}, T_{n-1,2}, \cdots, T_{1,n}$.}. Therefore, KMS constraints seem insufficient to specify all the coefficients of $T_{j,k}$ appearing in the Lagrangian function. However, after solving the existing equations for an appropriate set of $\left[\frac{n-1}{2}\right]+1$ coefficients \footnote{By linearly combining the equations, we have found that the most appropriate set of $T_{j,k}$'s at $n^{th}$ order is $\{T_{n,0}, T_{n-2,2}, T_{n-1,4}, T_{n-1,6}, \cdots, T_{n-1, 2k-2},  \cdots\}$. As it is seen, for $k>2$, the $k^{th}$ appropriate $T_{j,k}$ follows from a general form: $T_{n-1, 2k-2}$. Finally, in any order $n$, we choose the first $\left[\frac{n-1}{2}\right]+1$ elements of the set as the appropriate coefficients to solve from the KMS equations.}, and applying the solutions to the Lagrangian, it turns out that all $n-\left(\left[\frac{n-1}{2}\right]+1\right)$ unspecified $T_{j,k}$ coefficients appear as the coefficients of total derivative terms in the cubic Lagrangian $\mathcal{L}^{(3)}_{\text{eff}}$. This observation shows that we can actually start with an ansatz which is more constrained than \eqref{M_2_general} and \eqref{M_1_general}. However, we continue to use \eqref{M_2_general} and \eqref{M_1_general}.
	\end{enumerate}
We have derived the equations between $T_{j,k}$ coefficients to $40^{th}$ order in the derivative expansion. In Appendix \ref{KMS_App} , we display the corresponding equations only for the first $10$ orders. We have also defined $R_{j,k}=\alpha_1^{j}\alpha_2^{k}$ and calculated the equations between $R_{j,k}$ coefficients to the $40^{th}$ order. The equations can be displayed in a similar way as in Appendix \ref{KMS_App}; however, we will not show them in the paper.

Imposing all the above-mentioned KMS constraints to \eqref{L_EFT_3}, we have constructed the KMS invariant cubic Lagrangian $\mathcal{L}_{\text{eff}}^{(3)}[\epsilon, \epsilon_a]$ to $40^{th}$ order in the derivative expansion. However, as discussed in the Introduction,  our goal is to compute the energy density correlator defined by \eqref{EE}. For this reason, we find that it is more appropriate to rewrite the effective Lagrangian in terms of $\mathcal{E}$ and $\epsilon_a$, say  $\mathcal{L}_{\text{eff}}[\mathcal{E}, \epsilon_a]$. Therefore, in order to reduce complexity, we do not explicitly show the lengthy structure of $\mathcal{L}_{\text{eff}}^{(3)}[\epsilon, \epsilon_a]$. Instead, in the next section, we will discuss how to exchange $\epsilon$ with $\mathcal{E}$ and then explicitly display $\mathcal{L}^{(2)}_{\text{eff} }[\mathcal{E}, \epsilon_a]$ nd $\mathcal{L}^{(3)}_{\text{eff}}[\mathcal{E}, \epsilon_a]$.


\subsection{Effective action in terms of energy density}
 The off-shell hydrodynamic energy density current $J_E^{\mu}\equiv J^{\mu}_r=(\mathcal{E}, J_i)$ is defined as $J^{\mu}_E=\,\delta I_\text{EFT}/\delta A_{a\mu}$ \cite{Glorioso:2018wxw}.  Thus the energy density is given by
 \begin{equation}\label{mathcal_E_define}
\mathcal{E}=\,\frac{\delta I_\text{EFT}[B_{r\mu},B_{a\mu}]}{\delta A_{a 0}}=\,\frac{\delta I_\text{EFT}[B_{r\mu},B_{a\mu}]}{\delta \partial_t\epsilon_{a}}\,.
 \end{equation}
In the second equality we have used $B_{a0}=A_{a0}+\partial_t\epsilon_a$.
Calculating \eqref{mathcal_E_define}, we find \footnote{Note that $\partial_t\epsilon$ and $\partial_t\epsilon_a$ are dimensionless quantities, while $\mathcal{E}$ and $\mathfrak{h}$ have dimension of energy density.}
\begin{equation}\label{mathcal_E}
\mathcal{E}=\,\mathfrak{h}\left(\partial_t\epsilon -\frac{\partial^3_t\epsilon}{\lambda^2}\right)+\frac{\mathfrak{h}}{2}\left((\partial_t\epsilon)^2+\frac{3}{\lambda^2}(\partial^2_t\epsilon)^2+ \frac{2}{\lambda^2}\partial^3_t\epsilon\partial_t\epsilon\right)+\mathcal{O}(\epsilon^3)\,.
\end{equation}
Note that this expression is exact in the derivative expansion. However,  $\partial_t\epsilon$ in terms of $\mathcal{E}$ can only be found perturbatively in derivatives. For our later computations we need to go through the derivative expansion up to $40^{th}$ order. We have done this and the result related to the first $20$ orders are given in Appendix \ref{energy_density}.

The above discussion makes the statement below \eqref{sigma_X_a} more clear. In the linear regime and to first order in derivatives, \eqref{mathcal_E} takes the following form:
\begin{equation}\label{physical_meaning}
\mathcal{E}=\,\mathfrak{h}\,\partial_t\epsilon\,.
\end{equation}
Comparing this with $\mathcal{E}= c\, T\,\frac{\delta T}{T}$, we infer that $\partial_t \epsilon$ in our EFT, actually acts as a temperate disturbance in the system. Here, $c$ is the heat capacity.  In addition, $\mathfrak{h}$ is specified as $\mathfrak{h}= T\,c$.
\subsubsection{Quadratic action and the free propagators}
Substituting \eqref{patrtial_t_epsilon} into $\mathcal{L}_{\text{eff}}^{(2)}[\epsilon, \epsilon_a]$, we find $\mathcal{L}_{\text{eff}}^{(2)}[\mathcal{E}, \epsilon_a]$ as
\begin{equation}\label{L_2}
\mathcal{L}_{\text{eff}}^{(2)}=\,\,-\epsilon_a\,\partial_t\mathcal{E}+D_0\,\epsilon_a\,\partial_i^2\mathcal{E}\,+\,i\frac{D_0\,\mathfrak{h}}{\beta}\partial_i\epsilon_a\sum_{n=1}^{40}\chi_n\left(\frac{\partial_t}{\lambda}\right)^{n-1}\partial_i\epsilon_a\,.
\end{equation}
From the above quadratic Lagrangian we find the free propagators as \footnote{Note that in the original path integral, action appears as $\exp(i S)$. After integrating out the UV fields, this factor is conventionally written as  $\exp(-S_{\text{EFT}})$. Actually this is $S_{\text{EFT}}$ from which, we derive propagators and vertices. On the other hand, our EFT was given in terms of  $\mathcal{L}_{\text{eff}}$. Then it is clear that $S_{\text{EFT}}\equiv I_{\text{EFT}}=-i \int \mathcal{L}_{\text{eff}}$ ; the $-i$  factor, indeed, compensates the minus sign in $\exp(-S_{\text{EFT}})$ and gives the original $i$ factor of $\exp(i S)$. Thus in order to do further computations, we have to firstly multiply $\mathcal{L}_{\text{eff}}$ with the  factor $-i$. Doing so, the quadratic effective action can be written as $S_{\text{EFT}}=\frac{1}{2}\int_{\omega,\boldsymbol{k}}  \,\phi^{A\,*}_{\omega,\boldsymbol{k}} \,P_{AB}(\omega,\boldsymbol{k})\,\phi^{B}_{\omega,\boldsymbol{k}}$, with $\phi_A=(\varphi_a, \varepsilon)$. Then the free propagators are simply given by $G^{(0)}_{AB}=(P^{-1})_{AB}$. } 
\begin{equation}\label{tree_prop}
G^{0}_{\mathcal{E}\mathcal{E}}=\frac{2\, D_0\,T\,\mathfrak{h}\,k^2}{\omega^2+(D_0\,k^2)^2}\,\sum_{n=1} \chi_n\bigg(\frac{-i\,\omega}{\lambda}\bigg)^{n-1},\,\,\,\,\,\,G^{0}_{\mathcal{E}\epsilon_a}=\frac{1}{\omega+i D_0\,k^2},\,\,\,\,G^{0}_{\epsilon_a\mathcal{E}}=\frac{-1}{\omega-i D_0\,k^2}\,.
\end{equation}
It should be mentioned that at $\lambda\rightarrow \infty$  our quadratic Lagrangian simplifies to exactly that of Refs.\cite{Chen-Lin:2018kfl,Jain:2020fsm}. In this case, the above propagators are exactly the same as the corresponding propagators in the  the mentioned references \footnote{However, as mentioned in \cite{Jain:2020fsm}, $G^{0}_{\mathcal{E}\epsilon_a}$ and $G^{0}_{\epsilon_a\mathcal{E}}$ computed in \cite{Chen-Lin:2018kfl} are different from \eqref{tree_prop} with an overall sign.}.

In order to compute $G^{R}_{\mathcal{E}\mathcal{E}}$, we use the Kramers-Kronig relation (combined with KMS)\footnote{We follow the convention of Refs.\cite{Crossley:2015evo,Glorioso:2018wxw} to define $G^{R}(\omega,k)=G_{ra}(\omega,k)=\frac{-i \delta^2{W}}{\delta A_{a0}(\omega,k)\delta A_{r0}(-\omega,-k)}$. Then equation \eqref{Kremers}  matches with the equation (2.31) in \cite{Glorioso:2018wxw}.}
\begin{equation}\label{Kremers}
G^{0}_{\mathcal{E}\mathcal{E}}=\coth\left(\frac{\beta \omega}{2}\right)\text{Im}\,G^{R(0)}_{\mathcal{E}\mathcal{E}},
\end{equation}
we then find
\begin{equation}\label{G_R}
G^{R(0)}_{\mathcal{E}\mathcal{E}}=\,\frac{i\,\mathfrak{h}\,D_0\,k^2\,\left(1+\frac{\omega^2}{\lambda^2}\right)}{\omega+\,i\,D_0 k^2}\,,
\end{equation}
which agrees with \cite{Blake:2017ris} and \cite{Gu:2016oyy} \footnote{Let us denote that this relation is exact in the sense that it continues to hold to all orders in derivative. We would like to thank Hong Liu for pointing this out.}.

At this point we can specify the value of $\mathfrak{h}$ in terms of physical quantities in the system. Using \eqref{G_R}, we  compute the thermal conductivity:
 \begin{equation}\label{h}
\kappa=\lim_{\omega\to 0}\lim_{k\to 0}\frac{\omega \beta}{k^2}\text{Im} G^{R(0)}_{\mathcal{E}\mathcal{E}}=\,\mathfrak{h}\,\beta\,D_0\,\,\,\,\,\,\,\,\,\xrightarrow[]{}\,\,\,\,\,\,\,\,\,\boxed{\kappa=\,D_0\,c}
 \end{equation}
where we have used $\mathfrak{h}=T \,c$ found below \eqref{physical_meaning} \footnote{Note that $G_i$ in \eqref{G_i_M_2} can be rewritten in terms of temperature fluctuations: $G_i=-\kappa(\delta T)\partial_i T$ with $\kappa(\delta T)=\kappa + \kappa_1 \frac{\delta T}{T}+\cdots$. In terms of our $D_0$ and $D_1$,  one can easily show that $\kappa=D_0\,c$ and $\kappa_1=(D_0+D_1)c$.}. What we need to do in the following is to compute the loop correction to  \eqref{tree_prop} in order to find the one-loop corrected version of \eqref{G_R}. Doing so, we will be able to study the effect of nonlinear fluctuations on the diffusion pole.
 
 \subsubsection{Cubic action and coupling constants}
 As in the quadratic Lagrangian, we substitute \eqref{patrtial_t_epsilon} into $\mathcal{L}_{\text{eff}}^{(3)}[\epsilon, \epsilon_a]$ to find the cubic Lagrangian in terms of $\mathcal{E}$ and $\epsilon_a$, namely $\mathcal{L}_{\text{eff}}^{(3)}[\mathcal{E}, \epsilon_a]$ \footnote{Let us remind that as mentioned earlier, we do  not display $\mathcal{L}_3[\epsilon, \epsilon_a]$ in the paper. }. The result is formally written as 
 \begin{equation}\label{Final_1_loop_L}
 	\begin{split}
  \mathcal{L}_{\text{eff}}^{(3)}=\,\frac{\lambda_1}{2}\,\partial_i^2\epsilon_a\,\mathcal{E}^2+&i\,c\,T^2\,\lambda_2\,\sum_{n=1}\sum_{\ell=0}^{n-1}\frac{g^{(n)}_{\ell,n-\ell-1}}{\lambda^{n-1}}\,\partial_i\epsilon_a\partial_t^{\ell}(\partial_i\epsilon_a)\partial_t^{n-\ell-1}\mathcal{E}\\
  &+i\,T^3\,\lambda_3\sum_{n=3}\sum_{\ell=1}^{n-1}\frac{h^{(n)}_{\ell,n-\ell-1}}{\lambda^{n-2}}\,\partial_t\epsilon_a\partial_t^{\ell}\epsilon_a\partial_t^{n-\ell-1}\mathcal{E}\,,
   	\end{split}
 \end{equation}
with the three distinct coupling constants
 \begin{equation}\label{couplings}
\lambda_1=\frac{D_1}{T\,c},\,\,\,\,\,\,\,\,\,\lambda_2=\frac{D_0+D_1}{T\,c}\,,\,\,\,\,\,\,\,\,\,\,\lambda_3=\frac{6}{T^2\lambda}\,.
 \end{equation}
It should be noted that $\lambda_1$ and $\lambda_2$ couplings come from $M_2$  term in \eqref{The_most_general_L}, while, $\lambda_3$ coupling comes from $M_1$ term in  that equation. At first order in the derivative expansion, our $\lambda_1$ and $\lambda_2$ terms are exactly $\lambda$ and $\tilde{\lambda}$ terms of equation (B.1) in Ref.\cite{Chen-Lin:2018kfl}. Since this Reference does not exceed the first order in the derivative expansion, our $\lambda_3$ term has no analogues there. Note that in \eqref{Final_1_loop_L}, the $\lambda_3$ term starts to contribute from the third order of the derivative expansion.

In Appendices \ref{Coeff_3rd_partial_i} and \ref{Coeff_3rd_partial_t}, we have shown  $g^{(n)}_{\ell,n-\ell-1}$'s and $h^{(n)}_{\ell,n-\ell-1}$'s for the first  $20$ orders of the derivative expansion. Although  as any other part of the computations, we have computed these coefficients to  $40^{th}$ order in the derivative expansion.

 \section{Loop computations}
 \label{loop}
 In order to compute the one-loop renormalized Green's function, we use the standard textbook method:
 \begin{equation}\label{}
\langle\mathcal{E}\mathcal{E}\rangle=\,\langle\mathcal{E}\mathcal{E}\rangle_{0}+ i\langle\mathcal{E}S_{\text{int}}\mathcal{E}\rangle-\frac{1}{2}\langle\mathcal{E}S_{\text{int}}^2\mathcal{E}\rangle+\cdots\,.
 \end{equation}
 Then following \cite{Chen-Lin:2018kfl}, we parameterize the corrections to the Green's function as a numerator $C(\omega,\textbf{k})$ and  a self-energy $\Sigma(\omega, \textbf{k})$. The latter corresponds to diagrams that resum, shown by the gray blobs below, while the former corresponds to those that do not so, shown by shaded blobs in the following. Diagrammatically, we write
  \begin{equation}\label{}
   \begin{split}
 G_{\mathcal{E}\epsilon_a}(p)(-C(p))G_{\epsilon_a\mathcal{E}}(p)=&\,\feynmandiagram [small, horizontal=a to d] {
 	a -- [small] b};+\,\feynmandiagram [small, horizontal=a to d] {
 	a -- [small] b --[scalar] c [blob] d -- [scalar] e -- [small] f,
 };\,,\\
G_{\mathcal{E}\epsilon_a}(p)(-\Sigma(p))G_{\mathcal{E}\mathcal{E}}(p)=&\,\left[	\feynmandiagram [small, horizontal=a to d] {
	a -- [small] b --[scalar] c [blob,{/tikz/fill=gray}] d };+\feynmandiagram [small, horizontal=a to d] {
	a -- [small] b --[scalar] c [blob,{/tikz/fill=gray}] d };\feynmandiagram [small, horizontal=a to d] {
	a -- [small] b --[scalar] c [blob,{/tikz/fill=gray}] d };+\cdots\right]	\feynmandiagram [small, horizontal=a to d] {
	a -- [small] b};\\
=&\,\frac{\feynmandiagram [small, horizontal=a to b] {
		a -- [small] b --[scalar] c [blob,{/tikz/fill=gray}] d -- [small] e};\,}{1-\feynmandiagram [small, horizontal=a to d] {
		a -- [small] b --[scalar] c [blob,{/tikz/fill=gray}] d  };}\,.
  \end{split}
  \end{equation}
   Then the Green's functions can be written as
   \begin{equation}
   G_{\mathcal{E}\mathcal{E}}(p)=\frac{\feynmandiagram [small, horizontal=a to d] {
   		a -- [small] b};+\,	\feynmandiagram [small, horizontal=a to d] {
   		a -- [small] b --[scalar] c [blob] d -- [scalar] e -- [small] f,
   	};}{1-\left(\feynmandiagram [small, horizontal=a to d] {
   		a -- [small] b --[scalar] c [blob,{/tikz/fill=gray}] d };+\feynmandiagram [small, horizontal=a to d] {
   		a [blob,{/tikz/fill=gray}] b --[scalar] c -- [small] d};\right)}\\
   =\,\frac{G^0_{\mathcal{E}\epsilon_a}(p)(-C(p))G^0_{\epsilon_a\mathcal{E}}(p)}{1+\left(G^0_{\mathcal{E}\epsilon_a}\Sigma(p)-\Sigma^*(p)G^0_{\epsilon_a\mathcal{E}}(p)\right)}\,.
   \end{equation}
   By using the free propagators given in \eqref{tree_prop}, we simply find
   \begin{equation}\label{G_{EE}}
    G_{\mathcal{E}\mathcal{E}}(\omega,\textbf{k})=\,\frac{C(\omega,\textbf{k})}{\omega^2+D_0^2 \textbf{k}^4+2\,\omega\,\text{Re}\,\Sigma(\omega,\textbf{k})+2\,D_0 \,\textbf{k}^2\,\text{Im}\,\Sigma(\omega,\textbf{k})}
     \end{equation}
where 
\begin{equation}\label{C_expanded}
C(\omega,\textbf{k})=2\,T^2\,\kappa\,\textbf{k}^2\,\sum_{n=1}^{40} \chi_n\bigg(\frac{-i\,\omega}{\lambda}\bigg)^{n-1}+\text{loop corrections}\,.
\end{equation}
Using Kramers-Kronig relations, we can also parameterize  the one-loop retarded Green's function as the following:
\begin{equation}\label{G_R_loop}
    G^{R}_{\mathcal{E}\mathcal{E}}(\omega,\textbf{k})=\,\frac{i\bigg((\kappa+\delta \kappa)\big(1+\frac{\omega^2}{\lambda^2}\big)-\kappa\,\frac{2\delta\lambda}{\lambda}\, \frac{\omega^2}{\lambda^2}\bigg)T\,\textbf{k}^2}{\omega+i\,D_0\,\textbf{k}^2+\Sigma(\omega,\textbf{k})}\,.
\end{equation}
In the expression above, our focus will be on the denominator. We want to compute the self-energy, $\Sigma$, perturbatively, in the derivative expansion. Then we will be able to see the influence of fluctuations on the diffusion pole in the derivative expansion.
\subsection{Self-energy}
The detailed calculation of $\Sigma$ is given in the Appendix \ref{loop_App}.
We find the cutoff independent part of the self-energy as
\begin{equation}\label{self_energy}
\Sigma(\omega,k)=\frac{cT^2}{4D_0^2}\,k^2\,\lambda_1\,\left[k^2-\frac{2i\omega}{D_0}\right]^{-1/2}\left(\sigma_{1}(\omega,k)+\sigma_{2}(\omega,k)+\sigma_{3}(\omega,k)+\cdots\right)\,,
\end{equation}
where $\sigma_j$ function denotes the $j^{th}$ order correction in the derivative expansion. We have found that 
\begin{equation}\label{F_G}
\begin{split}
\sigma_{1}(\omega,k)=&\,\omega\,\lambda_1-(\omega+i D_0 k^2)\lambda_2\,,\\
\sigma_{2n}(\omega,k)=&\,0\,,\\
\sigma_{2n+1}(\omega,k)=&\,(\omega+i D_0 k^2)\,\bigg[F_{2n}(\omega, iD_0 k^2)\,\lambda_1+G_{2n}(\omega, iD_0 k^2)\,\lambda_2\bigg]\,,
\end{split}
\end{equation}
where $F_{2n}$ and $G_{2n}$ are polynomials of $\omega$ and $i D_0 k^2$ (see Appendix \ref{loop_App}). We have also defined $r=D_1/D_0$. Here, some important points should be mentioned:
\begin{itemize}
	\item The corrections at any \textit{even} order in the derivative expansion vanish.
	\item The $\lambda_3$-coupling in \eqref{Final_1_loop_L} does not contribute to $\Sigma$ at any order in the derivative expansion. The reason is that the corresponding Feynman diagrams are fully divergent after the renormalization, without any finite parts.
	\item Except for the first order in derivatives, at any other \textit{odd} order, the correction vanishes at the classical diffusion pole, namely at $\omega=-i D_0 k^2$.
	\end{itemize}

\subsection{Long time tails and breakdown of the derivative expansion}
The standard derivative expansion in our diffusion model is based on the assumption that $\omega\sim k^2$. On the other hand, our loop computations show  that\footnote{Note that the absence of even orders contributions to $\Sigma$ is just due to the special form of the Schwarzian. }
\begin{equation}\label{break_down}
\begin{split}
\text{at}\,\,1^{st}\,\text{order in derivative}:&\,\,\,\Sigma(\omega,k)\sim|\omega|^{3/2}\sim(k^2)^{3/2}\,,\\
\text{at}\,\,3^{rd}\,\text{order in derivative}:&\,\,\,\Sigma(\omega,k)\sim|\omega|^{3/2}\sim(k^2)^{7/2}\,,\\
\cdots\,\,\,\,\,\,\,\,\,\,\,\,\,\,\,\,\,\,\,\,&\\
\text{at}\,\,(n_{\text{odd}})^{th}\,\text{order in derivative}:&\,\,\,\Sigma(\omega,k)\sim|\omega|^{(2n_{\text{odd}}+1)/2}\sim(k^2)^{n_{\text{odd}}+1/2}\,.\\
\end{split}
\end{equation}
Obviously, these contributions are not consistent with our derivative counting scheme; the self-energy cannot be expressed by the natural powers of $\omega$ and $k^2$, but it shows some non-analyticities.

The appearance of  non-analytic contributions of the form $|\omega|^{n/2}$ is the result of
nonlinear interactions in $\mathcal{L}_{\text{EFT}}$. It is easy to see that the presence of $|\omega|^{n/2}$ in $ G^{R}_{\mathcal{E}\mathcal{E}}(\omega,\textbf{k}\rightarrow0)$ leads $G^{R}_{\mathcal{E}\mathcal{E}}(t,x)$ to show the power-law relaxation $\sim t^{-(n+2)/2}$. This non-exponential damping of the response function is obviously similar to the long-time tail effect in stochastic hydrodynamics \cite{Kovtun:2003vj,Chen-Lin:2018kfl}.

The nonlinear interactions in our model have another consequence; the breakdown of derivative expansion. In order to understand why this is the case, we only need to consider  \eqref{break_down} at first order. In the hydrodynamic limit, $|\omega|^{3/2}$ is smaller than the first order contribution $\mathcal{O}(\omega)$, but is larger than the second order contribution $\mathcal{O}(\omega^2)$. It simply indicates that the \textit{derivative expansion  in $1+1$ dimensional hydrodynamics breaks down beyond the first order} \footnote{See \cite{Kovtun:2012rj} for the discussion about the breakdown of derivative expansion in $(2+1)$ and $(3+1)$ dimensional cases.} \footnote{It has recently been discovered that the breakdown of hydrodynamics may also occur in certain condensed matter systems. For instance, the breakdown of diffusion at the edge of dirty Quantum Hall systems \cite{Delacretaz:2020jis} and breakdown of hydrodynamics in a fracton fluids \cite{Glorioso:2021bif}. }. Note that this statement is independent of the assumptions based on which, we have  studied the SYK chain in this work.
\section{Modified dispersion relation}
\label{modified}
It should be recalled that in our model, the linear hydro dispersion relation is exact to all order in the derivative expansion, i.e. $\omega=-i D_0 k^2$. In a more general case, we could consider the following dispersion relation\footnote{The truncation we choose to use is just to simplify the calculation. By performing tedious calculations, it can be proved that even if we use the non-truncated dispersion relation, \eqref{break_down} will be satisfied.}
\begin{equation}\label{linear}
\omega=-i D_{(1)} k^2-i D_{(2)} k^4-i D_{(3)} k^6+\,\mathcal{O}(k^8)=\,-i\sum_{n=1}D_{(n)}k^{2n}\,.
\end{equation}
The interesting point is that in any order in derivatives, the corresponding linear and nonlinear contributions, coming from \eqref{linear} and \eqref{break_down}, are two consecutive powers of $(k^2)^{1/2}$. For instance, at $3^{rd}$ order, the linear contribution $\sim (k^2)^3$ and the nonlinear contribution $\sim (k^2)^{7/2}$. This suggests that:
\newline\newline
\textit{Although the expansion \eqref{linear} fails to work due to nonlinear effects given in \eqref{break_down}, one can still represents $\omega$ as a power series\ in $(k^2)^{1/2}$}.  We find it convenient to write the series as follows 
\begin{equation}\label{New_Exp}
\begin{split}
\omega=&\,-i\sum_{n=1}\,k^{2n}\, \big(D_{(n,1)}\pm\,i D_{(n,2)}(k^{2})^{1/2}\big)\\
=&\,-i\,k^2\,\underbrace{ \big(D_{(1,1)}\pm\,i D_{(1,2)} (k^{2})^{1/2}\big)}_{1^{{st}} \,\text{order}}-i\, k^4\,\underbrace{ \big(D_{(2,1)}\pm\,i D_{(2,2)} (k^{2})^{1/2}\big)}_{2^{nd}\,\text{order}}\,+\,\mathcal{O}(k^6)\,.
\end{split}
\end{equation}
It is important to note that by ``$n^{th}$ order" in the right side of \eqref{New_Exp}, we mean the contribution to the dispersion relation that comes from the nonlinear Lagrangian at order $n$ in the derivative expansion. From now on we will refer to \eqref{New_Exp} as \textit{modified dispersion relation}. 
Let us emphasize that the above expansion is not actually a derivative expansion. The derivative expansion is meaningful only at the Lagrangian level. Our results show that at the level of dispersion relation, the derivative expansion should be replaced by \eqref{New_Exp}.
The modified dispersion relation has some properties: 
\begin{enumerate}
	\item The $\pm$ in \eqref{New_Exp} indicates that, due to the nonlinear effects,  the classical diffusion pole is split into two poles with opposite real parts (see Fig.\ref{modes} below). This is consistent with the same result in the first order of the derivative expansion found in Ref.\cite{Chen-Lin:2018kfl}.
	\item At any order in the derivative expansion, the modified dispersion relation is determined by twice the transport coefficients of the classical dispersion relation.
	\item As we will show below, it turns out that that $D_{(1,1)}=D_{(1)}$. On the other hand,  in our model $D_{(n)}=0, n>1$.  It would be interesting to see if there is a general relation between $D_{(n,1)}$ and $D_{(n)}$ in the extended model.
	\end{enumerate}
It is convenient to use dimensionless momentum and frequency at this time. To this end, we rearrange the series as
\begin{equation}\label{wn}
\wn=\,-i\sum_{n=1}\qn^{2n}\,\bigg(\tilde{D}_{(n,1)}(r,s)\,\pm\,i\tilde{D}_{(n,2)}(r,s)\,(\qn^2)^{1/2}\bigg)
\end{equation}
with    
\begin{equation}\label{rescale}
\wn=\,\frac{D_0\,\omega}{\kappa^2},\,\,\,\,\,\,\,\,\,\,\,\qn=\frac{D_0\,k}{\kappa},\,\,\,\,\,\,\,\,\,\,\,\tilde{D}_{(n,j)}=\,\frac{\kappa^{2n-3+j}}{D_0^{2n-2+j}}\,D_{(n,j)}:\,j=1,2\,.
\end{equation}
The parameters $r$ and $s$ are defined as
\begin{equation}\label{}
r=\frac{D_1}{D_0},\,\,\,\,\,\,s=\,\frac{\kappa^2}{D_0\,T}\,.
\end{equation}
For concreteness, we discuss the convergence of  \eqref{New_Exp} for $r=\frac{1}{2}$ and $s=250$.  The corresponding first forty coefficients have been given in Appendix \ref{coeff_App}. For now, let us limit the discussion to real values of momentum. Then \eqref{wn} takes the following simple form
\begin{equation}\label{2M}
\wn=\,-i\sum_{n=1}^{\mathcal{M}}\qn^{2n}\,\bigg(\tilde{D}_{(n,1)}(r,s)\,\pm\,i\tilde{D}_{(n,2)}(r,s)\,|\qn|\bigg)\equiv\, -i \sum_{n=1}^{2\mathcal{M}} c_{n} |\qn|^n\,.
\end{equation}
\begin{figure}[tb]
	\centering
	\includegraphics[width=0.5\textwidth]{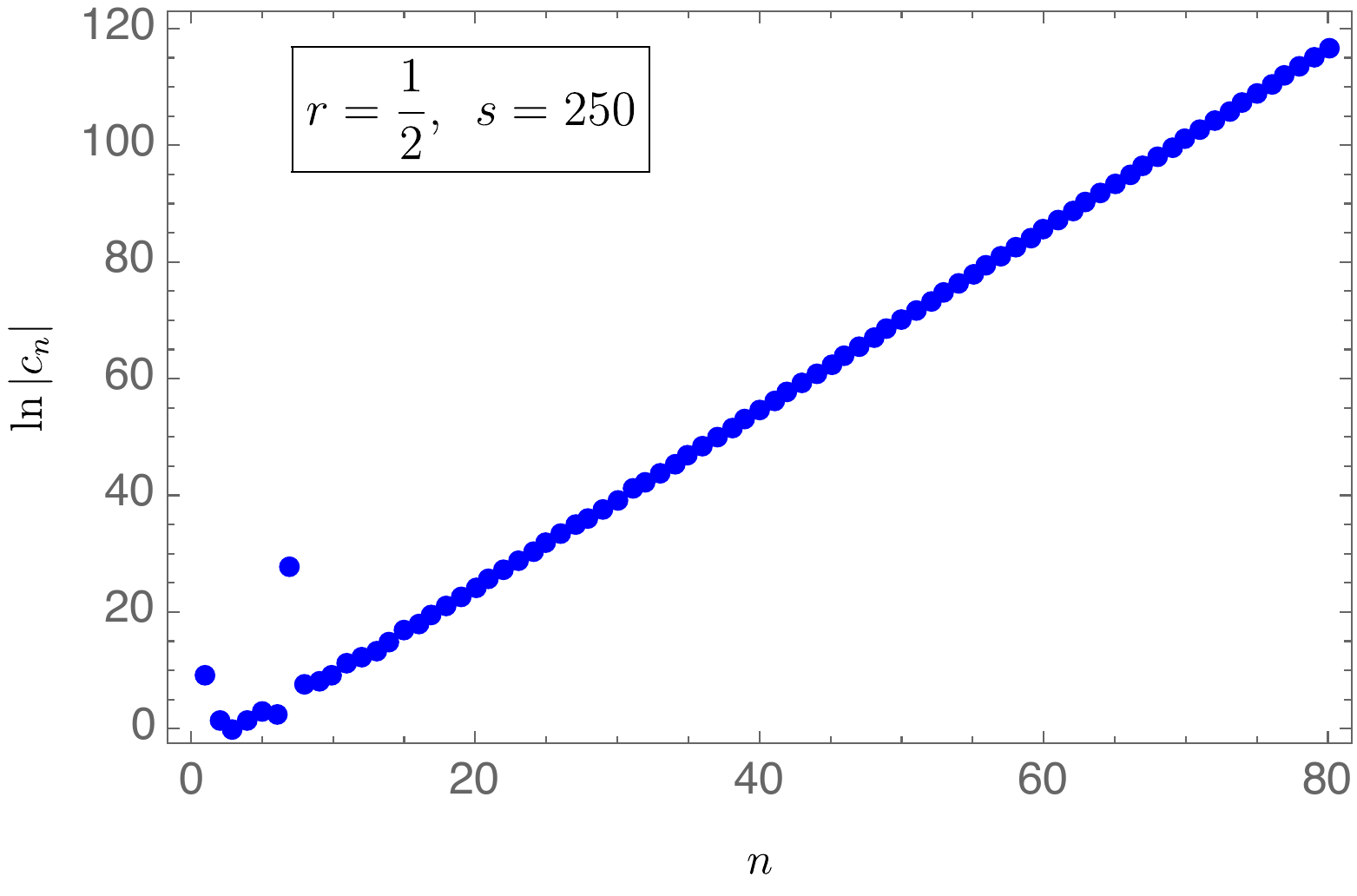}
	\caption{Illustration of coefficients of the series  \eqref{2M}. In our current case, $\mathcal{M}=40$, so the modified series is determined by  $80$ transport coefficients. }
	\label{log}
\end{figure}
In Fig.\ref{log} we have displayed the logarithm of the absolute value of the coefficients $c_n$.
For the mentioned value of the parameters $r$ and $s$, the slope of the plot at large $n$ tends to $4.720$. So the series approximately converges to 
\begin{equation}\label{}
\qn_c=\lim_{n\rightarrow\infty}\left(\ln |c_{n+1}|-\ln|c_{n}|\right)^{-1}=\,\lim_{n\rightarrow\infty}\left(\ln\bigg|\frac{ c_{n+1}}{c_n}\bigg|\right)^{-1}\approx 0.211\,.
\end{equation}
Let us denote that the value of $\qn_c$ only depends on the choice of $r=D_{1}/D_{0}$. In fact, as can be seen in Appendix \ref{Table_1}, at large $n$, all coefficients are proportional to $s^2$ resulting in $c_{n+1}/c_{n}$ independent of $s$. Then the only place where the effect of $s$ may appear is in the low-order terms in Fig.\ref{log}.  

In summary, we find that for a given value of $r$, the series \eqref{wn} converges to a finite value. Using \eqref{rescale}, the dimensionfull radius of convergence is given by
\begin{equation}\label{}
k_c=\,\qn_c\,\frac{\kappa}{D_0}=\,\#\frac{\kappa}{D_0}\,,
\end{equation}
where the numerical factor depends on the value of $r$.
\begin{figure}
	\centering
	\includegraphics[width=0.5\textwidth]{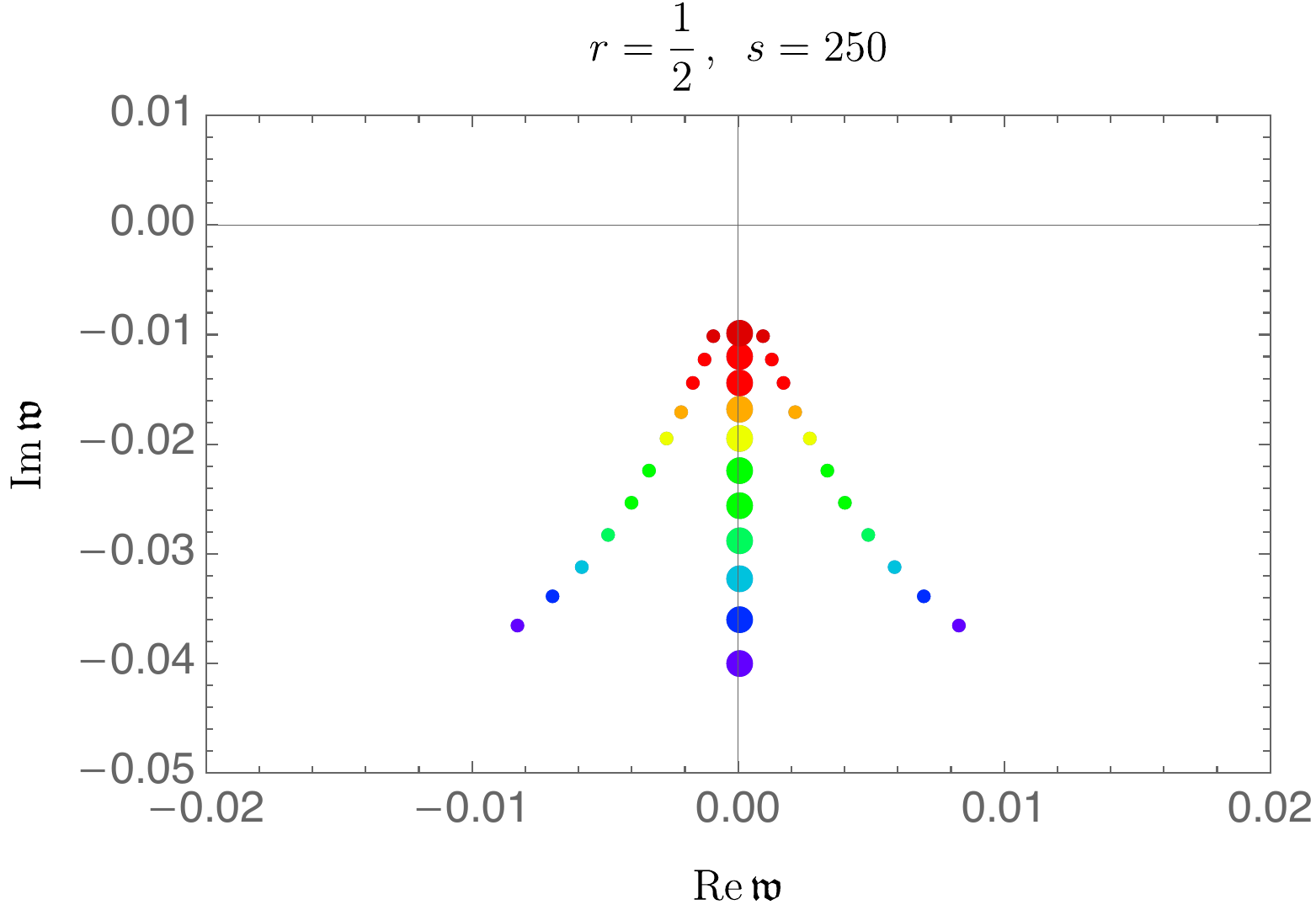}
	\caption{Splitting of the classical diffusion mode (represented by large colored dots) into two mode represented by the small dots. Each group of colored path-like points, starting with dark red and ending with purple, shows the change of a particular mode when $\qn$ discretely increases from $0.1$ to $0.2$. We have considered 10 regular steps by increments of $\Delta\qn=0.01$. }
	\label{modes}
\end{figure}

Before ending this section let us mention two points regarding the dispersion relations.
\begin{itemize} 
	\item As mentioned  earlier, one of the effects of nonlinear fluctuations is to split the classical diffusion pole into two poles with non-vanishing real parts. It can be seen from  Fig.\ref{modes} that the real parts of these modes are opposite to each other.
	\item It would be interesting to study these modes at complex momenta to find the reason behind the finite radius of convergence of the modified series.  Following \cite{Grozdanov:2019kge}, it should be determined whether there are any mode collisions that limit convergence \footnote{See also \cite{Withers:2018srf,Grozdanov:2019uhi,Abbasi:2020ykq,Abbasi:2020xli,Asadi:2021hds,Jeong:2021zhz,Grozdanov:2021gzh,Jeong:2021zsv,Wu:2021mkk,Baggioli:2020loj,Arean:2020eus,Heller:2020uuy,Baggioli:2021ujk,Jansen:2020hfd,Heller:2021oxl,Huh:2021ppg,Liu:2021qmt,Cartwright:2021qpp,Amado:2008ji} for recent studies on various aspects of mode collision and convergence radius of hydrodynamics.}. We leave more investigations on this issue for future work.
	\end{itemize}

\section{Discussion and outlook}
\label{conclusion}
In this paper we investigated the effect of hydrodynamic fluctuations in the SYK chain.  By constructing the interacting EFT of energy diffusion up to $40^{th}$ order in derivatives, as given by \eqref{L_2} and \eqref{Final_1_loop_L}, we computed the one-loop self-energy to the same order in derivatives (see \eqref{self_energy}).  The special form of derivative corrections in the self-energy led us to the following results:
\begin{enumerate}
	\item The dispersion relation of the energy diffusion does no longer follow the derivative expansion, namely the classical relation $\wn=\wn(\qn^2)$.  However,  the dispersion relation can be still represented  as a power series, but in terms of of $|\qn|$, say $\wn=\wn(|\qn|)$. We referred to it as the modified (or quantum) dispersion  relation. 
	\item Up to any order in the derivative expansion, the number of transport coefficients in the modified  dispersion relation  is twice of that in the classical dispersion relation. 
	\item The modified dispersion relation is convergent in the momentum space. The radius of convergent, $k_c$, is proportional to the ratio of the thermal conductivity to the diffusion constant. Since $\kappa= D_0\,c$, one immediately concludes that $k_c\sim c$, that is, it is proportional to the number of degrees of freedom. Therefore, in the large $N$ limit, $k_c$ has a large value. 
	\end{enumerate}

As the first generalization of our results, one might think of including the two-loop corrections along with derivative orders larger than $40$. Although the latter is a technical problem to obtain better convergence, the former one is physically important; it is actually equivalent to considering $\frac{1}{N^2}$ corrections to the self-energy.

To be more realistic, a larger EFT cut-off value should be considered to make room for the inclusion of non-hydrodynamic modes. In the simplest case, only the lowest quasi-normal mode can be included. The corresponding EFT on CTP contour then will have two additional dynamical variables, associated with this quasinormal mode. In this setting, the influence of nonlinear fluctuations on diffusion and  non-hydro mode should be studied.
Then it would be interesting to see what kind of singularity limits the  convergence of the modified dispersion relation.

More realistically, one can think of a 3+1 dimensional model. One way is to consider that the Schwarzian \eqref{H} depends on three spatial dimensions.  Repeating all the steps taken in the paper, one would arrive at the analogue of  \eqref{self_energy} in 3+1 dimension as follows
\begin{equation*}\label{}
\tilde{\Sigma}(\omega,k)=\frac{cT^2}{4D_0^2}\,k^2\,\lambda_1\,\left[k^2-\frac{2i\omega}{D_0}\right]^{1/2}\left(\tilde{\sigma}_{1}(\omega,k)+\tilde{\sigma}_{2}(\omega,k)+\tilde{\sigma}_{3}(\omega,k)+\cdots\right)\,.
\end{equation*}
As it is seen, at the $n^{th}$ order in derivatives, $\tilde{\Sigma}_{(n)}\sim (k^2)^{n+1/2}$. It can be concluded that for the 3+1 dimensional model, the expansion \eqref{New_Exp} should be changed to allow four transport coefficients at any $n$; i.e. $D_{(n,j)}: j=1,2,3,4$. However, it must be carefully considered; it turns out the coefficients of $(k^2)^{n}$ and $(k^2)^{n+1/2}$ will get contributions from two different orders of $n$:
\begin{equation*}\label{}
\begin{split}
\text{coefficient of } (k^2)^{n}\,\,=&\,D_{(n-1,3)}+D_{(n,1)}\,,\\
\text{coefficient of } (k^2)^{n+1/2}=&\,D_{(n-1,4)}+D_{(n,2)}\,.
\end{split}
\end{equation*}
Because of the above pairing between the coefficients $D_{(n,j)}$, one needs to use effectively $2n+4$ coefficients. Here, $4$ corresponds to $D_{(1,1)}$, $D_{(1,2)}$, $D_{(n,3)}$ and $D_{(n,4)}$; they are not paired with any other coefficients at order $n$.
We leave more investigation on 3+1 dimensional case to a future work.

Considering all aforementioned points, one may think of studying the same problem in the case of Quark Gluon Plasma (QGP). The hydrodynamic fluctuations in QGP have been studied previously on the static   \cite{Kovtun:2011np} and also on the evolutionary \cite{Akamatsu:2016llw} background. In order to extend the calculations of this paper to the QGP case, two important points should be considered. \textit{First}, the dynamics of energy density in our model is similar to that of 
charge density or transverse momentum density in QGP, which is a diffusion process. 
In QGP, however,  there are also sound modes. They actually propagate and diffuse into the medium.  \textit{Second},
QCD is not a large-N theory.  Therefore, to be applicable to QGP, our setting should be extended to include  relativistic hydrodynamic fluctuations and also to go beyond one-loop computations. 

Another direction to explore this problem is holography. In the pioneering work of Ref.\cite{Caron-Huot:2009kyg}, the one-loop calculation of Einstein's gravity in AdS space has been proven to reproduce the long-time tail effect in boundary quantum field theory at a finite temperature. Another interesting calculation can be found in \cite{Cheng:2021mop}, where the leading derivative contribution of the self energy is directly computed from gravity.  But one can do it in yet another way. The idea might be to use the  holographic prescription of the Schwinger-Keldysh contour \cite{Glorioso:2018mmw} \footnote{See \cite{deBoer:2018qqm,Skenderis:2008dg} for other holographic prescriptions.} and compute the boundary nonlinear effective action from gravity to higher orders in derivatives\footnote{According to the prescription of \cite{Glorioso:2018mmw}, the ``quadratic" effective action for the boundary theory has been computed in several different situations \cite{Ghosh:2020lel,Bu:2021clf,Bu:2020jfo}.}. The next step will be simple, calculating the loop correction of the energy density response function in the boundary. 

Lastly, it would be interesting to continue the loop computations in this paper and specify all parameters introduced in  \eqref{G_R_loop}.
Doing so, the full one-loop corrected response function  will be determined. This response function has an important characteristic. It is actually related to a maximally chaotic quantum system \cite{Blake:2021wqj}. Then it would be very interesting to find the pole-skipping phenomenon in this response function and extract the chaos point from it \cite{Blake:2017ris}. It is well-known that the chaos point might be located out of the regime  validity of hydrodynamics \cite{Grozdanov:2019uhi,Abbasi:2020ykq}. However, the fact that we have determined $\mathcal{L}_{\text{eff}}$ to high orders in the derivative expansion allows to probe the chaos point. Then the main question to be answered will be how the Lyapunov exponent and butterfly velocity are affected by hydrodynamic fluctuations.

\appendix
\section*{Acknowledgment}
We are grateful to  Ali Davody, Luca Delacr\'etaz, Paolo Glorioso, Sean Hartnoll, Matthias Kaminski, Pavel Kovtun, Hong Liu and Misha Stephanov for valuable discussions and comments. 
We  would  also like to thank Armin Ghazi and Omid Tavakol for discussing the different aspects of EFT of hydrodynamics.
This work was supported by grant number 561119208 ``Double First Class'' start-up funding of Lanzhou University, China.

\section{Dimensions}
The mass dimension of  quantities involved in our system is as follows:
\begin{equation}\label{}
\begin{split}
[H]=\,[\mathfrak{h}]=\,[\mathcal{E}]=&\,2\\
[\omega]=\,[k]=\,[c]=&\,1\\
[\kappa]=&\,0\\
[\epsilon]=\,[\epsilon_a]=\,[D_0]=\,[D_1]=&\,-1\\
[\lambda_1]=\,[\lambda_2]=\,[\lambda_3]=&\,-3\,.
\end{split}
\end{equation}

\section{Coefficients of $M_2$ term in the quadratic Lagrangian }
\label{chi_n}
\begingroup
\everymath{\scriptstyle}
\scriptsize
\begin{equation}
\begin{split}
\chi_1=1,\,\,\,\,\chi_3=\frac{1}{12} \left(-\beta ^2 \lambda ^2-12\right),\,\,\,\,\,\chi_5=-\frac{1}{720} \beta ^2 \lambda ^2 \left(\beta ^2 \lambda ^2-60\right),\,\,\,\,\,\,\,\,\,\chi_7=-\frac{\beta ^4 \lambda ^4 \left(\beta ^2 \lambda ^2-42\right)}{30240},\,\,\,\,\,\,\,\chi_9=\-\frac{\beta ^6 \lambda ^6 \left(\beta ^2 \lambda ^2-40\right)}{1209600}\\
\chi_{11}=-\frac{\beta ^8 \lambda ^8 \left(5 \beta ^2 \lambda ^2-198\right)}{239500800},\,\,\,\,\,\,\,\,\,\,\chi_{13}=-\frac{\beta ^{10} \lambda ^{10} \left(691 \beta ^2 \lambda ^2-27300\right)}{1307674368000},\,\,\,\,\,\,\chi_{15}=-\frac{\beta ^{12} \lambda ^{12} \left(35 \beta ^2 \lambda ^2-1382\right)}{2615348736000}\,\,\,\,\,\,\,\,\,\,\,\,\,\,\,\,\,\,\,\,\\
\chi_{17}=-\frac{\beta ^{14} \lambda ^{14} \left(3617 \beta ^2 \lambda ^2-142800\right)}{10670622842880000}\,\,\,\,\,\,\,\,\,\,\,
\chi_{19}-\frac{\beta ^{16} \lambda ^{16} \left(219335 \beta ^2 \lambda ^2-8659098\right)}{25545471085854720000}\,\,\,\,\,\,\,\,\,\,\,\,\,\,\,\,\,\,\,\,\,\,\,\,\,\,\,\,\,\,\,\,\,\,\,\,\,\,
\end{split}
\end{equation}
\endgroup

\section{Energy density in terms of reparametrization mode}
\label{energy_density}
In order to find the invert equation \eqref{mathcal_E}, we take an anstaz for $\partial_t\epsilon$ of the form
\begin{equation}\label{}
\partial_t\epsilon=\,\sum_{n=0}\frac{1}{\mathfrak{h}}e_n \frac{\partial_t^n\mathcal{E}}{\lambda^n}+\,\sum_{n,m=0}\frac{1}{\mathfrak{h}^2}e_{n,m} \frac{\partial_t^n\mathcal{E}\partial_t^m\mathcal{E}}{\lambda^{n+m}}\,.
\end{equation}
Substituting this into \eqref{mathcal_E}, we can then read the coefficients $e_n$ and $e_{n,m}$, order by order in the derivative expansion. Here is the result related the first $20$ orders: 
\begingroup
\everymath{\scriptstyle}
\scriptsize
\begin{equation}\label{patrtial_t_epsilon}
\begin{split}
\partial_t\epsilon&=\,\frac{1}{\mathfrak{h}}\left[\mathcal{E} +\frac{\partial^2_t{\mathcal{E}}}{\lambda^2}+\frac{\partial^4_t{\mathcal{E}}}{\lambda^4}+\cdots+\frac{\partial^{20}_t{\mathcal{E}}}{\lambda^{20}}+\mathcal{O}\left(\frac{1}{\mathfrak{h}}\frac{\partial_t^{21}\mathcal{E}}{\lambda^{21}}\right)\right]-\frac{1}{\mathfrak{h}^2}\bigg[\frac{\mathcal{E}^2}{2}+\frac{1}{\lambda^2}\left(\frac{5}{2}(\partial_t\mathcal{E})^2+ 3\,\mathcal{E}\partial^2_t\mathcal{E}\right)+\frac{1}{\lambda^4}\left(\frac{19}{2}(\partial^2_t\mathcal{E})^2+14\,\partial\mathcal{E}\partial^3_t\mathcal{E}+5\,\mathcal{E}\partial^4_t\mathcal{E}\right)\\
&+\frac{1}{\lambda^6}\left(\frac{69}{2}(\partial^3_t\mathcal{E})^2+55\,\partial_t^2\mathcal{E}\partial^4_t\mathcal{E}+27\,\partial_t\mathcal{E}\partial^5_t\mathcal{E},+7\,\mathcal{E}\partial^6_t\mathcal{E}\right)\\
&+\frac{1}{\lambda^8}\left(\frac{251}{2}(\partial^4_t\mathcal{E})^2+209\,\partial_t^3\mathcal{E}\partial^5_t\mathcal{E}+119\,\partial^2_t\mathcal{E}\partial^6_t\mathcal{E}\,+44\,\partial_t\mathcal{E}\partial^7_t\mathcal{E}+9\,\mathcal{E}\partial^8_t\mathcal{E}\right)\\
&+\frac{1}{\lambda^{10}}\left(\frac{923}{2}(\partial^5_t\mathcal{E})^2+791\,\partial_t^4\mathcal{E}\partial^6_t\mathcal{E}+494\,\partial^3_t\mathcal{E}\partial^7_t\mathcal{E}+219\,\partial^2_t\mathcal{E}\partial^8_t\mathcal{E}+65\,\partial_t\mathcal{E}\partial^9_t\mathcal{E}+11\,\mathcal{E}\partial^{10}_t\mathcal{E}\right)\\
&+\frac{1}{\lambda^{12}}\bigg(\frac{3431}{2}(\partial^6_t\mathcal{E})^2+3002\,\partial_t^5\mathcal{E}\partial^7_t\mathcal{E}+2001\,\partial^4_t\mathcal{E}\partial^8_t\mathcal{E}+1000\,\partial^3_t\mathcal{E}\partial^9_t\mathcal{E}+363\,\partial^2_t\mathcal{E}\partial^{10}_t\mathcal{E}+90\,\partial_t\mathcal{E}\partial^{11}_t\mathcal{E}+\,13\,\mathcal{E}\partial^{12}_t\mathcal{E}\bigg)\\
&+\frac{1}{\lambda^{14}}\bigg(\frac{12869}{2}(\partial^7_t\mathcal{E})^2+11439\,\partial_t^6\mathcal{E}\partial^8_t\mathcal{E}+8007\,\partial^5_t\mathcal{E}\partial^9_t\mathcal{E}+4367\,\partial^4_t\mathcal{E}\partial^{10}_t\mathcal{E}+1819\,\partial^3_t\mathcal{E}\partial^{11}_t\mathcal{E}+559\,\partial^2_t\mathcal{E}\partial^{12}_t\mathcal{E}+\,119\,\partial_t\mathcal{E}\partial^{13}_t\mathcal{E}+\,15\,\mathcal{E}\partial^{14}_t\mathcal{E}\bigg)\\
&+\frac{1}{\lambda^{16}}\bigg(\frac{48619}{2}(\partial^8_t\mathcal{E})^2+43757\,\partial_t^7\mathcal{E}\partial^9_t\mathcal{E}+31823\,\partial^6_t\mathcal{E}\partial^{10}_t\mathcal{E}+18563\,\partial^5_t\mathcal{E}\partial^{11}_t\mathcal{E}+8567\,\partial^4_t\mathcal{E}\partial^{12}_t\mathcal{E}\\
&\,\,\,\,\,\,\,\,\,\,\,\,\,\,\,\,\,\,\,\,\,\,\,\,\,\,\,\,\,\,+3059\,\partial^3_t\mathcal{E}\partial^{13}_t\mathcal{E}+\,815\,\partial^2_t\mathcal{E}\partial^{14}_t\mathcal{E}+\,152\,\partial_t\mathcal{E}\partial^{15}_t\mathcal{E}+\,17\,\mathcal{E}\partial^{16}_t\mathcal{E}\bigg)\\
&+\frac{1}{\lambda^{18}}\bigg(\frac{184755}{2}(\partial^9_t\mathcal{E})^2+167959\,\partial_t^8\mathcal{E}\partial^{10}_t\mathcal{E}+125969\,\partial^7_t\mathcal{E}\partial^{11}_t\mathcal{E}+77519\,\partial^6_t\mathcal{E}\partial^{12}_t\mathcal{E}+38759\,\partial^5_t\mathcal{E}\partial^{13}_t\mathcal{E}\\
&\,\,\,\,\,\,\,\,\,\,\,\,\,\,\,\,\,\,\,\,\,\,\,\,\,\,\,\,\,\,+15503\,\partial^4_t\mathcal{E}\partial^{14}_t\mathcal{E}+\,4844\,\partial^3_t\mathcal{E}\partial^{15}_t\mathcal{E}+\,1139\,\partial_t^2\mathcal{E}\partial^{16}_t\mathcal{E}+\,189\,\partial_t\mathcal{E}\partial^{17}_t\mathcal{E}+\,19\,\mathcal{E}\partial^{18}_t\mathcal{E}\bigg)\\
&+\frac{1}{\lambda^{20}}\bigg(\frac{705431}{2}(\partial^{10}_t\mathcal{E})^2+646645\,\partial_t^9\mathcal{E}\partial^{11}_t\mathcal{E}+497419\,\partial^8_t\mathcal{E}\partial^{12}_t\mathcal{E}+319769\,\partial^7_t\mathcal{E}\partial^{13}_t\mathcal{E}+170543\,\partial^6_t\mathcal{E}\partial^{14}_t\mathcal{E}\\
&\,\,\,\,\,\,\,\,\,\,\,\,\,\,\,\,\,\,\,\,\,\,\,\,\,\,\,\,\,\,+74612\,\partial^5_t\mathcal{E}\partial^{15}_t\mathcal{E}+\,26333\,\partial^4_t\mathcal{E}\partial^{16}_t\mathcal{E}+\,7314\,\partial_t^3\mathcal{E}\partial^{17}_t\mathcal{E}+\,1539\,\partial^2_t\mathcal{E}\partial^{18}_t\mathcal{E}+230\,\partial_t\mathcal{E}\partial^{19}_t\mathcal{E}+\,21\,\mathcal{E}\partial^{20}_t\mathcal{E}\bigg)\bigg]+\mathcal{O}\left(\frac{1}{\mathfrak{h}^2}\frac{\partial_t^{21}\mathcal{E}^2}{\lambda^{21}}\right)\,.
\end{split}
\end{equation}
\endgroup

\section{KMS constraints}
\label{KMS_App}
In the following table, we display the set of KMS constraints for the first ten orders in the derivative expansion.

\begin{landscape}
\begin{table}[!htb]
	\label{}
	\begin{center}
		\begin{tabular}{|c|c|}
			\hline
			\hline
			$n$ &KMS constraints at order $n$  of the derivative expansion \\
			\hline
			\hline	
		 $1$	&$ \frac{D_0+D_1}{2 \beta }-T_{0,0}=0$\\
		 	\hline	
		 $2$	&$ T_{0,1}-2 T_{1,0}=0$\\
		 		 	\hline	
		 $3$	&$ -12 \beta  \left(2 T_{0,0}-T_{0,2}+T_{1,1}-2 T_{2,0}\right)+\frac{1}{2} \beta ^2 D_0 \lambda ^2+\frac{1}{2} D_1 \left(\beta ^2 \lambda ^2+36\right)=0$\\
		 &$ 12 \beta  T_{0,2}+\frac{1}{2} D_0 \left(\beta ^2 \lambda ^2-12\right)+\frac{1}{2} D_1 \left(\beta ^2 \lambda ^2+12\right)=0$\\
		 	\hline	
		 $4$	&$ T_{0,1}-T_{0,3}-2 T_{1,0}+T_{1,2}-T_{2,1}+2 T_{3,0}=0$\\
		 &$ 2 T_{1,2}-3 T_{0,3}=0$\\
		 	\hline	
	$5$	&$\beta  \left(\frac{1}{2} \beta  D_0 \lambda ^2 \left(\beta ^2 \lambda ^2-60\right)-720 \left(T_{0,2}-T_{0,4}-T_{1,1}+T_{1,3}+2 T_{2,0}-T_{2,2}+T_{3,1}-2 T_{4,0}\right)\right)+\frac{1}{2} D_1 \left(\beta ^4 \lambda ^4-120 \beta ^2 \lambda ^2-720\right)=0$\\
	&$\beta  \left(\frac{1}{2} \beta  D_0 \lambda ^2 \left(\beta ^2 \lambda ^2-45\right)-180 \left(2 T_{0,2}-4 T_{0,4}+3 T_{1,3}-2 T_{2,2}\right)\right)+\frac{1}{2} D_1 \left(\beta ^4 \lambda ^4-90 \beta ^2 \lambda ^2-360\right)=0$\\	 
		&$720 T_{0,4}+\frac{1}{2} \beta  D_0 \lambda ^2 \left(\beta ^2 \lambda ^2+60\right)+\frac{1}{2} \beta  D_1 \lambda ^2 \left(\beta ^2 \lambda ^2-60\right)=0$\\	 
		 	\hline	
		$6$	&$T_{0,3}-T_{0,5}-T_{1,2}+T_{1,4}+T_{2,1}-T_{2,3}-2 T_{3,0}+T_{3,2}-T_{4,1}+2 T_{5,0}=0$\\
		&$3 T_{0,3}-5 T_{0,5}-2 T_{1,2}+4 T_{1,4}-3 T_{2,3}+2 T_{3,2}=0$\\	 
		&$5 T_{0,5}-2 T_{1,4}=0$\\	 
			 	\hline	
	$7$	&$\frac{ \left(-T_{0,4}+T_{0,6}+T_{1,3}-T_{1,5}-T_{2,2}+T_{2,4}+T_{3,1}-T_{3,3}-2 T_{4,0}+T_{4,2}-T_{5,1}+2 T_{6,0}\right)}{\lambda ^6}+\frac{ \left(\beta  D_1 \left(\beta ^4 \lambda ^4-84 \beta ^2 \lambda ^2+2520\right)+\beta ^3 D_0 \lambda ^2 \left(\beta ^2 \lambda ^2-42\right)\right)}{60480 \lambda ^4}=0$\\
	&$\frac{\left(-5040 \left(4 T_{0,4}-6 T_{0,6}-3 T_{1,3}+5 T_{1,5}+2 T_{2,2}-4 T_{2,4}+3 T_{3,3}-2 T_{4,2}\right)+\frac{1}{2} \beta  D_1 \lambda ^2 \left(\beta ^4 \lambda ^4-70 \beta ^2 \lambda ^2+1680\right)+\frac{1}{2} \beta ^3 D_0 \lambda ^4 \left(\beta ^2 \lambda ^2-35\right)\right)}{ \lambda ^6}=0$\\	 
	&$-10080 \left(2 T_{0,4}-9 T_{0,6}+5 T_{1,5}-2 T_{2,4}\right)+\frac{1}{2} \beta  D_1 \lambda ^2 \left(3 \beta ^4 \lambda ^4-154 \beta ^2 \lambda ^2+1680\right)+\frac{1}{2} \beta ^3 D_0 \lambda ^4 \left(3 \beta ^2 \lambda ^2-56\right)=0$\\	 
		&$30240 T_{0,6}+\frac{1}{2} \beta ^3 D_0 \lambda ^4 \left(\beta ^2 \lambda ^2+42\right)+\frac{1}{2} \beta ^3 D_1 \lambda ^4 \left(\beta ^2 \lambda ^2-42\right)=0$\\	 
	\hline	
	$8$	&$T_{0,5}-T_{0,7}-T_{1,4}+T_{1,6}+T_{2,3}-T_{2,5}-T_{3,2}+T_{3,4}+T_{4,1}-T_{4,3}-2 T_{5,0}+T_{5,2}-T_{6,1}+2 T_{7,0}=0$\\
	&$5 T_{0,5}-7 T_{0,7}-4 T_{1,4}+6 T_{1,6}+3 T_{2,3}-5 T_{2,5}-2 T_{3,2}+4 T_{3,4}-3 T_{4,3}+2 T_{5,2}=0$\\	 
	&$5 T_{0,5}-14 T_{0,7}-2 T_{1,4}+9 T_{1,6}-5 T_{2,5}+2 T_{3,4}=0$\\
		&$2 T_{1,6}-7 T_{0,7}=0$\\	 		 
			\hline	
		$9$	&$\frac{ \left(-T_{0,6}+T_{0,8}+T_{1,5}-T_{1,7}-T_{2,4}+T_{2,6}+T_{3,3}-T_{3,5}-T_{4,2}+T_{4,4}+T_{5,1}-T_{5,3}-2 T_{6,0}+T_{6,2}-T_{7,1}+2 T_{8,0}\right)}{\lambda ^8}+\frac{\left(\beta ^5 D_0 \lambda ^2 \left(\beta ^2 \lambda ^2-40\right)+\beta ^3 D_1 \left(\beta ^4 \lambda ^4-80 \beta ^2 \lambda ^2+1680\right)\right)}{2419200 \lambda ^4}=0$\\
		&$\frac{\left(-151200 \left(6 T_{0,6}-8 T_{0,8}-5 T_{1,5}+7 T_{1,7}+4 T_{2,4}-6 T_{2,6}-3 T_{3,3}+5 T_{3,5}+2 T_{4,2}-4 T_{4,4}+3 T_{5,3}-2 T_{6,2}\right)+\frac{1}{2} \beta ^5 D_0 \lambda ^6 \left(\beta ^2 \lambda ^2-35\right)+\frac{1}{2} \beta ^3 D_1 \lambda ^4 \left(\beta ^4 \lambda ^4-70 \beta ^2 \lambda ^2+1260\right)\right)}{ \lambda ^8}=0$\\	 
		&$\frac{\left(-60480 \left(9 T_{0,6}-20 T_{0,8}-5 T_{1,5}+14 T_{1,7}+2 T_{2,4}-9 T_{2,6}+5 T_{3,5}-2 T_{4,4}\right)+\frac{1}{2} \beta ^5 D_0 \lambda ^6 \left(\beta ^2 \lambda ^2-26\right)+\frac{1}{2} \beta ^3 D_1 \lambda ^4 \left(\beta ^4 \lambda ^4-58 \beta ^2 \lambda ^2+756\right)\right)}{ \lambda ^8}=0$\\
		&$-151200 \left(2 T_{0,6}-16 T_{0,8}+7 T_{1,7}-2 T_{2,6}\right)+\frac{1}{2} \beta ^5 D_0 \lambda ^6 \left(2 \beta ^2 \lambda ^2-15\right)+\beta ^3 D_1 \lambda ^4 \left(\beta ^4 \lambda ^4-45 \beta ^2 \lambda ^2+210\right)=0$\\	 	
				&$1209600 T_{0,8}+\frac{1}{2} \beta ^5 D_0 \lambda ^6 \left(\beta ^2 \lambda ^2+40\right)+\frac{1}{2} \beta ^5 D_1 \lambda ^6 \left(\beta ^2 \lambda ^2-40\right)=0$\\	 	
							\hline	
				$10$	&$-T_{0,7}+T_{0,9}+T_{1,6}-T_{1,8}-T_{2,5}+T_{2,7}+T_{3,4}-T_{3,6}-T_{4,3}+T_{4,5}+T_{5,2}-T_{5,4}-T_{6,1}+T_{6,3}+2 T_{7,0}-T_{7,2}+T_{8,1}-2 T_{9,0}=0$\\
				&$7 T_{0,7}-9 T_{0,9}-6 T_{1,6}+8 T_{1,8}+5 T_{2,5}-7 T_{2,7}-4 T_{3,4}+6 T_{3,6}+3 T_{4,3}-5 T_{4,5}-2 T_{5,2}+4 T_{5,4}-3 T_{6,3}+2 T_{7,2}=0$\\	 
				&$-14 T_{0,7}+27 T_{0,9}+9 T_{1,6}-20 T_{1,8}-5 T_{2,5}+14 T_{2,7}+2 T_{3,4}-9 T_{3,6}+5 T_{4,5}-2 T_{5,4}=0$\\
				&$7 T_{0,7}-30 T_{0,9}-2 T_{1,6}+16 T_{1,8}-7 T_{2,7}+2 T_{3,6}=0$\\	 	
				&$9 T_{0,9}-2 T_{1,8}=0$\\	 
\hline
\hline
\end{tabular}
\end{center}
\label{}
\end{table}
\end{landscape}

\section{Coefficients of $\nabla\epsilon_a\partial_t^{\ell}(\nabla\epsilon_a)\partial_t^{n-\ell-1}\mathcal{E}$ in the cubic action}
\label{Coeff_3rd_partial_i}
\begin{table}[!htb]
	\label{table one}
	\begin{center}
		\begin{tabular}{|c|c|}
			\hline
			\hline
			$(n)$ &non-vanishing $g^{(n)}_{\ell,n-\ell-1}$  to $20^{th}$ order\,\,\,\,\,\,\,\,\,$\left(r=\frac{D_1}{D_0},   \,\,\,\,\lambda=\frac{2\pi}{\beta},\,\,\,\,\,\beta=T^{-1}\right)$  \\
			\hline
			\hline	
			$(1)$  & $g^{(1)}_{0,0}=1$  \\
			\hline
			$(3)$ & $g^{(3)}_{0,2}=\frac{1}{2 r +2},\,\,g^{(3)}_{2,0}=-\frac{r  \left(\beta ^2 \lambda ^2+12\right)+\beta ^2 \lambda ^2-12}{12 (r +1)}$  \\
			\hline
			$(5)$   & $g^{(5)}_{0,4}=\frac{8-\beta ^2 \lambda ^2}{16 r +16},\,\,\,\,\,\,g^{(5)}_{2,2}=\frac{5 \beta ^2 \lambda ^2+24}{24 r +24},\,\,\,\,\,\,\,\,g^{(5)}_{4,0}=-\frac{\beta ^2 \lambda ^2 \left(r  \left(\beta ^2 \lambda ^2-60\right)+\beta ^2 \lambda ^2+60\right)}{720 (r +1)}$\\
			\hline
			$(7)$  & $g^{(7)}_{0,6}=\frac{\beta ^4 \lambda ^4-30 \beta ^2 \lambda ^2+240}{480 (r +1)},\,\,\,\,\,\,\,g^{(7)}_{2,4}=\frac{-\beta ^4 \lambda ^4+20 \beta ^2 \lambda ^2+96}{96 r +96}$ \\
			&$g^{(7)}_{4,2}=\frac{\beta^2\lambda^2(-120+11
				\beta^2\lambda^2)}{1440(1+r)},\,\,\,\,\,g^{(7)}_{6,0}=-\frac{\beta ^4 \lambda ^4 \left(r  \left(\beta ^2 \lambda ^2-42\right)+\beta ^2 \lambda ^2+42\right)}{30240 (r +1)}$\\
			\hline
			$(9)$  &  $g^{(9)}_{0,8}=\frac{-17 \beta ^6 \lambda ^6+168 \beta ^4 \lambda ^4-5040 \beta ^2 \lambda ^2+40320}{80640 (r +1)}$\\
			&$g^{(9)}_{2,6}=\frac{\beta ^6 \lambda ^6-10 \beta ^4 \lambda ^4+200 \beta ^2 \lambda ^2+960}{960 (r +1)},\,\,\,g^{(9)}_{4,4}=-\frac{\beta ^2 \lambda ^2 \left(5 \beta ^4 \lambda ^4-44 \beta ^2 \lambda ^2+480\right)}{5760 (r +1)}$\\
			&$g^{(9)}_{6,2}=\frac{\beta ^4 \lambda ^4 \left(17 \beta ^2 \lambda ^2-84\right)}{60480 (r +1)},\,\,\,\,g^{(9)}_{8,0}=-\frac{\beta ^6 \lambda ^6 \left(r  \left(\beta ^2 \lambda ^2-40\right)+\beta ^2 \lambda ^2+40\right)}{1209600 (r +1)}$\\
			\hline	
			$(11)$	&$g^{(11)}_{0,10}=\frac{31 \beta ^8 \lambda ^8-170 \beta ^6 \lambda ^6+1680 \beta ^4 \lambda ^4-50400 \beta ^2 \lambda ^2+403200}{806400 (r +1)},\,\,g^{(11)}_{2,8}=\frac{-51 \beta ^8 \lambda ^8+280 \beta ^6 \lambda ^6-2800 \beta ^4 \lambda ^4+56000 \beta ^2 \lambda ^2+268800}{268800 (r +1)}$\\
			& $g^{(11)}_{4,6}=\frac{\beta ^2 \lambda ^2 \left(9 \beta ^6 \lambda ^6-50 \beta ^4 \lambda ^4+440 \beta ^2 \lambda ^2-4800\right)}{57600 (r +1)},\,\,\,g^{(11)}_{6,4}=-\frac{\beta ^4 \lambda ^4 \left(63 \beta ^4 \lambda ^4-340 \beta ^2 \lambda ^2+1680\right)}{1209600 (r +1)}$\\
			&$g^{(11)}_{8,2}=\frac{\beta ^6 \lambda ^6 \left(23 \beta ^2 \lambda ^2-80\right)}{2419200 (r +1)},\,\,\,\,g^{(11)}_{10,0}-\frac{\beta ^8 \lambda ^8 \left(r  \left(5 \beta ^2 \lambda ^2-198\right)+5 \beta ^2 \lambda ^2+198\right)}{239500800 (r +1)}$\\
			\hline
			$(13)$  & $g^{(13)}_{0,12}=\frac{-3455 \beta ^{10} \lambda ^{10}+12276 \beta ^8 \lambda ^8-67320 \beta ^6 \lambda ^6+665280 \beta ^4 \lambda ^4-19958400 \beta ^2 \lambda ^2+159667200}{319334400 (r +1)}$\\
			&$g^{(13)}_{2,10}= \frac{775 \beta ^{10} \lambda ^{10}-2754 \beta ^8 \lambda ^8+15120 \beta ^6 \lambda ^6-151200 \beta ^4 \lambda ^4+3024000 \beta ^2 \lambda ^2+14515200}{14515200 (r +1)},\,\,\,g^{(13)}_{10,2}=\frac{\beta ^8 \lambda ^8 \left(145 \beta ^2 \lambda ^2-396\right)}{479001600 (r +1)}$\\
			&$g^{(13)}_{6,6}=\frac{\beta ^4 \lambda ^4 \left(35 \beta ^6 \lambda ^6-126 \beta ^4 \lambda ^4+680 \beta ^2 \lambda ^2-3360\right)}{2419200 (r +1)}$\\
			& $g^{(13)}_{4,8}=-\frac{\beta ^2 \lambda ^2 \left(425 \beta ^8 \lambda ^8-1512 \beta ^6 \lambda ^6+8400 \beta ^4 \lambda ^4-73920 \beta ^2 \lambda ^2+806400\right)}{9676800 (r +1)},\,\,\,\,g^{(13)}_{8,4}=-\frac{\beta ^6 \lambda ^6 \left(25 \beta ^4 \lambda ^4-92 \beta ^2 \lambda ^2+320\right)}{9676800 (r +1)}$\\
			&$g^{(13)}_{12,0}=-\frac{\beta ^{10} \lambda ^{10} \left(r  \left(691 \beta ^2 \lambda ^2-27300\right)+691 \beta ^2 \lambda ^2+27300\right)}{1307674368000 (r +1)}$\\
			\hline
			$(15)$  &$g^{(15)}_{0,14}=\frac{3773551 \beta ^{12} \lambda ^{12}-9432150 \beta ^{10} \lambda ^{10}+33513480 \beta ^8 \lambda ^8-183783600 \beta ^6 \lambda ^6+1816214400 \beta ^4 \lambda ^4-54486432000 \beta ^2 \lambda ^2+435891456000}{871782912000 (r +1)}$ \\
			&$g^{(15)}_{2,12}=\frac{-477481 \beta ^{12} \lambda ^{12}+1193500 \beta ^{10} \lambda ^{10}-4241160 \beta ^8 \lambda ^8+23284800 \beta ^6 \lambda ^6-232848000 \beta ^4 \lambda ^4+4656960000 \beta ^2 \lambda ^2+22353408000}{22353408000 (r +1)}$\\
			& $g^{(15)}_{4,10}=\frac{\beta ^2 \lambda ^2 \left(21421 \beta ^{10} \lambda ^{10}-53550 \beta ^8 \lambda ^8+190512 \beta ^6 \lambda ^6-1058400 \beta ^4 \lambda ^4+9313920 \beta ^2 \lambda ^2-101606400\right)}{1219276800 (r +1)}$\\
			&$g^{(15)}_{6,8}=-\frac{\beta ^4 \lambda ^4 \left(11747 \beta ^8 \lambda ^8-29400 \beta ^6 \lambda ^6+105840 \beta ^4 \lambda ^4-571200 \beta ^2 \lambda ^2+2822400\right)}{2032128000 (r +1)}$\\
			&$g^{(15)}_{8,6}=\frac{\beta ^6 \lambda ^6 \left(691 \beta ^6 \lambda ^6-1750 \beta ^4 \lambda ^4+6440 \beta ^2 \lambda ^2-22400\right)}{677376000 (r +1)}$\\
			&$g^{(15)}_{10,4}=-\frac{\beta ^8 \lambda ^8 \left(7601 \beta ^4 \lambda ^4-20300 \beta ^2 \lambda ^2+55440\right)}{67060224000 (r +1)}$\\
			& $g^{(15)}_{12,2}=\frac{\beta ^{10} \lambda ^{10} \left(691 \beta ^2 \lambda ^2-1560\right)}{74724249600 (\alpha +1)}$\\
			&$g^{(15)}_{14,0}=-\frac{\beta ^{12} \lambda ^{12} \left(r  \left(35 \beta ^2 \lambda ^2-1382\right)+35 \beta ^2 \lambda ^2+1382\right)}{2615348736000 (r +1)}$\\
			\hline
			\hline
		\end{tabular}
	\end{center}
	\label{tdpvfd}
\end{table}
\begin{landscape}
	\begin{table}[!htb]
		\label{table one}
		\begin{center}
			\begin{tabular}{|c|c|}
				
				\hline 			 $(n)$ &non-vanishing $g^{(n)}_{\ell,n-\ell-1}$ to $20^{th}$ order\\
				\hline
				\hline
				&  \\ 
				$(17)$ & $g^{(17)}_{0,16}=\frac{-32534915 \beta ^{14} \lambda ^{14}+60376816 \beta ^{12} \lambda ^{12}-150914400 \beta ^{10} \lambda ^{10}+536215680 \beta ^8 \lambda ^8-2940537600 \beta ^6 \lambda ^6+29059430400 \beta ^4 \lambda ^4-871782912000 \beta ^2 \lambda ^2+6974263296000}{13948526592000 (r +1)}$  \\
				&$g^{(17)}_{2,14}=\frac{6689725 \beta ^{14} \lambda ^{14}-12414506 \beta ^{12} \lambda ^{12}+31031000 \beta ^{10} \lambda ^{10}-110270160 \beta ^8 \lambda ^8+605404800 \beta ^6 \lambda ^6-6054048000 \beta ^4 \lambda ^4+121080960000 \beta ^2 \lambda ^2+581188608000}{581188608000 (r +1)}$\\
				&$g^{(17)}_{4,12}=-\frac{\beta ^2 \lambda ^2 \left(507885 \beta ^{12} \lambda ^{12}-942524 \beta ^{10} \lambda ^{10}+2356200 \beta ^8 \lambda ^8-8382528 \beta ^6 \lambda ^6+46569600 \beta ^4 \lambda ^4-409812480 \beta ^2 \lambda ^2+4470681600\right)}{53648179200 (r +1)}$\\
				&$g^{(17)}_{6,10}=\frac{\beta ^4 \lambda ^4 \left(37975 \beta ^{10} \lambda ^{10}-70482 \beta ^8 \lambda ^8+176400 \beta ^6 \lambda ^6-635040 \beta ^4 \lambda ^4+3427200 \beta ^2 \lambda ^2-16934400\right)}{12192768000 (r +1)}$\\
				&$g^{(17)}_{8,8}=-\frac{\beta ^6 \lambda ^6 \left(2975 \beta ^8 \lambda ^8-5528 \beta ^6 \lambda ^6+14000 \beta ^4 \lambda ^4-51520 \beta ^2 \lambda ^2+179200\right)}{5419008000 (r +1)}$\\
				&$g^{(17)}_{10,6}=\frac{\beta ^8 \lambda ^8 \left(8085 \beta ^6 \lambda ^6-15202 \beta ^4 \lambda ^4+40600 \beta ^2 \lambda ^2-110880\right)}{134120448000 (r +1)}$\\
				&$g^{(17)}_{12,4}=-\frac{\beta ^{10} \lambda ^{10} \left(1365 \beta ^4 \lambda ^4-2764 \beta ^2 \lambda ^2+6240\right)}{298896998400 (r +1)}$\\
				&$g^{(17)}_{14,2}=\frac{\beta ^{12} \lambda ^{12} \left(1435 \beta ^2 \lambda ^2-2764\right)}{5230697472000 (r +1)}$\\
				&$g^{(17)}_{16,0}=-\frac{\beta ^{14} \lambda ^{14} \left(r  \left(3617 \beta ^2 \lambda ^2-142800\right)+3617 \beta ^2 \lambda ^2+142800\right)}{10670622842880000 (r +1)}$\\
				\hline
				&\\ 
				$(19)$ &$g^{(19)}_{0,18}=\frac{1}{7113748561920000 (r +1)}[11582686547 \beta ^{16} \lambda ^{16}-16592806650 \beta ^{14} \lambda ^{14}+30792176160 \beta ^{12} \lambda ^{12}-76966344000 \beta ^{10} \lambda ^{10}$ \\
				& $+273469996800 \beta ^8 \lambda ^8-1499674176000 \beta ^6 \lambda ^6+14820309504000 \beta ^4 \lambda ^4-444609285120000 \beta ^2 \lambda ^2+3556874280960000]$\\
				&$g^{(19)}_{2,16}=\frac{1}{418455797760000 (r +1)}[-3362251073 \beta ^{16} \lambda ^{16}+4816602000 \beta ^{14} \lambda ^{14}-8938444320 \beta ^{12} \lambda ^{12}+22342320000 \beta ^{10} \lambda ^{10}$\\
				&$-79394515200 \beta ^8 \lambda ^8+435891456000 \beta ^6 \lambda ^6-4358914560000 \beta ^4 \lambda ^4+87178291200000 \beta ^2 \lambda ^2+418455797760000]$\\
				&$g^{(19)}_{4,14}=\frac{\beta ^2 \lambda ^2 \left(138267059 \beta ^{14} \lambda ^{14}-198075150 \beta ^{12} \lambda ^{12}+367584360 \beta ^{10} \lambda ^{10}-918918000 \beta ^8 \lambda ^8+3269185920 \beta ^6 \lambda ^6-18162144000 \beta ^4 \lambda ^4+159826867200 \beta ^2 \lambda ^2-1743565824000\right)}{20922789888000 (r +1)}$\\
				&$g^{(19)}_{6,12}=-\frac{\beta ^4 \lambda ^4 \left(17495429 \beta ^{12} \lambda ^{12}-25063500 \beta ^{10} \lambda ^{10}+46518120 \beta ^8 \lambda ^8-116424000 \beta ^6 \lambda ^6+419126400 \beta ^4 \lambda ^4-2261952000 \beta ^2 \lambda ^2+11176704000\right)}{8047226880000 (r +1)}$\\
				&$g^{(19)}_{8,10}=\frac{\beta ^6 \lambda ^6 \left(112127 \beta ^{10} \lambda ^{10}-160650 \beta ^8 \lambda ^8+298512 \beta ^6 \lambda ^6-756000 \beta ^4 \lambda ^4+2782080 \beta ^2 \lambda ^2-9676800\right)}{292626432000 (r +1)}$\\
				&$g^{(19)}_{10,8}=-\frac{\beta ^8 \lambda ^8 \left(676379 \beta ^8 \lambda ^8-970200 \beta ^6 \lambda ^6+1824240 \beta ^4 \lambda ^4-4872000 \beta ^2 \lambda ^2+13305600\right)}{16094453760000 (r +1)}$\\
				&$g^{(19)}_{12,6}=\frac{\beta ^{10} \lambda ^{10} \left(47021 \beta ^6 \lambda ^6-68250 \beta ^4 \lambda ^4+138200 \beta ^2 \lambda ^2-312000\right)}{14944849920000 (r +1)}$\\
				&$g^{(19)}_{14,4}=-\frac{\beta ^{12} \lambda ^{12} \left(3617 \beta ^4 \lambda ^4-5740 \beta ^2 \lambda ^2+11056\right)}{20922789888000 (\alpha +1)}$\\
				&$g^{(19)}_{16,2}=\frac{\beta ^{14} \lambda ^{14} \left(169999 \beta ^2 \lambda ^2-285600\right)}{21341245685760000 (r +1)}$\\
				&$g^{(19)}_{18,0}=-\frac{\beta ^{16} \lambda ^{16} \left(r  \left(219335 \beta ^2 \lambda ^2-8659098\right)+219335 \beta ^2 \lambda ^2+8659098\right)}{25545471085854720000 (r +1)}$\\
				\hline
			\end{tabular}
		\end{center}
		\label{}
	\end{table}
\end{landscape}
\section{Coefficients of $\partial_t\epsilon_a\partial_t^{\ell}\epsilon_a\partial_t^{n-\ell-1}\mathcal{E}$ in the cubic action}
\label{Coeff_3rd_partial_t}
\begin{table}[!htb]
	\label{table one}
	\begin{center}
		\begin{tabular}{|c|c|}
			
			\hline 			 $(n)$ &non-vanishing $h^{(n)}_{\ell,n-\ell-1}$ to $20^{th}$ order\\
			\hline
			\hline
			$(3)$ & $h^{(3)}_{1,1}=1$  \\
			\hline
			$(5)$   & $h^{(5)}_{1,3}=\frac{1}{72} \left(72-\beta ^2 \lambda ^2\right),\,\,\,\,\,\,\,h^{(5)}_{3,1}=-\frac{1}{36} \beta ^2 \lambda ^2$\\			
			\hline
			$(7)$  & $h^{(7)}_{1,5}=\frac{\beta ^4 \lambda ^4-30 \beta ^2 \lambda ^2+2160}{2160 },\,\,\,\,\,h^{(7)}_{3,3}=-\frac{1}{432} \beta ^2 \lambda ^2\left(\beta ^2 \lambda ^2+12\right),\,\,\,\,\,\,h^{(7)}_{5,1}=\frac{\beta ^4 \lambda ^4}{2160}$\\
			\hline
			$(9)$  & $h^{(9)}_{1,7}=\frac{-17 \beta ^6 \lambda ^6+168 \beta ^4 \lambda ^4-5040 \beta ^2 \lambda ^2+362880}{362880},\,\,\,h^{(9)}_{3,5}=\frac{\beta ^2 \lambda ^2 \left(\beta ^4 \lambda ^4-10 \beta ^2 \lambda ^2-120\right)}{4320}$\\
			& $h^{(9)}_{5,3}=-\frac{\beta ^4 \lambda ^4 \left(5 \beta ^2 \lambda ^2-12\right)}{25920},\,\,\,\,h^{(9)}_{7,1}=\frac{\beta ^6 \lambda ^6}{30240}$\\
			\hline
			$(11)$  & $h^{(11)}_{1,9}=\frac{31 \beta ^8 \lambda ^8-170 \beta ^6 \lambda ^6+1680 \beta ^4 \lambda ^4-50400 \beta ^2 \lambda ^2+3628800}{3628800}$\\
			&$h^{(11)}_{3,7}=-\frac{\beta ^2 \lambda ^2 \left(51 \beta ^6 \lambda ^6-280 \beta ^4 \lambda ^4+2800 \beta ^2 \lambda ^2+33600\right)}{1209600},\,\,\,\,h^{(11)}_{5,5}=\frac{\beta ^4 \lambda ^4 \left(9 \beta ^4 \lambda ^4-50 \beta ^2 \lambda ^2+120\right)}{259200}$\\
			&$h^{(11)}_{7,3}=\frac{120 \beta ^6 \lambda ^6-42 \beta ^8 \lambda ^8}{3628800},\,\,\,\,h^{(11)}_{9,1}=\frac{\beta ^8 \lambda ^8}{725760}$\\
			\hline
			$(13)$  & $h^{(13)}_{1,11}=\frac{-3455 \beta ^{10} \lambda ^{10}+12276 \beta ^8 \lambda ^8-67320 \beta ^6 \lambda ^6+665280 \beta ^4 \lambda ^4-19958400 \beta ^2 \lambda ^2+1437004800}{1437004800}$\\
			&$h^{(13)}_{3,9}=\frac{17050 \beta ^{10} \lambda ^{10}-60588 \beta ^8 \lambda ^8+332640 \beta ^6 \lambda ^6-3326400 \beta ^4 \lambda ^4-39916800 \beta ^2 \lambda ^2}{1437004800}$\\
			&$h^{(13)}_{5,7}=-\frac{\beta ^4 \lambda ^4 \left(425 \beta ^6 \lambda ^6-1512 \beta ^4 \lambda ^4+8400 \beta ^2 \lambda ^2-20160\right)}{43545600},\,\,\,\,h^{(13)}_{7,5}=\frac{\beta ^6 \lambda ^6 \left(35 \beta ^4 \lambda ^4-126 \beta ^2 \lambda ^2+360\right)}{10886400}$\\
			&$h^{(13)}_{9,3}=\frac{1980 \beta ^8 \lambda ^8-825 \beta ^{10} \lambda ^{10}}{1437004800},\,\,\,\,h^{(13)}_{11,1}=\frac{\beta ^{10} \lambda ^{10}}{20528640}$\\
			\hline
			$(15)$  &$h^{(15)}_{1,13}=\frac{3773551 \beta ^{12} \lambda ^{12}-9432150 \beta ^{10} \lambda ^{10}+33513480 \beta ^8 \lambda ^8-183783600 \beta ^6 \lambda ^6+1816214400 \beta ^4 \lambda ^4-54486432000 \beta ^2 \lambda ^2+3923023104000}{3923023104000}$\\
			&$h^{(15)}_{3,11}=\frac{-18621759 \beta ^{12} \lambda ^{12}+46546500 \beta ^{10} \lambda ^{10}-165405240 \beta ^8 \lambda ^8+908107200 \beta ^6 \lambda ^6-9081072000 \beta ^4 \lambda ^4-108972864000 \beta ^2 \lambda ^2}{3923023104000}$\\
			&$h^{(15)}_{5,9}=\frac{\beta ^4 \lambda ^4 \left(21421 \beta ^8 \lambda ^8-53550 \beta ^6 \lambda ^6+190512 \beta ^4 \lambda ^4-1058400 \beta ^2 \lambda ^2+2540160\right)}{5486745600}$\\
			&$h^{(15)}_{7,7}=-\frac{\beta ^6 \lambda ^6 \left(11747 \beta ^6 \lambda ^6-29400 \beta ^4 \lambda ^4+105840 \beta ^2 \lambda ^2-302400\right)}{9144576000}$\\
			&$h^{(15)}_{9,5}=\frac{\beta ^8 \lambda ^8 \left(691 \beta ^4 \lambda ^4-1750 \beta ^2 \lambda ^2+4200\right)}{3048192000}$\\
			&$h^{(15)}_{11,3}=\frac{\beta ^{10} \lambda ^{10} \left(14700-7601 \beta ^2 \lambda ^2\right)}{301771008000}$\\
			&$h^{(15)}_{13,1}=\frac{691 \beta ^{12} \lambda ^{12}}{435891456000}$\\
			\hline
			$(17)$  &$h^{(17)}_{1,15}=-\frac{929569 \beta ^{14} \lambda ^{14}}{1793381990400}+\frac{3773551 \beta ^{12} \lambda ^{12}}{3923023104000}-\frac{691 \beta ^{10} \lambda ^{10}}{287400960}+\frac{31 \beta ^8 \lambda ^8}{3628800}-\frac{17 \beta ^6 \lambda ^6}{362880}+\frac{\beta ^4 \lambda ^4}{2160}-\frac{\beta ^2 \lambda ^2}{72}+1$\\
			&$h^{(17)}_{3,13}=\frac{5461 \beta ^{14} \lambda ^{14}}{2134978560}-\frac{477481 \beta ^{12} \lambda ^{12}}{100590336000}+\frac{31 \beta ^{10} \lambda ^{10}}{2612736}-\frac{17 \beta ^8 \lambda ^8}{403200}+\frac{\beta ^6 \lambda ^6}{4320}-\frac{\beta ^4 \lambda ^4}{432}-\frac{\beta ^2 \lambda ^2}{36}$\\
			&$h^{(17)}_{5,11}=\frac{\beta ^4 \lambda ^4 \left(-507885 \beta ^{10} \lambda ^{10}+942524 \beta ^8 \lambda ^8-2356200 \beta ^6 \lambda ^6+8382528 \beta ^4 \lambda ^4-46569600 \beta ^2 \lambda ^2+111767040\right)}{241416806400}$\\
			&$h^{(17)}_{7,9}=\frac{\beta ^6 \lambda ^6 \left(37975 \beta ^8 \lambda ^8-70482 \beta ^6 \lambda ^6+176400 \beta ^4 \lambda ^4-635040 \beta ^2 \lambda ^2+1814400\right)}{54867456000}$\\
			&$h^{(17)}_{9,7}=-\frac{\beta ^8 \lambda ^8 \left(2975 \beta ^6 \lambda ^6-5528 \beta ^4 \lambda ^4+14000 \beta ^2 \lambda ^2-33600\right)}{24385536000}$\\
			&$h^{(17)}_{11,5}=\frac{\beta ^{10} \lambda ^{10} \left(8085 \beta ^4 \lambda ^4-15202 \beta ^2 \lambda ^2+29400\right)}{603542016000}$\\
			&$h^{(17)}_{13,3}=\frac{\beta ^{12} \lambda ^{12} \left(24876-15925 \beta ^2 \lambda ^2\right)}{15692092416000}$\\
			&$h^{(17)}_{15,1}=\frac{\beta ^{14} \lambda ^{14}}{20379340800}$\\
			\hline
			\hline
		\end{tabular}
	\end{center}
	\label{}
\end{table}
\begin{table}[!htb]
	\label{table one}
	\begin{center}
		\begin{tabular}{|c|c|}
			
			\hline 			 $(n)$ &non-vanishing $h^{(n)}_{\ell,n-\ell-1}$ to $20^{th}$ order\\
			\hline
			\hline
			&\\
			$(19)$ & $h^{(19)}_{1,17}=\frac{11582686547 \beta ^{16} \lambda ^{16}}{32011868528640000}-\frac{929569 \beta ^{14} \lambda ^{14}}{1793381990400}+\frac{3773551 \beta ^{12} \lambda ^{12}}{3923023104000}-\frac{691 \beta ^{10} \lambda ^{10}}{287400960}+\frac{31 \beta ^8 \lambda ^8}{3628800}$  \\
			&$-\frac{17 \beta ^6 \lambda ^6}{362880}+\frac{\beta ^4 \lambda ^4}{2160}-\frac{\beta ^2 \lambda ^2}{72}+1$\\
			&$h^{(19)}_{3,15}=-\frac{3362251073 \beta ^{16} \lambda ^{16}}{1883051089920000}+\frac{5461 \beta ^{14} \lambda ^{14}}{2134978560}-\frac{477481 \beta ^{12} \lambda ^{12}}{100590336000}+\frac{31 \beta ^{10} \lambda ^{10}}{2612736}-\frac{17 \beta ^8 \lambda ^8}{403200}+\frac{\beta ^6 \lambda ^6}{4320}-\frac{\beta ^4 \lambda ^4}{432}-\frac{\beta ^2 \lambda ^2}{36}$\\
			&$h^{(19)}_{5,13}=\frac{1}{94152554496000}[\beta ^4 \lambda ^4 (138267059 \beta ^{12} \lambda ^{12}-198075150 \beta ^{10} \lambda ^{10}+367584360 \beta ^8 \lambda ^8 $ \\
			&$-918918000 \beta ^6 \lambda ^6+3269185920 \beta ^4 \lambda ^4-18162144000 \beta ^2 \lambda ^2+43589145600)-198075150 \beta ^{10} \lambda ^{10}$\\
			&$+367584360 \beta ^8 \lambda ^8-918918000 \beta ^6 \lambda ^6+3269185920 \beta ^4 \lambda ^4-18162144000 \beta ^2 \lambda ^2+43589145600]$\\
			&$h^{(19)}_{7,11}=\frac{\beta ^6 \lambda ^6 \left(-17495429 \beta ^{10} \lambda ^{10}+25063500 \beta ^8 \lambda ^8-46518120 \beta ^6 \lambda ^6+116424000 \beta ^4 \lambda ^4-419126400 \beta ^2 \lambda ^2+1197504000\right)}{36212520960000}$\\
			&$h^{(19)}_{9,9}=\frac{\beta ^8 \lambda ^8 \left(112127 \beta ^8 \lambda ^8-160650 \beta ^6 \lambda ^6+298512 \beta ^4 \lambda ^4-756000 \beta ^2 \lambda ^2+1814400\right)}{1316818944000}$\\
			&$h^{(19)}_{11,7}=\frac{\beta ^{10} \lambda ^{10} \left(-676379 \beta ^6 \lambda ^6+970200 \beta ^4 \lambda ^4-1824240 \beta ^2 \lambda ^2+3528000\right)}{72425041920000}$\\
			&$h^{(19)}_{13,5}= \frac{\beta ^{12} \lambda ^{12} \left(329147 \beta ^4 \lambda ^4-477750 \beta ^2 \lambda ^2+746280\right)}{470762772480000}$\\
			&$h^{(19)}_{15,3}=\frac{\beta ^{14} \lambda ^{14} \left(4620-3617 \beta ^2 \lambda ^2\right)}{94152554496000}$\\
			&$h^{(19)}_{17,1}=\frac{3617 \beta ^{16} \lambda ^{16}}{2462451425280000}$\\
			\hline
			\hline
		\end{tabular}
	\end{center}
	\label{}
\end{table}
\section{Loop computations}
\label{loop_App}
The numerator $C(p)$ is given by the following diagram with $p=(\omega , k)$
	\begin{equation*}
	\begin{split}
&G_{\mathcal{E}\epsilon_a}(p)(-C(p))G_{\epsilon_a\mathcal{E}}(p)=\\
&\feynmandiagram[scale=0.5,transform shape] [inline=(a.base),horizontal=a to b] {a --[thick]c-- [thick]b};
+
\scalebox{0.65}{	\begin{tikzpicture}[baseline=(a.base)]
	\begin{feynman}
	\vertex (a) ;
	\vertex [right=of a] (a1) ;
	\vertex [right=of a1] (a2);
	\vertex [above=of a2] (a3);
	\vertex [ right=of a2] (a4);
	\vertex [ right=of a4] (a5); 
	\diagram* {
		(a) -- [very thick](a1)-- [boson,very thick] (a2) -- [out=140, in=180, loop, min distance=1.cm,very thick] (a3)--[out=0, in=40, loop, min distance=1.cm,very thick](a2) ,
		(a2) -- [boson,very thick] (a4)--   [very thick](a5),
	};
	\end{feynman}
	\end{tikzpicture}}
+
\scalebox{0.5}{	\begin{tikzpicture}[baseline=(a.base)]
	\begin{feynman}
	\vertex (a) ;
	\vertex [right=of a] (a1) ;
	\vertex [right=of a1] (a2);
	\vertex [above right=of a2] (a3);
	\vertex [below right=of a2] (a4);
	\vertex [above right=of a4] (a5); 
	\vertex [ right=of a5] (a6);
	\vertex [ right=of a6] (a7);
	\diagram* {
		(a) -- [ very thick](a1)-- [boson,very thick] (a2) -- [quarter left, very thick] (a3)--[quarter left, very thick](a5)--[quarter left, very thick](a4)--[quarter left, very thick](a2) ,
		(a5) -- [boson,very thick] (a6)--   [very thick](a7),
	};
	\end{feynman}
	\end{tikzpicture}}
+
\left [\, \scalebox{0.5}{	\begin{tikzpicture}[baseline=(a.base)]
	\begin{feynman}
	\vertex (a) ;
	\vertex [right=of a] (a1) ;
	\vertex [right=of a1] (a2);
	\vertex [above right=of a2] (a3);
	\vertex [below right=of a2] (a4);
	\vertex [above right=of a4] (a5); 
	\vertex [ right=of a5] (a6);
	\vertex [ right=of a6] (a7);
	\diagram* {
		(a) --[very thick](a1)-- [boson,very thick] (a2) -- [quarter left, very thick] (a3)--[quarter left,boson,very thick](a5)--[quarter left, very thick](a4)--[quarter left, very thick](a2) ,
		(a5) -- [boson,very thick] (a6)--  [very thick](a7),
	};
	\end{feynman}
	\end{tikzpicture}}
+\text{c.c} \right]\,\,.
\end{split}
\end{equation*}
Since  we are not interested in computing $C(p)$ we do not write the corresponding loop integrals.  The self-energy $\Sigma(p)$ is what we want to compute. Diagrammatically:
\begin{equation}
\begin{split}
&G_{\mathcal{E}\epsilon_a}(p)(-\Sigma(p))G_{\mathcal{E}\mathcal{E}}(p)=\\
&\scalebox{0.65}{	\begin{tikzpicture}[baseline=(a.base)]
	\begin{feynman}
	\vertex (a) ;
	\vertex [right=of a] (a1) ;
	\vertex [right=of a1] (a2);
	\vertex [above=of a2] (a3);
	\vertex [ right=of a2] (a4);
	\vertex [ right=of a4] (a5); 
	\diagram* {
			(a) -- [very thick](a1)-- [boson,very thick] (a2) -- [out=140, in=180, loop, min distance=1.cm,very thick] (a3)--[out=0, in=40, loop, min distance=1.cm,very thick](a2) ,
	(a2) -- [very thick] (a4)--   [very thick](a5),
	};
	\end{feynman}
	\end{tikzpicture}}
+
\scalebox{0.53}{	\begin{tikzpicture}[baseline=(a.base)]
	\begin{feynman}
	\vertex (a) ;
	\vertex [right=of a] (a1) ;
	\vertex [right=of a1] (a2);
	\vertex [above right=of a2] (a3);
	\vertex [below right=of a2] (a4);
	\vertex [above right=of a4] (a5); 
	\vertex [ right=of a5] (a6);
	\vertex [ right=of a6] (a7);
	\diagram* {
		(a) --[very thick](a1)-- [boson,very thick] (a2) -- [quarter left, very thick] (a3)--[boson,quarter left,very thick](a5)--[quarter left, very thick](a4)--[quarter left,very  thick](a2) ,
		(a5) -- [very thick] (a6)--  [very thick](a7),
	};
	\end{feynman}
	\end{tikzpicture}}
+
\scalebox{0.53}{	\begin{tikzpicture}[baseline=(a.base)]
	\begin{feynman}
	\vertex (a) ;
	\vertex [right=of a] (a1) ;
	\vertex [right=of a1] (a2);
	\vertex [above right=of a2] (a3);
	\vertex [below right=of a2] (a4);
	\vertex [above right=of a4] (a5); 
	\vertex [ right=of a5] (a6);
	\vertex [ right=of a6] (a7);
	\diagram* {
		(a) --[very thick](a1)-- [boson,very thick] (a2) -- [quarter left, very thick] (a3)--[quarter left,boson,very thick](a5)--[quarter left,boson,very thick](a4)--[quarter left, very thick](a2) ,
		(a5) --  [very thick](a6)--  [very thick](a7),
	};\,.
	\end{feynman}
	\end{tikzpicture}}
\end{split}
\end{equation}
which leads to
\begin{equation}\label{Sigma_Appendix}
\begin{split}
\Sigma(p)=&\,\text{quartic}+\,\lambda_1^2k^2\int_{p'}k'^2G_{\epsilon_a\mathcal{E}}(p')G_{\mathcal{E}\mathcal{E}}(p+p')-\,i c T^2\lambda_1k^2\int_{p'}\,\tilde{\lambda}_2\,(k'^2+ k k')G_{\epsilon_a\mathcal{E}}(p')G_{\mathcal{E}\epsilon_a}(p+p')\\
&\,-\,i  T^3\lambda_1k^2\int_{p'}\,\tilde{\lambda}_3\,(\omega'^2+ \omega \omega')G_{\epsilon_a\mathcal{E}}(p')\,G_{\mathcal{E}\epsilon_a}(p+p')
\end{split}
\end{equation}
In the above expression 
\begin{equation}\label{lambda_tilde}
\begin{split}
\tilde{\lambda}_2=&\frac{1}{2}\lambda_2\,\sum_{n=1}\sum_{\ell=0}^{n-1}\frac{g^{(n)}_{\ell,n-\ell-1}}{\lambda^{n-1}}(-i)^{n-1}\left[\omega'^{\ell}+(\omega+\omega')^{\ell}\right]\omega^{n-\ell-1}\,,\\
\tilde{\lambda}_3=&\frac{1}{2}\lambda_3\,\sum_{n=3}\sum_{\ell=1}^{n-1}\frac{h^{(n)}_{\ell,n-\ell-1}}{\lambda^{n-2}}(-i)^{n-1}\left[\omega'^{\ell}+(\omega+\omega')^{\ell}\right]\omega^{n-\ell-1}\,,
\end{split}
\end{equation}
with $\lambda_1$, $\lambda_2$ and $\lambda_3$ given in \eqref{couplings}. Let us recall that $h$ and $g$ coefficients have been given in Appendix .
In eq. \eqref{Sigma_Appendix}, ``quartic" denotes the contribution coming from the quartic Lagrangian. We have found that this term does not have any cutoff independent part.

From  \eqref{lambda_tilde}, we see that  at first order in derivatives, namely $n=1$,  $\tilde{\lambda}_2=\lambda_2$ and $\tilde{\lambda}_3=0$. This is exactly the case considered in \cite{Chen-Lin:2018kfl}. In this case, the frequency integrals are UV finite and one can easily perform the momentum integrals by considering a hard cutoff momentum. Beyond $n=1$, however, the frequency integrals are UV divergent, as can be seen by substituting \eqref{lambda_tilde} into \eqref{Sigma_Appendix}.
In order to regularize them, we follow the ideas developed in \cite{Gao:2018bxz} and  use  the following regularization for the three integrals in \eqref{Sigma_Appendix}, respectively:
\begin{equation}\label{regularization}
G_{\mathcal{E}\mathcal{E}}(\omega',k')\rightarrow \left(\frac{\Lambda}{\Lambda+i \omega}\right)^{n-1}G_{\mathcal{E}\mathcal{E}}(\omega',k'),\,\,\,\,\,\,\,\,\,\tilde{\lambda}_{2,3}\rightarrow \left(\frac{\Lambda}{\Lambda-i \omega}\right)^{n-1}\tilde{\lambda}_{2,3}
\end{equation}
where $\Lambda$ is a UV energy cutoff and $n$ denotes the order of derivative, which is actually $40$ in our case.  Note that to perform the integral in the $\lambda_1^2$ term, we close the contour along the semicircle in the lower half of the complex $\omega$-plane, while for the next two integrals in \eqref{Sigma_Appendix}, we close the contour along the upper semicircle in the complex plane. These choices are consistent with the regularizations introduced in \eqref{regularization}.
After performing the frequency integrals, we consider a hard cutoff momentum $\Lambda'$ and evaluate the momentum integrals. 

Another point that should be noted is the inclusion of ghosts. As mentioned in \cite{Crossley:2015evo}, some anti-commuting ghosts variables and BRST symmetry are necessary to protect the unitarity condition \eqref{conditions_1} at the loop level. However, it was shown that they do not play any role in low energy dynamics \cite{Gao:2018bxz,Glorioso:2018wxw}.

\begin{landscape}
	\begin{table}[!htb]
		\label{table one}
		\begin{center}
			\begin{tabular}{|c|c|}
				
				\hline 		
									&\\
					 $(2n+1)$ &$F_{2n}$ and $G_{2n}$ functions defined in \eqref{F_G}\\
					 					&\\
				\hline
				\hline
				$(3)$ & \begingroup
				\everymath{\scriptstyle}
				\scriptsize$F_2=\frac{\omega  \left(\beta ^2 \lambda ^2+12\right) \left(\omega +i D_0 k^2\right)}{48 \lambda ^2}$\endgroup\\
				&\begingroup
				\everymath{\scriptstyle}
				\scriptsize$G_2=-\frac{D_0^2 k^4 \left(\beta ^2 \lambda ^2+r \left(\beta ^2 \lambda ^2+12\right)-12\right)-2 i D_0 k^2 \omega  \left(\beta ^2 \lambda ^2+r \left(\beta ^2 \lambda ^2+12\right)-12\right)+\omega ^2 \left(\beta ^2 \lambda ^2+r \left(\beta ^2 \lambda ^2+12\right)-36\right)}{48 \lambda ^2 (r+1)}$\endgroup\\
				\hline
				\hline
				$(5)$ & 
				\begingroup
				\everymath{\scriptstyle}
				\scriptsize$F_4=-\frac{\beta ^2 \omega  \left(\beta ^2 \lambda ^2-60\right) \left(\omega +i D_0 k^2\right) \left(3 D_0^2 k^4-6 i D_0 k^2 \omega +\omega ^2\right)}{11520 \lambda ^2}$\endgroup\\
				& \begingroup
				\everymath{\scriptstyle}
				\scriptsize$G_4=\frac{720 \omega ^4 \left(\beta ^2 \lambda ^2-8\right)-120 \omega ^2 \left(5 \beta ^2 \lambda ^2+24\right) \left(D_0^2 k^4-2 i D_0 k^2 \omega +\omega ^2\right)+\beta ^2 \lambda ^2 \left(D_0^4 k^8-4 i D_0^3 k^6 \omega +2 D_0^2 k^4 \omega ^2-12 i D_0 k^2 \omega ^3+\omega ^4\right) \left(\beta ^2 \lambda ^2+r \left(\beta ^2 \lambda ^2-60\right)+60\right)}{11520 \lambda ^4 (r+1)}$\endgroup\\
				\hline
				\hline
				$(7)$&\begingroup
				\everymath{\scriptstyle}
				\scriptsize $F_6=\frac{\beta ^4 \omega  \left(\beta ^2 \lambda ^2-42\right) \left(\omega +i D_0 k^2\right) \left(5 D_0^4 k^8-20 i D_0^3 k^6 \omega -10 D_0^2 k^4 \omega ^2-20 i D_0 k^2 \omega ^3+\omega ^4\right)}{1935360 \lambda ^2}$\endgroup\\
				&\begingroup
				\everymath{\scriptstyle}
				\scriptsize $G_6=\frac{1}{1935360 \lambda ^6 (r+1)}\bigg[4032 \omega ^6 \left(\beta ^4 \lambda ^4-30 \beta ^2 \lambda ^2+240\right)-5040 \omega ^4 \left(\beta ^2 \lambda ^2-24\right) \left(\beta ^2 \lambda ^2+4\right) \left(D_0^2 k^4-2 i D_0 k^2 \omega +\omega ^2\right)+84 \beta ^2 \lambda ^2 \omega ^2 \left(11 \beta ^2 \lambda ^2-120\right) \left(D_0^4 k^8-4 i D_0^3 k^6 \omega +2 D_0^2 k^4 \omega ^2-12 i D_0 k^2 \omega ^3+\omega ^4\right)$ \endgroup\\
				&\begingroup
				\everymath{\scriptstyle}
				\scriptsize$-\beta ^4 \lambda ^4 \left(D_0^2 k^4-2 i D_0 k^2 \omega +\omega ^2\right) \left(D_0^4 k^8-4 i D_0^3 k^6 \omega +10 D_0^2 k^4 \omega ^2-28 i D_0 k^2 \omega ^3+\omega ^4\right) \left(\beta ^2 \lambda ^2+r \left(\beta ^2 \lambda ^2-42\right)+42\right)\bigg]$\endgroup\\
				\hline
				\hline
				$(9)$&\begingroup
				\everymath{\scriptstyle}
				\scriptsize$F_8=-\frac{\beta ^6 \omega  \left(\beta ^2 \lambda ^2-40\right) \left(\omega +i D_0 k^2\right) \left(7 D_0^6 k^{12}-42 i D_0^5 k^{10} \omega -49 D_0^4 k^8 \omega ^2-84 i D_0^3 k^6 \omega ^3-119 D_0^2 k^4 \omega ^4-42 i D_0 k^2 \omega ^5+\omega ^6\right)}{309657600 \lambda ^2}$\endgroup\\
				&\begingroup
				\everymath{\scriptstyle}
				\scriptsize$G_8=\frac{1}{309657600 \lambda ^8 (r+1)}\bigg[3840 \omega ^8 \left(17 \beta ^6 \lambda ^6-168 \beta ^4 \lambda ^4+5040 \beta ^2 \lambda ^2-40320\right)-80640 \omega ^6 \left(\beta ^6 \lambda ^6-10 \beta ^4 \lambda ^4+200 \beta ^2 \lambda ^2+960\right) \left(D_0^2 k^4-2 i D_0 k^2 \omega +\omega ^2\right)$\endgroup\\
				&\begingroup
				\everymath{\scriptstyle}
				\scriptsize$+3360 \beta ^2 \lambda ^2 \omega ^4 \left(5 \beta ^4 \lambda ^4-44 \beta ^2 \lambda ^2+480\right) \left(D_0^4 k^8-4 i D_0^3 k^6 \omega +2 D_0^2 k^4 \omega ^2-12 i D_0 k^2 \omega ^3+\omega ^4\right)-80 \beta ^4 \lambda ^4 \omega ^2 \left(17 \beta ^2 \lambda ^2-84\right) \left(D_0^2 k^4-2 i D_0 k^2 \omega +\omega ^2\right) \left(D_0^4 k^8-4 i D_0^3 k^6 \omega +10 D_0^2 k^4 \omega ^2-28 i D_0 k^2 \omega ^3+\omega ^4\right)$\endgroup\\
				&\begingroup
				\everymath{\scriptstyle}
				\scriptsize$+\beta ^6 \lambda ^6 \left(D_0^8 k^{16}-8 i D_0^7 k^{14} \omega +4 D_0^6 k^{12} \omega ^2-136 i D_0^5 k^{10} \omega ^3-250 D_0^4 k^8 \omega ^4-56 i D_0^3 k^6 \omega ^5-252 D_0^2 k^4 \omega ^6-56 i D_0 k^2 \omega ^7+\omega ^8\right)\times  \left(\beta ^2 \lambda ^2+r \left(\beta ^2 \lambda ^2-40\right)+40\right)\bigg]$\endgroup\\
				\hline
				\hline
				$(11)$&\begingroup
				\everymath{\scriptstyle}
				\scriptsize$F_{10}=\frac{\beta ^8 \omega  \left(5 \beta ^2 \lambda ^2-198\right) \left(\omega +i D_0 k^2\right) \left(3 D_0^2 k^4-6 i D_0 k^2 \omega +\omega ^2\right) \left(3 D_0^6 k^{12}-18 i D_0^5 k^{10} \omega -9 D_0^4 k^8 \omega ^2-84 i D_0^3 k^6 \omega ^3-75 D_0^2 k^4 \omega ^4-66 i D_0 k^2 \omega ^5+\omega ^6\right)}{245248819200 \lambda ^2}$\endgroup\\
				&\begingroup
				\everymath{\scriptstyle}
				\scriptsize$G_{10}=\frac{1}{245248819200 \lambda ^{10} (r+1)}\bigg[304128 \omega ^{10} \left(31 \beta ^8 \lambda ^8-170 \beta ^6 \lambda ^6+1680 \beta ^4 \lambda ^4-50400 \beta ^2 \lambda ^2+403200\right)+228096 \omega ^8 \left(-51 \beta ^8 \lambda ^8+280 \beta ^6 \lambda ^6-2800 \beta ^4 \lambda ^4+56000 \beta ^2 \lambda ^2+268800\right) \left(D_0^2 k^4-2 i D_0 k^2 \omega +\omega ^2\right)$\endgroup\\
				&\begingroup
				\everymath{\scriptstyle}
				\scriptsize$-3168 \beta ^4 \lambda ^4 \omega ^4 \left(63 \beta ^4 \lambda ^4-340 \beta ^2 \lambda ^2+1680\right) \left(D_0^2 k^4-2 i D_0 k^2 \omega +\omega ^2\right) \left(D_0^4 k^8-4 i D_0^3 k^6 \omega +10 D_0^2 k^4 \omega ^2-28 i D_0 k^2 \omega ^3+\omega ^4\right)$\endgroup\\
				&\begingroup
				\everymath{\scriptstyle}
				\scriptsize$+266112 \beta ^2 \lambda ^2 \omega ^6 \left(9 \beta ^6 \lambda ^6-50 \beta ^4 \lambda ^4+440 \beta ^2 \lambda ^2-4800\right) \left(D_0^4 k^8-4 i D_0^3 k^6 \omega +2 D_0^2 k^4 \omega ^2-12 i D_0 k^2 \omega ^3+\omega ^4\right)+396 \beta ^6 \lambda ^6 \omega ^2 \left(23 \beta ^2 \lambda ^2-80\right) $\endgroup\\
				& \begingroup
				\everymath{\scriptstyle}
				\scriptsize$\times \left(D_0^8 k^{16}-8 i D_0^7 k^{14} \omega +4 D_0^6 k^{12} \omega ^2-136 i D_0^5 k^{10} \omega ^3-250 D_0^4 k^8 \omega ^4-56 i D_0^3 k^6 \omega ^5-252 D_0^2 k^4 \omega ^6-56 i D_0 k^2 \omega ^7+\omega ^8\right)$\endgroup\\
				& \begingroup
				\everymath{\scriptstyle}
				\scriptsize$-\beta ^8 \lambda ^8 \left(D_0^2 k^4-2 i D_0 k^2 \omega +\omega ^2\right)  \left(5 \beta ^2 \lambda ^2+r \left(5 \beta ^2 \lambda ^2-198\right)+198\right)\left(D_0^8 k^{16}-8 i D_0^7 k^{14} \omega +20 D_0^6 k^{12} \omega ^2-232 i D_0^5 k^{10} \omega ^3-346 D_0^4 k^8 \omega ^4-312 i D_0^3 k^6 \omega ^5-620 D_0^2 k^4 \omega ^6-88 i D_0 k^2 \omega ^7+\omega ^8\right)\bigg]$\endgroup\\
				\hline
				\hline
				$(13)$	& \begingroup
				\everymath{\scriptstyle}
				\scriptsize$F_{12}=-\frac{\beta ^{10} \omega  \left(691 \beta ^2 \lambda ^2-27300\right) \left(\omega +i D_0 k^2\right) \left(11 D_0^{10} k^{20}-110 i D_0^9 k^{18} \omega -275 D_0^8 k^{16} \omega ^2-440 i D_0^7 k^{14} \omega ^3-2618 D_0^6 k^{12} \omega ^4+2156 i D_0^5 k^{10} \omega ^5-2574 D_0^4 k^8 \omega ^6+2376 i D_0^3 k^6 \omega ^7-1265 D_0^2 k^4 \omega ^8-110 i D_0 k^2 \omega ^9+\omega ^{10}\right)}{5356234211328000 \lambda ^2}$\endgroup\\
				&      \begingroup
				\everymath{\scriptstyle}
				\scriptsize $G_{12}=\frac{1}{5356234211328000 (r+1) \lambda ^{12}}\big[16773120 \left(\beta ^2 \lambda ^2 \left(3455 \beta ^8 \lambda ^8-12276 \beta ^6 \lambda ^6+67320 \beta ^4 \lambda ^4-665280 \beta ^2 \lambda ^2+19958400\right)-159667200\right) \omega ^{12}$\endgroup\\
				& \begingroup
				\everymath{\scriptstyle}
				\scriptsize $-92252160 \left(\beta ^2 \left(775 \beta ^8 \lambda ^8-2754 \beta ^6 \lambda ^6+15120 \beta ^4 \lambda ^4-151200 \beta ^2 \lambda ^2+3024000\right) \lambda ^2+14515200\right) \left(D_0^2 k^4-2 i \omega  D_0 k^2+\omega ^2\right) \omega ^{10}$\endgroup\\
					& \begingroup
				\everymath{\scriptstyle}
				\scriptsize $+34594560 \beta ^2 \lambda ^2 \left(425 \beta ^8 \lambda ^8-1512 \beta ^6 \lambda ^6+8400 \beta ^4 \lambda ^4-73920 \beta ^2 \lambda ^2+806400\right) \left(D_0^4 k^8-4 i \omega  D_0^3 k^6+2 \omega ^2 D_0^2 k^4-12 i \omega ^3 D_0 k^2+\omega ^4\right) \omega ^8$\endgroup\\
					& \begingroup
				\everymath{\scriptstyle}
				\scriptsize $-34594560 \beta ^4 \lambda ^4 \left(35 \beta ^6 \lambda ^6-126 \beta ^4 \lambda ^4+680 \beta ^2 \lambda ^2-3360\right) \left(D_0^2 k^4-2 i \omega  D_0 k^2+\omega ^2\right) \left(D_0^4 k^8-4 i \omega  D_0^3 k^6+10 \omega ^2 D_0^2 k^4-28 i \omega ^3 D_0 k^2+\omega ^4\right) \omega ^6$\endgroup\\
			& \begingroup
			\everymath{\scriptstyle}
			\scriptsize $+2162160 \beta ^6 \lambda ^6 \left(25 \beta ^4 \lambda ^4-92 \beta ^2 \lambda ^2+320\right) \left(D_0^8 k^{16}-8 i \omega  D_0^7 k^{14}+4 \omega ^2 D_0^6 k^{12}-136 i \omega ^3 D_0^5 k^{10}-250 \omega ^4 D_0^4 k^8-56 i \omega ^5 D_0^3 k^6-252 \omega ^6 D_0^2 k^4-56 i \omega ^7 D_0 k^2+\omega ^8\right) \omega ^4$
			\endgroup\\
						& \begingroup
			\everymath{\scriptstyle}
			\scriptsize $-10920 \beta ^8 \lambda ^8 \left(145 \beta ^2 \lambda ^2-396\right) \left(D_0^2 k^4-2 i \omega  D_0 k^2+\omega ^2\right) \left(D_0^8 k^{16}-8 i \omega  D_0^7 k^{14}+20 \omega ^2 D_0^6 k^{12}-232 i \omega ^3 D_0^5 k^{10}-346 \omega ^4 D_0^4 k^8-312 i \omega ^5 D_0^3 k^6-620 \omega ^6 D_0^2 k^4-88 i \omega ^7 D_0 k^2+\omega ^8\right) \omega ^2$
			\endgroup\\
			& \begingroup
			\everymath{\scriptstyle}
			\scriptsize
			$+\beta ^{10} \lambda ^{10} \left(691 \beta ^2 \lambda ^2+r \left(691 \beta ^2 \lambda ^2-27300\right)+27300\right) \left(D_0^4 k^8-4 i \omega  D_0^3 k^6+2 \omega ^2 D_0^2 k^4-12 i \omega ^3 D_0 k^2+\omega ^4\right)$	\endgroup\\
				& \begingroup
			\everymath{\scriptstyle}
			\scriptsize
			$\times  \left(D_0^8 k^{16}-8 i \omega  D_0^7 k^{14}+36 \omega ^2 D_0^6 k^{12}-328 i \omega ^3 D_0^5 k^{10}-570 \omega ^4 D_0^4 k^8-56 i \omega ^5 D_0^3 k^6-476 \omega ^6 D_0^2 k^4-120 i \omega ^7 D_0 k^2+\omega ^8\right)\big]$	\endgroup\\
				\hline
				\hline
			\end{tabular}
		\end{center}
		\label{}
	\end{table}
\end{landscape}

\begin{landscape}
	\begin{table}[!htb]
		\label{table one}
		\begin{center}
			\begin{tabular}{|c|c|}
				\hline 		
									&\\
										 $(2n+1)$ &$F_{2n}$ and $G_{2n}$ functions defined in \eqref{F_G}\\
				&\\
				\hline
				\hline
				$(15)$ & \begingroup
				\everymath{\scriptstyle}
				\scriptsize$F_{14}=\frac{1}{42849873690624000 \lambda ^2}\big[\beta ^{12} \omega  \left(35 \beta ^2 \lambda ^2-1382\right) \left(\omega +i D_0 k^2\right) \big(13 D_0^{12} k^{24}-156 i D_0^{11} k^{22} \omega -494 D_0^{10} k^{20} \omega ^2-780 i D_0^9 k^{18} \omega ^3-7033 D_0^8 k^{16} \omega ^4$ \endgroup\\
				& \begingroup
				\everymath{\scriptstyle}
				\scriptsize$+10088 i D_0^7 k^{14} \omega ^5-7124 D_0^6 k^{12} \omega ^6+21736 i D_0^5 k^{10} \omega ^7+715 D_0^4 k^8 \omega ^8+10868 i D_0^3 k^6 \omega ^9-2782 D_0^2 k^4 \omega ^{10}-156 i D_0 k^2 \omega ^{11}+\omega ^{12}\big)\big]$\endgroup\\
					& \begingroup
				\everymath{\scriptstyle}
				\scriptsize$
				G_{14}=\frac{1}{42849873690624000 (r+1) \lambda ^{14}}\big[49152 \left(\beta ^4 \left(3773551 \beta ^8 \lambda ^8-9432150 \beta ^6 \lambda ^6+33513480 \beta ^4 \lambda ^4-183783600 \beta ^2 \lambda ^2+1816214400\right) \lambda ^4-54486432000 \beta ^2 \lambda ^2+435891456000\right) \omega ^{14}$ \endgroup\\
					& \begingroup
				\everymath{\scriptstyle}
				\scriptsize$+479232 \left(\beta ^4 \left(-477481 \beta ^8 \lambda ^8+1193500 \beta ^6 \lambda ^6-4241160 \beta ^4 \lambda ^4+23284800 \beta ^2 \lambda ^2-232848000\right) \lambda ^4+4656960000 \beta ^2 \lambda ^2+22353408000\right) \left(D_0^2 k^4-2 i \omega  D_0 k^2+\omega ^2\right) \omega ^{12}$\endgroup\\
					& \begingroup
				\everymath{\scriptstyle}
				\scriptsize$+2196480 \beta ^2 \lambda ^2 \left(\beta ^2 \lambda ^2 \left(21421 \beta ^8 \lambda ^8-53550 \beta ^6 \lambda ^6+190512 \beta ^4 \lambda ^4-1058400 \beta ^2 \lambda ^2+9313920\right)-101606400\right) \left(D_0^4 k^8-4 i \omega  D_0^3 k^6+2 \omega ^2 D_0^2 k^4-12 i \omega ^3 D_0 k^2+\omega ^4\right) \omega ^{10}$\endgroup\\
				& \begingroup
				\everymath{\scriptstyle}
				\scriptsize$
				-329472 \beta ^4 \lambda ^4 \left(11747 \beta ^8 \lambda ^8-29400 \beta ^6 \lambda ^6+105840 \beta ^4 \lambda ^4-571200 \beta ^2 \lambda ^2+2822400\right) \left(D_0^2 k^4-2 i \omega  D_0 k^2+\omega ^2\right) \left(D_0^4 k^8-4 i \omega  D_0^3 k^6+10 \omega ^2 D_0^2 k^4-28 i \omega ^3 D_0 k^2+\omega ^4\right) \omega ^8$\endgroup\\
				& \begingroup
				\everymath{\scriptstyle}
				\scriptsize$+247104 \beta ^6 \lambda ^6 \left(691 \beta ^6 \lambda ^6-1750 \beta ^4 \lambda ^4+6440 \beta ^2 \lambda ^2-22400\right) \left(D_0^8 k^{16}-8 i \omega  D_0^7 k^{14}+4 \omega ^2 D_0^6 k^{12}-136 i \omega ^3 D_0^5 k^{10}-250 \omega ^4 D_0^4 k^8-56 i \omega ^5 D_0^3 k^6-252 \omega ^6 D_0^2 k^4-56 i \omega ^7 D_0 k^2+\omega ^8\right) \omega ^6$\endgroup\\
				& \begingroup
				\everymath{\scriptstyle}
				\scriptsize$
				-624 \beta ^8 \lambda ^8 \left(7601 \beta ^4 \lambda ^4-20300 \beta ^2 \lambda ^2+55440\right) \left(D_0^2 k^4-2 i \omega  D_0 k^2+\omega ^2\right) \left(D_0^8 k^{16}-8 i \omega  D_0^7 k^{14}+20 \omega ^2 D_0^6 k^{12}-232 i \omega ^3 D_0^5 k^{10}-346 \omega ^4 D_0^4 k^8-312 i \omega ^5 D_0^3 k^6-620 \omega ^6 D_0^2 k^4-88 i \omega ^7 D_0 k^2+\omega ^8\right) \omega ^4$\endgroup\\
				& \begingroup
				\everymath{\scriptstyle}
				\scriptsize$
				+140 \beta ^{10} \lambda ^{10} \left(691 \beta ^2 \lambda ^2-1560\right) \left(D_0^4 k^8-4 i \omega  D_0^3 k^6+2 \omega ^2 D_0^2 k^4-12 i \omega ^3 D_0 k^2+\omega ^4\right) \left(D_0^8 k^{16}-8 i \omega  D_0^7 k^{14}+36 \omega ^2 D_0^6 k^{12}-328 i \omega ^3 D_0^5 k^{10}-570 \omega ^4 D_0^4 k^8-56 i \omega ^5 D_0^3 k^6-476 \omega ^6 D_0^2 k^4-120 i \omega ^7 D_0 k^2+\omega ^8\right) \omega ^2$\endgroup\\
				& \begingroup
				\everymath{\scriptstyle}
				\scriptsize$				
				-\beta ^{12} \lambda ^{12} \left(35 \beta ^2 \lambda ^2+r \left(35 \beta ^2 \lambda ^2-1382\right)+1382\right) \left(D_0^2 k^4-2 i \omega  D_0 k^2+\omega ^2\right) $\endgroup\\
				& \begingroup
				\everymath{\scriptstyle}
				\scriptsize$	
                \times				
				\left(D_0^{12} k^{24}-12 i \omega  D_0^{11} k^{22}+30 \omega ^2 D_0^{10} k^{20}-740 i \omega ^3 D_0^9 k^{18}-2449 \omega ^4 D_0^8 k^{16}-280 i \omega ^5 D_0^7 k^{14}-12636 \omega ^6 D_0^6 k^{12}+13720 i \omega ^7 D_0^5 k^{10}-9617 \omega ^8 D_0^4 k^8+13092 i \omega ^9 D_0^3 k^6-3554 \omega ^{10} D_0^2 k^4-180 i \omega ^{11} D_0 k^2+\omega ^{12}\right)\big]$\endgroup\\
				\hline
\hline
$(17)$& \begingroup
\everymath{\scriptstyle}
\scriptsize$
F_{16}=-\frac{1}{699309938630983680000 \lambda ^2}\big[\beta ^{14} \omega  \left(3617 \beta ^2 \lambda ^2-142800\right) \left(\omega +i D_0 k^2\right) \left(3 D_0^2 k^4-6 i D_0 k^2 \omega +\omega ^2\right) \left(5 D_0^4 k^8-20 i D_0^3 k^6 \omega -10 D_0^2 k^4 \omega ^2-20 i D_0 k^2 \omega ^3+\omega ^4\right)$ \endgroup\\
& \begingroup
\everymath{\scriptstyle}
\scriptsize$
\times\left(D_0^8 k^{16}-8 i D_0^7 k^{14} \omega +4 D_0^6 k^{12} \omega ^2-136 i D_0^5 k^{10} \omega ^3-186 D_0^4 k^8 \omega ^4-312 i D_0^3 k^6 \omega ^5-444 D_0^2 k^4 \omega ^6-184 i D_0 k^2 \omega ^7+\omega ^8\right)$ \endgroup\\
& \begingroup
\everymath{\scriptstyle}
\scriptsize$G_{16}=\frac{1}{699309938630983680000 (r+1) \lambda ^{16}}\big[50135040 \left(\beta ^6 \left(32534915 \beta ^8 \lambda ^8-60376816 \beta ^6 \lambda ^6+150914400 \beta ^4 \lambda ^4-536215680 \beta ^2 \lambda ^2+2940537600\right) \lambda ^6-29059430400 \beta ^4 \lambda ^4+871782912000 \beta ^2 \lambda ^2-6974263296000\right) \omega ^{16}
$ \endgroup\\
& \begingroup
\everymath{\scriptstyle}
\scriptsize$
-300810240 \left(\beta ^6 \left(6689725 \beta ^8 \lambda ^8-12414506 \beta ^6 \lambda ^6+31031000 \beta ^4 \lambda ^4-110270160 \beta ^2 \lambda ^2+605404800\right) \lambda ^6-6054048000 \beta ^4 \lambda ^4+121080960000 \beta ^2 \lambda ^2+581188608000\right) \left(D_0^2 k^4-2 i \omega  D_0 k^2+\omega ^2\right) \omega ^{14}$ \endgroup\\
& \begingroup
\everymath{\scriptstyle}
\scriptsize$+814694400 \beta ^2 \lambda ^2 \left(\beta ^4 \left(507885 \beta ^8 \lambda ^8-942524 \beta ^6 \lambda ^6+2356200 \beta ^4 \lambda ^4-8382528 \beta ^2 \lambda ^2+46569600\right) \lambda ^4-409812480 \beta ^2 \lambda ^2+4470681600\right) \left(D_0^4 k^8-4 i \omega  D_0^3 k^6+2 \omega ^2 D_0^2 k^4-12 i \omega ^3 D_0 k^2+\omega ^4\right) \omega ^{12}$ \endgroup\\
& \begingroup
\everymath{\scriptstyle}
\scriptsize$-896163840 \beta ^4 \lambda ^4 \left(\beta ^2 \lambda ^2 \left(37975 \beta ^8 \lambda ^8-70482 \beta ^6 \lambda ^6+176400 \beta ^4 \lambda ^4-635040 \beta ^2 \lambda ^2+3427200\right)-16934400\right) \left(D_0^2 k^4-2 i \omega  D_0 k^2+\omega ^2\right) \left(D_0^4 k^8-4 i \omega  D_0^3 k^6+10 \omega ^2 D_0^2 k^4-28 i \omega ^3 D_0 k^2+\omega ^4\right) \omega ^{10}$ \endgroup\\
& \begingroup
\everymath{\scriptstyle}
\scriptsize$+504092160 \beta ^6 \lambda ^6 \left(2975 \beta ^8 \lambda ^8-5528 \beta ^6 \lambda ^6+14000 \beta ^4 \lambda ^4-51520 \beta ^2 \lambda ^2+179200\right) \left(D_0^8 k^{16}-8 i \omega  D_0^7 k^{14}+4 \omega ^2 D_0^6 k^{12}-136 i \omega ^3 D_0^5 k^{10}-250 \omega ^4 D_0^4 k^8-56 i \omega ^5 D_0^3 k^6-252 \omega ^6 D_0^2 k^4-56 i \omega ^7 D_0 k^2+\omega ^8\right) \omega ^8$ \endgroup\\
& \begingroup
\everymath{\scriptstyle}
\scriptsize$-5091840 \beta ^8 \lambda ^8 \left(8085 \beta ^6 \lambda ^6-15202 \beta ^4 \lambda ^4+40600 \beta ^2 \lambda ^2-110880\right) \left(D_0^2 k^4-2 i \omega  D_0 k^2+\omega ^2\right)$ \endgroup\\
& \begingroup
\everymath{\scriptstyle}
\scriptsize$\times \left(D_0^8 k^{16}-8 i \omega  D_0^7 k^{14}+20 \omega ^2 D_0^6 k^{12}-232 i \omega ^3 D_0^5 k^{10}-346 \omega ^4 D_0^4 k^8-312 i \omega ^5 D_0^3 k^6-620 \omega ^6 D_0^2 k^4-88 i \omega ^7 D_0 k^2+\omega ^8\right) \omega ^6$ \endgroup\\
& \begingroup
\everymath{\scriptstyle}
\scriptsize$+571200 \beta ^{10} \lambda ^{10} \left(1365 \beta ^4 \lambda ^4-2764 \beta ^2 \lambda ^2+6240\right) \left(D_0^4 k^8-4 i \omega  D_0^3 k^6+2 \omega ^2 D_0^2 k^4-12 i \omega ^3 D_0 k^2+\omega ^4\right)$ \endgroup\\
& \begingroup
\everymath{\scriptstyle}
\scriptsize$\times \left(D_0^8 k^{16}-8 i \omega  D_0^7 k^{14}+36 \omega ^2 D_0^6 k^{12}-328 i \omega ^3 D_0^5 k^{10}-570 \omega ^4 D_0^4 k^8-56 i \omega ^5 D_0^3 k^6-476 \omega ^6 D_0^2 k^4-120 i \omega ^7 D_0 k^2+\omega ^8\right) \omega ^4$ \endgroup\\
& \begingroup
\everymath{\scriptstyle}
\scriptsize$-8160 \beta ^{12} \lambda ^{12} \left(1435 \beta ^2 \lambda ^2-2764\right) \left(D_0^2 k^4-2 i \omega  D_0 k^2+\omega ^2\right) \big(D_0^{12} k^{24}-12 i \omega  D_0^{11} k^{22}+30 \omega ^2 D_0^{10} k^{20}-740 i \omega ^3 D_0^9 k^{18}-2449 \omega ^4 D_0^8 k^{16}-280 i \omega ^5 D_0^7 k^{14}$ \endgroup\\
& \begingroup
\everymath{\scriptstyle}
\scriptsize$-12636 \omega ^6 D_0^6 k^{12}+13720 i \omega ^7 D_0^5 k^{10}-9617 \omega ^8 D_0^4 k^8+13092 i \omega ^9 D_0^3 k^6-3554 \omega ^{10} D_0^2 k^4-180 i \omega ^{11} D_0 k^2+\omega ^{12}\big) \omega ^2$ \endgroup\\
& \begingroup
\everymath{\scriptstyle}
\scriptsize$+\beta ^{14} \lambda ^{14} \left(3617 \beta ^2 \lambda ^2+r \left(3617 \beta ^2 \lambda ^2-142800\right)+142800\right) big(D_0^{16} k^{32}-16 i \omega  D_0^{15} k^{30}+8 \omega ^2 D_0^{14} k^{28}-1232 i \omega ^3 D_0^{13} k^{26}-7140 \omega ^4 D_0^{12} k^{24}+9968 i \omega ^5 D_0^{11} k^{22}-35784 \omega ^6 D_0^{10} k^{20}+131504 i \omega ^7 D_0^9 k^{18}$ \endgroup\\
& \begingroup
\everymath{\scriptstyle}
\scriptsize$+75846 \omega ^8 D_0^8 k^{16}+203600 i \omega ^9 D_0^7 k^{14}+223288 \omega ^{10} D_0^6 k^{12}+107536 i \omega ^{11} D_0^5 k^{10}+111644 \omega ^{12} D_0^4 k^8+56784 i \omega ^{13} D_0^3 k^6-7160 \omega ^{14} D_0^2 k^4-240 i \omega ^{15} D_0 k^2+\omega ^{16}\big)\big]$
\endgroup\\
	\hline
\hline
$(19)$ & We do not present $F_{18}$ ans $G_{18}$ because the corresponding expressions are terribly  complicated.\\
	\hline
\hline
\end{tabular}
\end{center}
\label{}
\end{table}
\end{landscape}

\section{Series coefficients and convergence of series}
\label{coeff_App}
The first forty coefficients of series \eqref{wn} for $r=\frac{1}{2}$ and $s=250$ are shown in the following table:
\begin{table}[!htb]
	\begin{center}
		\begin{tabular}{|c||c|c|}
			\hline
			\hline
			$n$& $D_{(n,1)}$ &  $D_{(n,2)}$ \\
			\hline
			$1$	&	$1$	& $1$ \\ 
			$2$	&	$0.75$	&$-1.8125$\\ 
			$3$	&	$-5.35938$	& $1.38778 \times 10^{-17} s^2$ \\ 
			$4$	&	$-0.03125 s^2$	&$0.0427909 s^2$\\ 
			$5$	&	$0.15915 s^2$	& $-0.642968 s^2$ \\ 
			$6$	&	$-2.69663 s^2$	&$11.5966 s^2$\\ 
			$7$	&	$50.7156 s^2$	& $-224.481 s^2$ \\ 
			$8$	&	$-1002.63 s^2$	&$4509.76 s^2$\\ 
			$9$	&	$20398.8 s^2$	& $-92694.3 s^2$ \\ 
			$10$	&	$-422827.0 s^2$	&$1.93498\times 10^6 s^2$\\ 
			$11$	&	$8.87966\times 10^6 s^2$	&$-4.08468\times 10^7 s^2$\\ 
			$12$	&	$-1.88292\times 10^8 s^2$	& $8.69585\times 10^8 s^22$ \\ 
			$13$	&	$4.02259\times 10^9 s^2$	&$-1.86355\times 10^{10} s^2$\\ 
			$14$	&	$-8.64469\times 10^{10} s^2$	& $4.01494\times 10^{11} s^2$ \\ 
			$15$	&	$1.86673\times 10^{12} s^2$	&$-8.68793\times 10^{12} s^2$\\ 
			$16$	&	$-8.80382\times 10^{14} s^2$	& $1.88687\times 10^{14} s^2$ \\ 
			$17$	&	$1.92062\times 10^{16} s^2$	&$-4.11067\times 10^{15} s^2$\\ 
			$18$	&	$-4.2004\times 10^{17} s^2$	& $8.97928\times 10^{16} s^2$ \\ 
			$19$	&	$9.20617\times 10^{18} s^2$	&$-1.96596\times 10^{18} s^2$\\ 
			$20$	&	$-2.0215\times10^{20} s^2$	&$4.3131\times 10^{19} s^2$\\ 
			\hline
			\hline
		\end{tabular}
	\end{center}
	\label{Table_1}
\end{table}

\bibliographystyle{utphys}

\begin{thebibliography}{10}
	
	\bibitem{Landau:fluid}
L. D.~ Landau, E. M.~ Lifshitz, 
	``Fluid Mechanics",
	 Course of Theoretical Physics, Vol. 6,
	 (Elsevier Science, 2013).

	\bibitem{Landau:stat}
L. D.~ Landau, E. M.~ Lifshitz, 
``Statistical Physics Part I",
 Course of Theoretical Physics, Vol. 5,
 Pergamon Press, Oxford (1958).

\bibitem{Crossley:2015evo}
M.~Crossley, P.~Glorioso and H.~Liu,
``Effective field theory of dissipative fluids,''
JHEP \textbf{09} (2017), 095
doi:10.1007/JHEP09(2017)095
[arXiv:1511.03646 [hep-th]].

\bibitem{Haehl:2015pja}
F.~M.~Haehl, R.~Loganayagam and M.~Rangamani,
``Adiabatic hydrodynamics: The eightfold way to dissipation,''
JHEP \textbf{05} (2015), 060
[arXiv:1502.00636 [hep-th]].

\bibitem{Jensen:2017kzi}
K.~Jensen, N.~Pinzani-Fokeeva and A.~Yarom,
``Dissipative hydrodynamics in superspace,''
JHEP \textbf{09} (2018), 127
[arXiv:1701.07436 [hep-th]].

\bibitem{Haehl:2018lcu}
F.~M.~Haehl, R.~Loganayagam and M.~Rangamani,
``Effective Action for Relativistic Hydrodynamics: Fluctuations, Dissipation, and Entropy Inflow,''
JHEP \textbf{10} (2018), 194
[arXiv:1803.11155 [hep-th]].


\bibitem{Jensen:2018hse}
K.~Jensen, R.~Marjieh, N.~Pinzani-Fokeeva and A.~Yarom,
``A panoply of Schwinger-Keldysh transport,''
SciPost Phys. \textbf{5} (2018) no.5, 053
[arXiv:1804.04654 [hep-th]].

\bibitem{Dubovsky:2011sj}
S.~Dubovsky, L.~Hui, A.~Nicolis and D.~T.~Son,
``Effective field theory for hydrodynamics: thermodynamics, and the derivative expansion,''
Phys. Rev. D \textbf{85} (2012), 085029
[arXiv:1107.0731 [hep-th]].

\bibitem{Endlich:2012vt}
S.~Endlich, A.~Nicolis, R.~A.~Porto and J.~Wang,
``Dissipation in the effective field theory for hydrodynamics: First order effects,''
Phys. Rev. D \textbf{88} (2013), 105001
[arXiv:1211.6461 [hep-th]].

\bibitem{Grozdanov:2013dba}
S.~Grozdanov and J.~Polonyi,
``Viscosity and dissipative hydrodynamics from effective field theory,''
Phys. Rev. D \textbf{91} (2015) no.10, 105031
[arXiv:1305.3670 [hep-th]].


\bibitem{Kovtun:2014hpa}
P.~Kovtun, G.~D.~Moore and P.~Romatschke,
``Towards an effective action for relativistic dissipative hydrodynamics,''
JHEP \textbf{07} (2014), 123
doi:10.1007/JHEP07(2014)123
[arXiv:1405.3967 [hep-ph]].


\bibitem{Harder:2015nxa}
M.~Harder, P.~Kovtun and A.~Ritz,
``On thermal fluctuations and the generating functional in relativistic hydrodynamics,''
JHEP \textbf{07} (2015), 025
doi:10.1007/JHEP07(2015)025
[arXiv:1502.03076 [hep-th]].



\bibitem{Glorioso:2018wxw}
H.~Liu and P.~Glorioso,
``Lectures on non-equilibrium effective field theories and fluctuating hydrodynamics,''
PoS \textbf{TASI2017} (2018), 008
[arXiv:1805.09331 [hep-th]].



\bibitem{Gao:2018bxz}
P.~Gao, P.~Glorioso and H.~Liu,
``Ghostbusters: Unitarity and Causality of Non-equilibrium Effective Field Theories,''
JHEP \textbf{03} (2020), 040
[arXiv:1803.10778 [hep-th]].



\bibitem{Chen-Lin:2018kfl}
X.~Chen-Lin, L.~V.~Delacr\'etaz and S.~A.~Hartnoll,
``Theory of diffusive fluctuations,''
Phys. Rev. Lett. \textbf{122} (2019) no.9, 091602
doi:10.1103/PhysRevLett.122.091602
[arXiv:1811.12540 [hep-th]].


\bibitem{Jain:2020fsm}
A.~Jain and P.~Kovtun,
``Non-universality of hydrodynamics,''
[arXiv:2009.01356 [hep-th]].

\bibitem{Sogabe:2021wqk}
N.~Sogabe, N.~Yamamoto and Y.~Yin,
``Positive magnetoresistance induced by hydrodynamic fluctuations in chiral media,''
[arXiv:2105.10271 [hep-th]].


\bibitem{Blake:2017ris}
M.~Blake, H.~Lee and H.~Liu,
``A quantum hydrodynamical description for scrambling and many-body chaos,''
JHEP \textbf{10} (2018), 127
doi:10.1007/JHEP10(2018)127
[arXiv:1801.00010 [hep-th]].

\bibitem{Blake:2021wqj}
M.~Blake and H.~Liu,
``On systems of maximal quantum chaos,''
[arXiv:2102.11294 [hep-th]].

\bibitem{Maldacena:2015waa} 
J.~Maldacena, S.~H.~Shenker and D.~Stanford,
``A bound on chaos,''
JHEP {\bf 1608}, 106 (2016)
[arXiv:1503.01409 [hep-th]].

\bibitem{Blake:2019otz} 
M.~Blake, R.~A.~Davison and D.~Vegh,
``Horizon constraints on holographic Green's functions,''
arXiv:1904.12883 [hep-th].

\bibitem{Grozdanov:2018kkt} 
S.~Grozdanov,
``On the connection between hydrodynamics and quantum chaos in holographic theories with stringy corrections,''
JHEP {\bf 1901}, 048 (2019)
[arXiv:1811.09641 [hep-th]].



\bibitem{Natsuume:2019sfp} 
M.~Natsuume and T.~Okamura,
``Holographic chaos, pole-skipping, and regularity,''
arXiv:1905.12014 [hep-th].


\bibitem{Natsuume:2019xcy} 
M.~Natsuume and T.~Okamura,
``Nonuniqueness of Green's functions at special points,''
arXiv:1905.12015 [hep-th].

\bibitem{Natsuume:2019vcv} 
M.~Natsuume and T.~Okamura,
``Pole-skipping with finite-coupling corrections,''
arXiv:1909.09168 [hep-th].


\bibitem{Wu:2019esr} 
X.~Wu,
``Higher curvature corrections to pole-skipping,''
arXiv:1909.10223 [hep-th].


\bibitem{Ahn:2019rnq} 
Y.~Ahn, V.~Jahnke, H.~S.~Jeong and K.~Y.~Kim,
``Scrambling in Hyperbolic Black Holes: shock waves and pole-skipping,''
arXiv:1907.08030 [hep-th].


\bibitem{Li:2019bgc} 
W.~Li, S.~Lin and J.~Mei,
``Thermal diffusion and quantum chaos in neutral magnetized plasma,''
Phys.\ Rev.\ D {\bf 100}, no. 4, 046012 (2019)
[arXiv:1905.07684 [hep-th]].



\bibitem{Ceplak:2019ymw} 
N.~Ceplak, K.~Ramdial and D.~Vegh,
``Fermionic pole-skipping in holography,''
arXiv:1910.02975 [hep-th].



\bibitem{Das:2019tga} 
S.~Das, B.~Ezhuthachan and A.~Kundu,
``Real Time Dynamics in Low Point Correlators,''
arXiv:1907.08763 [hep-th].

\bibitem{Abbasi:2019rhy}
N.~Abbasi and J.~Tabatabaei,
``Quantum chaos, pole-skipping and hydrodynamics in a holographic system with chiral anomaly,''
JHEP \textbf{03}, 050 (2020)
[arXiv:1910.13696 [hep-th]].


\bibitem{Liu:2020yaf}
Y.~Liu and A.~Raju,
``Quantum Chaos in Topologically Massive Gravity,''
[arXiv:2005.08508 [hep-th]].


\bibitem{Ahn:2020bks}
Y.~Ahn, V.~Jahnke, H.~S.~Jeong, K.~Y.~Kim, K.~S.~Lee and M.~Nishida,
``Pole-skipping of scalar and vector fields in hyperbolic space: conformal blocks and holography,''
[arXiv:2006.00974 [hep-th]].

\bibitem{Ahn:2020baf}
Y.~Ahn, V.~Jahnke, H.~S.~Jeong, K.~S.~Lee, M.~Nishida and K.~Y.~Kim,
``Classifying pole-skipping points,''
JHEP \textbf{03} (2021), 175
[arXiv:2010.16166 [hep-th]].

\bibitem{Kim:2020url}
K.~Y.~Kim, K.~S.~Lee and M.~Nishida,
``Holographic scalar and vector exchange in OTOCs and pole-skipping phenomena,''
JHEP \textbf{04} (2021), 092
[erratum: JHEP \textbf{04} (2021), 229]
[arXiv:2011.13716 [hep-th]].

\bibitem{Sil:2020jhr}
K.~Sil,
``Pole skipping and chaos in anisotropic plasma: a holographic study,''
JHEP \textbf{03} (2021), 232
[arXiv:2012.07710 [hep-th]].

\bibitem{Yuan:2020fvv}
H.~Yuan and X.~H.~Ge,
``Pole-skipping and hydrodynamic analysis in Lifshitz, AdS$_{2}$ and Rindler geometries,''
JHEP \textbf{06} (2021), 165
[arXiv:2012.15396 [hep-th]].

\bibitem{Abbasi:2020xli}
N.~Abbasi and M.~Kaminski,
``Constraints on quasinormal modes and bounds for critical points from pole-skipping,''
JHEP \textbf{03} (2021), 265
[arXiv:2012.15820 [hep-th]].

\bibitem{Ceplak:2021efc}
N.~Ceplak and D.~Vegh,
``Pole-skipping and Rarita-Schwinger fields,''
Phys. Rev. D \textbf{103} (2021) no.10, 106009
[arXiv:2101.01490 [hep-th]].

\bibitem{Jeong:2021zhz}
H.~S.~Jeong, K.~Y.~Kim and Y.~W.~Sun,
``Bound of diffusion constants from pole-skipping points: spontaneous symmetry breaking and magnetic field,''
JHEP \textbf{07} (2021), 105
[arXiv:2104.13084 [hep-th]].

\bibitem{Yuan:2021ets}
H.~Yuan and X.~H.~Ge,
``Analogue of the pole-skipping phenomenon in acoustic black holes,''
[arXiv:2110.08074 [hep-th]].

\bibitem{Blake:2021hjj}
M.~Blake and R.~A.~Davison,
``Chaos and pole-skipping in rotating black holes,''
[arXiv:2111.11093 [hep-th]].

\bibitem{Kim:2021xdz}
K.~Y.~Kim, K.~S.~Lee and M.~Nishida,
``Construction of bulk solutions for towers of pole-skipping points,''
[arXiv:2112.11662 [hep-th]].

\bibitem{Haehl:2018izb} 
F.~M.~Haehl and M.~Rozali,
``Effective Field Theory for Chaotic CFTs,''
JHEP {\bf 1810}, 118 (2018)
[arXiv:1808.02898 [hep-th]].

\bibitem{Ramirez:2020qer}
D.~M.~Ramirez,
``Chaos and pole skipping in CFT$_{2}$,''
JHEP \textbf{12} (2021), 006
[arXiv:2009.00500 [hep-th]].



\bibitem{Jain:2020hcu}
A.~Jain, P.~Kovtun, A.~Ritz and A.~Shukla,
``Hydrodynamic effective field theory and the analyticity of hydrostatic correlators,''
JHEP \textbf{02} (2021), 200
[arXiv:2011.03691 [hep-th]].





\bibitem{Chao:2020kcf}
J.~Chao and T.~Schaefer,
``Multiplicative noise and the diffusion of conserved densities,''
JHEP \textbf{01} (2021), 071
[arXiv:2008.01269 [hep-th]].

\bibitem{Sogabe:2021svv}
N.~Sogabe and Y.~Yin,
``Off-equilibrium non-Gaussian fluctuations near the QCD critical point: an effective field theory perspective,''
[arXiv:2111.14667 [nucl-th]].

\bibitem{Baggioli:2020haa}
M.~Baggioli and M.~Landry,
``Effective Field Theory for Quasicrystals and Phasons Dynamics,''
SciPost Phys. \textbf{9} (2020) no.5, 062
[arXiv:2008.05339 [hep-th]].

\bibitem{Delacretaz:2021qqu}
L.~V.~Delacr\'etaz, B.~Gout\'eraux and V.~Ziogas,
``Damping of Pseudo-Goldstone Fields,''
[arXiv:2111.13459 [hep-th]].

\bibitem{Heller:2013fn}
M.~P.~Heller, R.~A.~Janik and P.~Witaszczyk,
``Hydrodynamic Gradient Expansion in Gauge Theory Plasmas,''
Phys. Rev. Lett. \textbf{110}, no.21, 211602 (2013)
[arXiv:1302.0697 [hep-th]].

\bibitem{Withers:2018srf}
B.~Withers,
``Short-lived modes from hydrodynamic dispersion relations,''
JHEP \textbf{06}, 059 (2018)
[arXiv:1803.08058 [hep-th]].


\bibitem{Grozdanov:2019kge} 
S.~Grozdanov, P.~K.~Kovtun, A.~O.~Starinets and P.~Tadić,
``Convergence of the Gradient Expansion in Hydrodynamics,''
Phys.\ Rev.\ Lett.\  {\bf 122}, no. 25, 251601 (2019)
[arXiv:1904.01018 [hep-th]].

\bibitem{Grozdanov:2019uhi} 
S.~Grozdanov, P.~K.~Kovtun, A.~O.~Starinets and P.~Tadić,
``The complex life of hydrodynamic modes,''
JHEP {\bf 1911}, 097 (2019)
[arXiv:1904.12862 [hep-th]].

\bibitem{Kovtun:2003vj}
P.~Kovtun and L.~G.~Yaffe,
``Hydrodynamic fluctuations, long time tails, and supersymmetry,''
Phys. Rev. D \textbf{68} (2003), 025007
[arXiv:hep-th/0303010 [hep-th]].

\bibitem{Kovtun:2011np}
P.~Kovtun, G.~D.~Moore and P.~Romatschke,
``The stickiness of sound: An absolute lower limit on viscosity and the breakdown of second order relativistic hydrodynamics,''
Phys. Rev. D \textbf{84} (2011), 025006
[arXiv:1104.1586 [hep-ph]].

\bibitem{Kovtun:2012rj}
P.~Kovtun,
``Lectures on hydrodynamic fluctuations in relativistic theories,''
J. Phys. A \textbf{45} (2012), 473001
doi:10.1088/1751-8113/45/47/473001
[arXiv:1205.5040 [hep-th]].

\bibitem{Shukla:2021ksb}
A.~Shukla,
``Hydrodynamic fluctuations and long-time tails in a fluid on an anisotropic background,''
Nucl. Phys. B \textbf{968} (2021), 115442
[arXiv:2101.10000 [hep-th]].



\bibitem{Delacretaz:2020nit}
L.~V.~Delacretaz,
``Heavy Operators and Hydrodynamic Tails,''
SciPost Phys. \textbf{9} (2020) no.3, 034
[arXiv:2006.01139 [hep-th]].


\bibitem{Gu:2016oyy}
Y.~Gu, X.~L.~Qi and D.~Stanford,
``Local criticality, diffusion and chaos in generalized Sachdev-Ye-Kitaev models,''
JHEP \textbf{05} (2017), 125
[arXiv:1609.07832 [hep-th]].



\bibitem{Herzog:2007ij}
C.~P.~Herzog, P.~Kovtun, S.~Sachdev and D.~T.~Son,
``Quantum critical transport, duality, and M-theory,''
Phys. Rev. D \textbf{75} (2007), 085020
[arXiv:hep-th/0701036 [hep-th]].




\bibitem{Choi:2020tdj}
C.~Choi, M.~Mezei and G.~S\'arosi,
``Pole skipping away from maximal chaos,''
[arXiv:2010.08558 [hep-th]].


\bibitem{Kitaev}
A. Kitaev, 
``A simple model of quantum holography."
http://online.kitp.ucsb.edu/online/entangled15/kitaev/,http:
//online.kitp.ucsb.edu/online/entangled15/kitaev2/. Talks at KITP, April
7, 2015 and May 27, 2015.


\bibitem{Sachdev}
S. Sachdev and J.-w. Ye, 
``Gapless spin liquid ground state in a random, quantum Heisenberg magnet," 
Phys. Rev. Lett. 70 (1993) 3339, arXiv:cond-mat/9212030
[cond-mat].



\bibitem{Maldacena:2016hyu}
J.~Maldacena and D.~Stanford,
``Remarks on the Sachdev-Ye-Kitaev model,''
Phys. Rev. D \textbf{94} (2016) no.10, 106002
[arXiv:1604.07818 [hep-th]].



\bibitem{Jensen:2016pah}
K.~Jensen,
``Chaos in AdS$_2$ Holography,''
Phys. Rev. Lett. \textbf{117} (2016) no.11, 111601
[arXiv:1605.06098 [hep-th]].

\bibitem{Maldacena:2016upp}
J.~Maldacena, D.~Stanford and Z.~Yang,
``Conformal symmetry and its breaking in two dimensional Nearly Anti-de-Sitter space,''
PTEP \textbf{2016} (2016) no.12, 12C104
[arXiv:1606.01857 [hep-th]].

\bibitem{Wang:1998wg}
E.~Wang and U.~W.~Heinz,
``A Generalized fluctuation dissipation theorem for nonlinear response functions,''
Phys. Rev. D \textbf{66} (2002), 025008
[arXiv:hep-th/9809016 [hep-th]].



\bibitem{Delacretaz:2020jis}
L.~V.~Delacretaz and P.~Glorioso,
``Breakdown of Diffusion on Chiral Edges,''
Phys. Rev. Lett. \textbf{124} (2020) no.23, 236802
[arXiv:2002.08365 [cond-mat.str-el]].

\bibitem{Glorioso:2021bif}
P.~Glorioso, J.~Guo, J.~F.~Rodriguez-Nieva and A.~Lucas,
``Breakdown of hydrodynamics below four dimensions in a fracton fluid,''
[arXiv:2105.13365 [cond-mat.str-el]].



\bibitem{Abbasi:2020ykq}
N.~Abbasi and S.~Tahery,
``Complexified quasinormal modes and the pole-skipping in a holographic system at finite chemical potential,''
JHEP \textbf{10} (2020), 076
[arXiv:2007.10024 [hep-th]].

\bibitem{Jansen:2020hfd}
A.~Jansen and C.~Pantelidou,
``Quasinormal modes in charged fluids at complex momentum,''
JHEP \textbf{10} (2020), 121
[arXiv:2007.14418 [hep-th]].

\bibitem{Baggioli:2020loj}
M.~Baggioli,
``How small hydrodynamics can go,''
Phys. Rev. D \textbf{103} (2021) no.8, 086001
[arXiv:2010.05916 [hep-th]].

\bibitem{Arean:2020eus}
D.~Arean, R.~A.~Davison, B.~Gout\'eraux and K.~Suzuki,
``Hydrodynamic Diffusion and Its Breakdown near AdS2 Quantum Critical Points,''
Phys. Rev. X \textbf{11} (2021) no.3, 031024
[arXiv:2011.12301 [hep-th]].



\bibitem{Heller:2020uuy}
M.~P.~Heller, A.~Serantes, M.~Spali\'nski, V.~Svensson and B.~Withers,
``Hydrodynamic gradient expansion in linear response theory,''
Phys. Rev. D \textbf{104}, no.6, 066002 (2021)
[arXiv:2007.05524 [hep-th]].



\bibitem{Asadi:2021hds}
M.~Asadi, H.~Soltanpanahi and F.~Taghinavaz,
``Critical behaviour of hydrodynamic series,''
JHEP \textbf{05} (2021), 287
[arXiv:2102.03584 [hep-th]].

\bibitem{Baggioli:2021ujk}
M.~Baggioli, U.~Gran and M.~Torns\"o,
``Collective modes of polarizable holographic media in magnetic fields,''
JHEP \textbf{06} (2021), 014
[arXiv:2102.09969 [hep-th]].



\bibitem{Wu:2021mkk}
N.~Wu, M.~Baggioli and W.~J.~Li,
``On the universality of AdS$_{2}$ diffusion bounds and the breakdown of linearized hydrodynamics,''
JHEP \textbf{05} (2021), 014
[arXiv:2102.05810 [hep-th]].



\bibitem{Grozdanov:2021gzh}
S.~Grozdanov, A.~O.~Starinets and P.~Tadi\'c,
``Hydrodynamic dispersion relations at finite coupling,''
JHEP \textbf{06} (2021), 180
[arXiv:2104.11035 [hep-th]].



\bibitem{Heller:2021oxl}
M.~P.~Heller, A.~Serantes, M.~Spali\'nski, V.~Svensson and B.~Withers,
``The hydrodynamic gradient expansion diverges beyond Bjorken flow,''
[arXiv:2110.07621 [hep-th]].

\bibitem{Jeong:2021zsv}
H.~S.~Jeong, K.~Y.~Kim and Y.~W.~Sun,
``The breakdown of magneto-hydrodynamics near AdS$_2$ fixed point and energy diffusion bound,''
[arXiv:2105.03882 [hep-th]].

\bibitem{Huh:2021ppg}
K.~B.~Huh, H.~S.~Jeong, K.~Y.~Kim and Y.~W.~Sun,
``Upper bound of the charge diffusion constant in holography,''
[arXiv:2111.07515 [hep-th]].


\bibitem{Liu:2021qmt}
Y.~Liu and X.~M.~Wu,
``Breakdown of hydrodynamics from holographic pole collision,''
[arXiv:2111.07770 [hep-th]].

\bibitem{Cartwright:2021qpp}
C.~Cartwright, M.~G.~Amano, M.~Kaminski, J.~Noronha and E.~Speranza,
``Convergence of hydrodynamics in rapidly spinning strongly coupled plasma,''
[arXiv:2112.10781 [hep-th]].

\bibitem{Amado:2008ji}
I.~Amado, C.~Hoyos-Badajoz, K.~Landsteiner and S.~Montero,
``Hydrodynamics and beyond in the strongly coupled N=4 plasma,''
JHEP \textbf{07} (2008), 133
[arXiv:0805.2570 [hep-th]].

\bibitem{Akamatsu:2016llw}
Y.~Akamatsu, A.~Mazeliauskas and D.~Teaney,
``A kinetic regime of hydrodynamic fluctuations and long time tails for a Bjorken expansion,''
Phys. Rev. C \textbf{95} (2017) no.1, 014909
[arXiv:1606.07742 [nucl-th]].

\bibitem{Caron-Huot:2009kyg}
S.~Caron-Huot and O.~Saremi,
``Hydrodynamic Long-Time tails From Anti de Sitter Space,''
JHEP \textbf{11} (2010), 013
[arXiv:0909.4525 [hep-th]].

\bibitem{Glorioso:2018mmw}
P.~Glorioso, M.~Crossley and H.~Liu,
``A prescription for holographic Schwinger-Keldysh contour in non-equilibrium systems,''
[arXiv:1812.08785 [hep-th]].

\bibitem{Cheng:2021mop}
G.~Cheng and B.~Swingle,
``Scrambling with conservation laws,''
JHEP \textbf{11} (2021), 174
[arXiv:2103.07624 [cond-mat.stat-mech]].

\bibitem{deBoer:2018qqm}
J.~de Boer, M.~P.~Heller and N.~Pinzani-Fokeeva,
``Holographic Schwinger-Keldysh effective field theories,''
JHEP \textbf{05} (2019), 188
doi:10.1007/JHEP05(2019)188
[arXiv:1812.06093 [hep-th]].


\bibitem{Skenderis:2008dg}
K.~Skenderis and B.~C.~van Rees,
``Real-time gauge/gravity duality: Prescription, Renormalization and Examples,''
JHEP \textbf{05} (2009), 085
doi:10.1088/1126-6708/2009/05/085
[arXiv:0812.2909 [hep-th]].

\bibitem{Ghosh:2020lel}
J.~K.~Ghosh, R.~Loganayagam, S.~G.~Prabhu, M.~Rangamani, A.~Sivakumar and V.~Vishal,
``Effective field theory of stochastic diffusion from gravity,''
JHEP \textbf{05} (2021), 130
[arXiv:2012.03999 [hep-th]].



\bibitem{Bu:2021clf}
Y.~Bu, M.~Fujita and S.~Lin,
``Ginzburg-Landau effective action for a fluctuating holographic superconductor,''
[arXiv:2106.00556 [hep-th]].

\bibitem{Bu:2020jfo}
Y.~Bu, T.~Demircik and M.~Lublinsky,
``All order effective action for charge diffusion from Schwinger-Keldysh holography,''
JHEP \textbf{05} (2021), 187
[arXiv:2012.08362 [hep-th]].












\end{thebibliography}
\providecommand{\href}[2]{#2}\begingroup\raggedright\endgroup

\end{document}